# Superconductivity

## The Structure Scale of the Universe

## (Elastic Resonant Symmetric Medium by Self-Energy)

Richard D. Saam

525 Louisiana Ave

Corpus Christi, Texas 78404 USA

e-mail: rdsaam@att.net

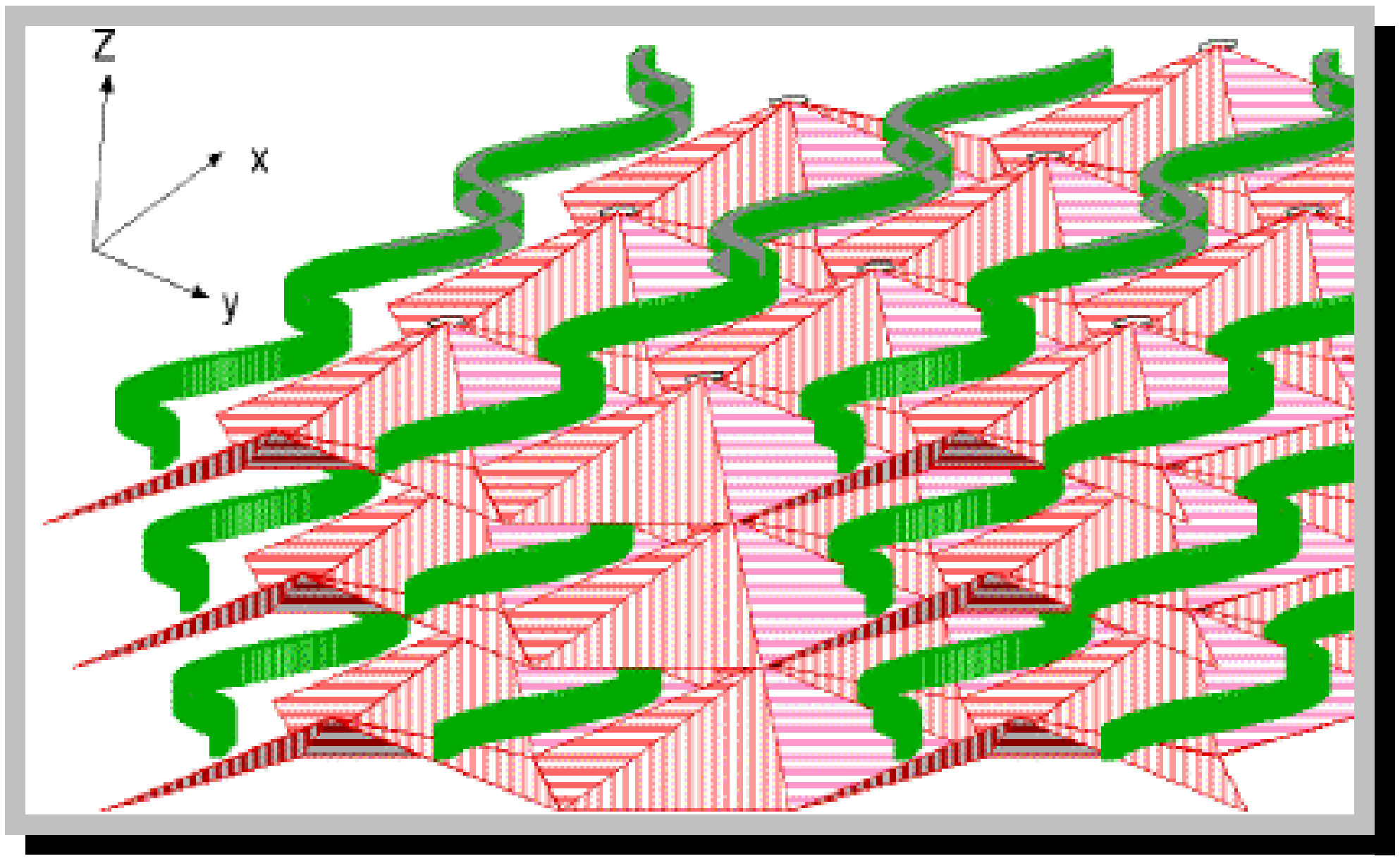

## ABSTRACT

A dimensional analysis correlation framework supported by reported experimental evidence (Homes, Harshman along with Voyager and EGRET space platforms and others) is presented which indicates that superconductivity is a self-energy phenomenon and congruent with the concept of the Charge Conjugation, Parity Change and Time Reversal (CPT) theorem (discrete Noether's theorem). A resonant structure is proposed as an extension of Bardeen Cooper and Schrieffer (BCS) theory, which suspends Lorentz transforms at superluminal velocities in the context of the de Broglie hypothesis. A momentum and energy conserving (elastic) CPT resonant structural lattice scalable over 15 orders of magnitude from nuclear to universe dimensions and associated superconducting theory is postulated whereby nuclear (quark) weak and strong forces, electromagnetic and gravitational forces are mediated by a particle of resonant transformed mass ($m_t$) (110.12275343 x electron mass or $56.2726/c^2$ Mev) and unit charge congruent with Heisenberg Uncertainty, such that the electron and proton mass and charge are maintained, nuclear density is maintained at 2.34E14 $g/cm^3$, proton charge radius is maintained at 8.75E-14 cm, the universe radius is 2.25E28 cm, the universe mass is 3.02E56 gram, the universe density is 6.38E-30 $g/cm^3$ or 2/3 the critical density and the universe escape velocity is c. Standard model up, down, strange, charm, bottom and top particles are correlated to the developed model at the nuclear scale. The universe time or age is 1.37E10 years and the universe Hubble constant is 2.31E-18/sec (71.2 km/sec-million parsec), which is gravitationally related to the proton charge radius by Planck's constant. The calculated universe mass and density are based on an isotropic homogeneous media filling the vacuum of space analogous to the 'aether' referred to in the $19^{th}$ century (but still in conformance with Einstein's relativity theory and extended to nuclear dimensions) and could be considered a candidate for the 'dark matter/energy' in present universe theories. Each particle of mass $m_t$ in the proposed dark matter is contained in a volume of 15733 $cm^3$. In this context the universe cosmic microwave background radiation (CMBR) black body temperature is linked to universe dark matter/energy superconducting temperature. The model predicts an deceleration value with observed universe expansion and consistent with observed Pioneer 10 &11 deep space translational and rotational deceleration and consistent with the notion that:

**An object moving through momentum space will slow down.**

This is evidenced by 56 Mev space sources as detected by Voyager and The Compton Gamma Ray Observatory Mission (EGRET) spacecraft. Also, a reasonable value for the cosmological constant is derived having dimensions of the known universe. Also, a 93 K superconductor should loose .04% of its weight while 100% in superconducting mode, which is close to, reported results by Podkletnov and Nieminen (.05%). Also this trisine model predicts 1, 2 and 3 dimensional superconductors, which has been verified by observed magnesium diboride ($MgB_2$) superconductor critical data. Also dimensional guidelines are provided for design of room temperature superconductors and beyond and which are related to practical goals such as fabricating a superconductor with the energy content equivalent to the combustion energy of gasoline. These dimensional guidelines have been experimentally verified by Homes' Law and generally fit the parameters of a superconductor operating in the "dirty" limit. Confirmation of the trisine approach is represented by correlation to Koide Lepton relation at nuclear dimensions. Also, the concept of critical volume is introduced to study characteristics associated with CPT symmetry Critical Optical Volume Energy (COVE). A $200,000,000 - 10 year experimental plan is proposed to study this concept.

## PREVIOUS PUBLICATION

**Table of Contents**

# 1. Introduction

The trisine structure is a geometrical lattice model for the superconductivity phenomena based on Gaussian surfaces (within the context of Maxwell Displacement (D) concept) encompassing superconducting Cooper CPT Charge conjugated pairs (discrete Noether's theorem entities). This Gaussian surface has the same configuration as a particular matrix geometry as defined in reference [1], and essentially consists of mirror image non-parallel surface pairs. Originally, the main purpose of the engineered lattice [1] was to control the flow of particles suspended in a fluid stream and generally has been fabricated on a macroscopic scale to remove particles in industrial fluid streams on the order of $m^3$/sec. In the particular configuration discussed in this report, a more generalized conceptual lattice is described wherein there is a perfect elastic resonant character to fluid and particle flow. In other words there is 100 percent conservation of energy and momentum (elastic or resonant condition), which by definition we will assume to describe the phenomenon of superconductivity.

The superconducting model presented herein is a logical translation of this geometry [1] in terms of classical and quantum theory to provide a basis for explaining aspects of the superconducting phenomenon and integration of this phenomenon with nuclear, electromagnetic and gravitational forces within the context of references[1-80].

This approach is an attempt to articulate a geometrical model for superconductivity in order to anticipate dimensional regimes in which one could look for higher performance materials. This approach does not address specific particles such as polarons, exitons, magnons, plasmons that may be necessary for the expression of superconductivity in a given material. This is similar to acknowledging the existence of fundamental standard model nuclear particles such as quarks and gluons but only being able to observe resulting unstable (inherently inelastic) particles such as pions. Indeed, the dimensionally correct (mass, length, time) Trisine structure or lattice as presented in this report may be expressed in terms of something very profound and elementary and that is the Charge Conjugate Parity Change Time Reversal (CPT) Theorem as a basis for creating resonant structures. The validity of the Trisine model is related to its dimensional (mass, length, time) correctness or numerical consistency with fundamental constants ($\hbar$, $G$, $c$ & $k_b$). Model numerical coherency is maintained at less than 1 part in 10,000 on a spreadsheet with name plate calculation precision at 15 significant digits.

Superconductivity or resonance is keyed to a critical temperature$(T_c)$, which is representative of the energy at which the superconductive resonant property takes place in a material or medium. The material or medium that is characterized as a superconductor has this property at all temperatures below the critical temperature.

In terms of the trisine model, any critical temperature can be inserted, but for the purposes of this report, six critical temperatures $(T_c)$ are selected as follows:

1.1 $T_c$ 8.95E+13 $^oK$ $T_b$ 9.07E14 $^oK$

The upper bounds of the superconductive phenomenon as dictated by the proton density 2.34E14 g/cm$^3$ and radius B of 6.65E-14 cm (8.750E-14 cm NIST reported charge radius value) which is defined at a critical temperature $T_c$ of

$$m_t \hbar c^3/(e^2 k_b)$$

which is a consequence of a superluminal condition:

$$v_{dx} > c$$

which is justifiable in elastic resonant condition (see equation 2.1.6 and explanation) and

$$\hbar c/e^2 = m_r/m_t = v_{dx}^2/(2c^2) = \text{fine structure constant}$$

Which is consistent with the proton ($m_p$) state

$$m_p c^2 = \hbar K_B c = \hbar(\pi/B)c$$

composed of standard model up quarks (2) and down quark (1) with respective masses of

$$(2/3)(m_e/m_t)(\hbar K_B c) \text{ and } (1/3)(m_e/m_t)(\hbar K_B c)$$

(5.64 Mev and 2.82 Mev).

Also the nuclear magneton identity exists as

$$(2/5)e\hbar/(2m_p c) = 2.785(g_s^2/\cos(\theta))e\hbar/(2m_t v_{dx})$$

with the ratio 2/5 being the sphere moment of inertia factor and the constant 2.785 being equal to the experimentally observed as the nucleus of hydrogen atom, which has a magnetic moment of 2.79 nuclear magnetons.

At this proton dimension B, gravity energy is related to Hubble constant$(H_U)$

$$hH_U = (3/2)Gm_t^2 / B_{proton}$$

This proton dimensional condition conforms to Universe conditions at hundredths of second after the Big Bang.

1.2 $T_c$ 2,135 $^oK$ $T_b$ 4.43E+09 $^oK$

The critical temperature of an anticipated superconducting material with the assumed density (6.39 g/cm$^3$) [17] of $YBa_2Cu_3O_{7-x}$ which would express gravitational effects to the extent that it would be weightless in earth's gravitational field (See Table 2.9.1).

1.3 $T_c$ 1190 $^oK$ $T_b$ 3.31E+09 $^oK$

The critical temperature of an anticipated material, which would have an energy content equivalent to the combustion energy content of gasoline(see Table 2.13.1).

1.4 $T_c$ 447 $^oK$ $T_b$ 2.03E+09 $^oK$

The critical temperature of an anticipated room temperature superconductor, which is calculated to be 1.5 x 298 degree Kelvin such that the superconducting material may have substantial superconducting properties at an operating temperature of 298 degree Kelvin.

1.5 $T_c$ 93 $^oK$ $T_b$ 9.24E+08 $^oK$

The critical temperature of the $YBa_2Cu_3O_{7-x}$

superconductor as discovered by Paul Chu at the University of Houston and which was the basis for the gravitational shielding effect of .05% as observed by Podkletnov and Nieminen(see Table 2.9.1).

1.6 $T_c$ 8.11E-16 $^oK$ $T_b$ 2.729 $^oK$

The background temperature of the Universe represented by the black body cosmic microwave background radiation (CMBR) of 2.728 ± .004 $^oK$ as detected in all directions of space in 60 - 630 GHz frequency range by the FIRAS instrument in the Cobe satellite and 2.730 ± .014 $^oK$ as measured at 10.7 GHz from a balloon platform [26]. The most recent NASA estimation of CMBR temperature is 2.725 ± .002 K. The particular temperature of 2.729 $^oK$ indicated above which is essentially equal to the CMBR temperature was established by the trisine model wherein the superconducting Cooper CPT Charge conjugated pair density (see table 2.9.1) (6.38E-30 g/cm$^3$) equals the density of the universe (see equation 2.11.5 and 2.11.16). Also, this temperature is congruent with an reported interstellar magnetic field of 2E-6 gauss [78] (see table 2.6.2].

As a general comment, a superconductor resonator should be operated (or destroyed) at about 2/3 of $(T_c)$ for maximum effect of most properties, which are usually expressed as when extrapolated to $T = 0$ (see section 2.12). In this light, critical temperatures $(T_c)$ in items 1.4 & 1.5 could be operated (or destroyed) up to 2/3 of the indicated temperature (793 and 1423 degree Kelvin respectively). Also, the present development is for one-dimensional superconductivity. Consideration should be given to 2 and 3 dimensional superconductivity by 2 or 3 multiplier to $T_c$'s contained herein interpret experimental results.

## 2. Trisine Model Development

In this report, dimensionally (mass, length and time) mathematical relationships which link trisine geometry and superconducting theory are developed and then numerical values are presented in accordance with these relationships. Centimeter, gram and second (cgs) as well as Kelvin temperature units are used unless otherwise specified. The actual model is developed in spreadsheet format with all model equations computed simultaneously, iteratively and interactively as required.

The general approach is: given a particular lattice form (in this case trisine), determine lattice momentum wave vectors $K_1$, $K_2$, $K_3$ & $K_4$ so that lattice cell energy ($\sim K^2$) per volume is at a minimum. The procedure is analogous to the variational principle used in quantum mechanics. More than four wave vectors could be evaluated at once, but four is deemed sufficient.

### 2.1 Defining Model Relationships

The four defining model equations within the superconductor conventional tradition are presented as 2.1.1 - 2.1.4 below:

$$\iint d(Area) = 4\pi(\mathbb{C}\ e_\pm) \tag{2.1.1}$$

$$\frac{2m_t k_b T_c}{\hbar^2} = \frac{(KK)}{e^{trisine}-1} = \frac{e^{Euler}}{\pi}\frac{(KK)}{\sinh(trisine)} = K_B^2 \tag{2.1.2}$$

$$trisine = \frac{1}{D(\epsilon_t)\cdot V} \tag{2.1.3}$$

$$f\left(\frac{m_t}{m_e}\right)\frac{e_\pm\hbar}{2m_e v_\varepsilon} = cavity\ H_c \tag{2.1.4}$$

Equation 2.1.1 is a representation of Gauss's law with a Cooper CPT Charge conjugated pair charge contained within a bounding surface area defining a cell volume or cell *cavity*. Equation 2.1.2 and 2.1.3 define the superconducting model as developed by Bardeen, Cooper and Schrieffer in reference [2] and Kittel in reference [4].

Recognizing this approach and developing it further with a resonant concept, it is the further objective herein to select a particular *(wave vector)$^2$ or (KK)* based on trisine geometry (see figures 2.2.1-2.2.5) that fits the model relationship as indicated in equation 2.1.2 with $K_B^2$ being defined as the trisine *(wave vector)$^2$ or (KK)* associated with superconductivity of Cooper CPT Charge conjugated pairs through the lattice. This procedure establishes the trisine geometrical dimensions, then equation 2.1.4 is used to establish an effective mass function $f(m_t/m_e)$ of the particles in order for particles to flip in spin when going from trisine *cavity* to *cavity* while in thermodynamic equilibrium with critical field $(H_c)$. The implementation of this method assumes the conservation of momentum ($p$) (2.1.5) and energy (E) (2.1.6) such that:

$$\sum_{n=1,2,3,4} \Delta p_n \Delta x_n = 0 \qquad \pm K_1 \pm K_2 \rightleftarrows \pm K_3 \pm K_4 \tag{2.1.5}$$

$$\sum_{n=1,2,3,4} \Delta E_n \Delta t_n = 0 \qquad K_1K_1 + K_2K_2 = K_3K_3 + K_4K_4 + Q \tag{2.1.6}$$

Where $Q = 0$ (reversible process)

The numerical value of Q is associated with the equation 2.1.6 is associated with the non-elastic nature of a process. Q is a measure of the heat exiting the system. By stating that Q=0, the elastic nature of the system is defined. An effective mass must be introduced to maintain the system elastic character. Essentially, a reversible process or reaction is defined recognizing that momentum is a vector and energy is a scalar.

These conditions of conservation of momentum and energy provide the necessary condition of perfect elastic character, which must exist in the superconducting resonant state. In addition, the superconducting resonant state may be viewed as boiling of states $\Delta E_n, \Delta t_n, \Delta p_n, \Delta x_n$ on top of the zero point state in a coordinated manner.

In conjunction with the de Broglie hypothesis [77] providing momentum $K$ and energy $KK$ change with Lorentz transform written as

$$K\left(1/\sqrt{1-v^2/c^2}\right) = K\beta$$
$$KK\left(1/\sqrt{1-v^2/c^2}-1\right) = KK(\beta-1)$$

congruent with:

$$m\left(v^2/c^2\right)c^2\left(1/\sqrt{1-v^2/c^2}-1\right) = m\left(v^2/c^2\right)c^2(\beta-1)$$

where the energy defined as '$KK$' contains the factor $v^2/c^2$, the superluminal $(i\beta)$ and subluminal $(\beta)$ Lorentz transform conditions are allowed because in a resonant elastic condition these Lorentz transforms cancel

$$\pm K_1\beta_1 \pm K_2\beta_2 \rightleftarrows \pm K_3\beta_3 \pm K_4\beta_4$$
$$K_1K_1(\beta_1-1)+K_2K_2(\beta_2-1) = K_3K_3(\beta_3-1)+K_4K_4(\beta_4-1)$$

and

$$\pm K_1 i\beta_1 \pm K_2 i\beta_2 \rightleftarrows \pm K_3 i\beta_3 \pm K_4 i\beta_4$$
$$K_1K_1(i\beta_1-1)+K_2K_2(i\beta_2-1) = K_3K_3(i\beta_3-1)+K_4K_4(i\beta_4-1)$$

for all values of velocity $(|v|>0)$ under normalization

where $K_1 \neq K_2 \neq K_3 \neq K_4$ and

where $\pm K_1 \pm K_2 \rightleftarrows \pm K_3 \pm K_4$

and: $K_1K_1 + K_2K_2 \cong K_3K_3 + K_4K_4$

since $(\beta/\beta = i/i = c/c = \infty/\infty = 1)$.

Equations 2.1.5 and 2.1.6 are proven generally valid. In essence, conservation of momentum and energy (elastic condition) is a fundamental property in a resonant elastic state overriding Lorentz transform considerations.

Conceptually, this model is defined within the context of the CPT theorem and generalized 3 dimensional parity of Cartesian coordinates (x,y,z) as follows:

$$\begin{vmatrix} x_{11} & x_{12} & x_{13} \\ y_{11} & y_{12} & y_{13} \\ z_{11} & z_{12} & z_{13} \end{vmatrix} = -\begin{vmatrix} x_{21} & x_{22} & x_{23} \\ y_{21} & y_{22} & y_{23} \\ z_{21} & z_{22} & z_{23} \end{vmatrix} \qquad (2.1.6a)$$

Given this parity condition defined within trisine geometrical constraints (with dimensions A and B as defined in Section 2.2) which necessarily fulfills determinate identity in equation 2.1.6b.

$$\begin{vmatrix} -B & 0 & 0 \\ 0 & \frac{B}{\sqrt{3}} & -\frac{B}{\sqrt{3}} \\ -\frac{A}{2} & 0 & \frac{A}{2} \end{vmatrix} = -\begin{vmatrix} 0 & 0 & B \\ 0 & \frac{B}{\sqrt{3}} & -\frac{B}{\sqrt{3}} \\ -\frac{A}{2} & 0 & \frac{A}{2} \end{vmatrix} \qquad (2.1.6b)$$

This geometry fills space lattice with hexagonal cells, each of volume (*cavity* or $2\sqrt{3}AB^2$) and having property that lattice is equal to its reciprocal lattice. Now each cell contains a charge pair $(e_+)$ and $(e_-)$ in conjunction with cell dimensions $(A)$ and $(B)$ defining a permittivity $(\varepsilon)$ and permeability $(k_m)$ such that:

$$v^2 = c^2/(k_m\varepsilon).$$

Then v is necessarily + or – and $(+k_m+\varepsilon) = (-k_m-\varepsilon)$

Now time $(time)$ is defined as interval to traverse cell:

$$time = 2B/v$$

then $time$ is necessarily + or -

Further: given the Heisenberg Uncertainty

$$\Delta p\Delta x \geq \hbar/2 \quad \text{and} \quad \Delta E\Delta time \geq \hbar/2$$

but defined for this model which is equivalent and justified by trisine geometry:

$$\Delta p\Delta x \geq h/2 \quad \text{and} \quad \Delta E\Delta time \geq h/2$$

where: $\hbar = h/(2\pi)$

The parity geometry satisfies following condition:

$$\Delta p\Delta x = \Delta E\Delta time \geq h/2$$

where: $\Delta x = 2B, \quad \Delta v = v, \quad$ and $\quad \Delta time = time$

$m_t$ = constant and is always +

$$\Delta p = m_t\Delta v$$
$$\Delta E = m_t\Delta v^2/2 = \frac{2}{3}\frac{e_\pm^2}{B} = \frac{h}{2\ time} = kT_c \text{ (always +)}$$
$$\Delta time = \Delta x/\Delta v$$

the $(B/A)$ ratio must have a specific ratio (2.379760996) and mass$(m_t)$ a constant + value of 110.12275343 x electron mass$(m_e)$ or 56.2726/$c^2$ Mev to make this work. (see Appendix E for background derivation based on Heisenberg Uncertainty and de Broglie condition).

The trisine symmetry (as visually seen in figures 2.2.1-2.2.5) allows movement of Cooper CPT Charge conjugated pairs with center of mass wave vector equal to zero as required by superconductor theory as presented in references [2, 4].

Wave vectors$(K)$ as applied to trisine are assumed to be free particles (ones moving in the absence of any potential field) and are solutions to the one dimensional Schrödinger equation 2.1.7 as originally presented in reference [3]:

$$\frac{d^2\psi}{ds^2} + \frac{2m_t}{\hbar^2}Energy\ \psi = 0 \qquad (2.1.7)$$

The well known solution to the one dimensional Schrödinger equation is in terms of a wave function$(\psi)$ is:

$$\psi = Amplitude\ \sin\left(\left(\frac{2m_t}{\hbar^2}Energy\right)^{\frac{1}{2}} s\right) \qquad (2.1.8)$$
$$= Amplitude\ \sin(K \cdot s)$$

Wave vector solutions $(KK)$ are constrained to fixed trisine cell boundaries such that energy eigen values are established by the condition $s = S$ and $\psi = 0$ such that

$$\left(\frac{2m_t}{\hbar^2}Energy\right)^{\frac{1}{2}} S = K\ S = n\pi \qquad (2.1.9)$$

Energy eigen values are then described in terms of what is generally called the particle in the box relationship as follows:

$$Energy = \frac{n^2\hbar^2}{2m_t}\left(\frac{\pi}{S}\right)^2 \qquad (2.1.10)$$

For our model development, we assume the quantum number $n$ = 2 and this quantum number is incorporated into the wave vectors $(KK)$ as presented as follows:

$$Energy = \frac{2^2\hbar^2}{2m_t}\left(\frac{\pi}{S}\right)^2 = \frac{\hbar^2}{2m_t}\left(\frac{2\pi}{S}\right)^2 = \frac{\hbar^2}{2m_t}(KK) \qquad (2.1.11)$$

Trisine geometry is described in Figures 2.2.1 – 2.2.5 is in general characterized by resonant dimensions A and B with a characteristic ratio B/A of 2.379760996 and corresponding characteristic angle $(\theta)$ where:

$$\theta = \tan^{-1}\left(\frac{A}{B}\right) = \tan^{-1}\left(\frac{1}{\mathrm{e}}\frac{2}{\sqrt{3}}\right) = 22.80^o \qquad (2.1.12)$$

Note that 22.8 ~ 360/8 which indicates a hint to trisine resonant lattice model similarity with reported octal (8) nature of the nuclear particle numbers.

The superconducting model is described with equations 2.1.13 - 2.1.19, with variable definitions described in the remaining equations. The model again converges around a particular resonant transformed mass $(m_t)$ (110.12275343 x electron mass$(m_e)$ or 56.2726/c$^2$ Mev) and dimensional ratio$(B/A)$ of 2.379760996.

$$k_bT_c = \frac{m_t v_{dx}^2}{2} = \frac{1}{2}\hbar\omega \qquad (2.1.13a)$$

$$k_bT_c = \left\{\frac{\dfrac{\hbar^2 K_B^2}{2m_t}}{\dfrac{\hbar^2}{\dfrac{m_e m_t}{m_e+m_t}(2B)^2 + \dfrac{m_t m_p}{m_t+m_p}(A)^2}}\right\} \qquad (2.1.13b)$$

$$k_bT_c = \left\{\frac{\dfrac{\hbar^2\left(K_C^2+K_{Ds}^2\right)}{2m_t}\dfrac{1}{\mathrm{e}^{trisine}-1}}{\dfrac{\hbar^2\left(K_C^2+K_{Ds}^2\right)}{2m_t}\dfrac{\mathrm{e}^{Euler}}{\pi\sinh(trisine)}}\right\} \qquad (2.1.13c)$$

$$k_bT_c = \left\{\frac{\left(\dfrac{1}{g_s^2}\right)\dfrac{cavity}{8\pi}DE}{\dfrac{chain}{cavity}\dfrac{e_\pm^2}{\varepsilon B}}\right\} \qquad (2.1.13d)$$

$$k_bT_c = \frac{4\hbar v_{dz}}{time_\pm}\frac{1}{2\pi v_{dx}} \quad (\text{Unruh}) \qquad (2.1.13e)$$

Also, Bose Einstein Condensate criteria is satisfied as:

$$cavity^{\frac{1}{3}} = \frac{h}{\sqrt{2\pi m_t k_b T_c}} \qquad (2.1.13f)$$

Bardeen, Cooper and Schrieffer (BCS) [2] theory is additionally adhered to by equations 2.1.14 – 2.1.18 as verified by equation 2.1.13c, array equation 2 in conjunction with equations A.3, A.4, and A.5 as presented in Appendix A.

$$trisine = \frac{K_B^2+K_P^2}{K_P K_B} = \frac{1}{D(\in)\ \mathrm{V}} = -\ln\left(\frac{2}{\pi}\ \mathrm{e}^{Euler}-1\right) \qquad (2.1.14)$$

$$density\ of\ states\ D(\in_T) = \frac{3\ cavity\ m_t K_C}{4\pi^3\hbar^2} = \frac{1}{k_bT_c} \qquad (2.1.15)$$

The attractive Cooper CPT Charge conjugated pair energy$(V)$ is expressed in equation 2.1.16:

$$V = \frac{K_P K_B}{K_B^2+K_P^2}\ k_bT_c \qquad (2.1.16)$$

**Table 2.1.1 with Density of States**

| | | | | | | |
|---|---|---|---|---|---|---|
| $T_c\ (^oK)$ | 8.95E+13 | 2,135 | 1,190 | 447 | 93 | 8.11E-16 |
| $T_b\ (^oK)$ | 9.07E+14 | 4.43E+09 | 3.31E+09 | 2.03E+09 | 9.24E+08 | 2.729 |
| $V$ | 8.09E+01 | 1.46E-13 | 8.13E-14 | 3.05E-14 | 6.35E-15 | 5.54E-32 |
| $D(\in)$ | 6.11E-03 | 3.39E+12 | 6.09E+12 | 1.62E+13 | 7.79E+13 | 8.93E+30 |

The equation 2.1.13c array equation elements 1 and 2 along with equations 2.1.14, 2.1.15, 2.1.16, 2.1.17 and 2.1.18 are consistent with the BCS weak link relationship:

$$k_bT_c = \frac{2e^{Euler}}{\pi}\frac{\hbar^2}{2m_t}\left(K_C^2+K_{Ds}^2\right)e^{\frac{-1}{D(\in)V}} \qquad (2.1.17)$$

$$k_bT_c = \left(\frac{1}{g_s}\right)\left(\frac{chain}{cavity}\right)\frac{\hbar^2}{2m_t}\left(\frac{K_A^2}{\mathbb{C}}\right)e^{\frac{-1}{D(\in)V}} \qquad (2.1.18)$$

The relationship in equation 2.1.19 is based on the Bohr magneton $e\hbar/2m_e v_\varepsilon$ (which has dimensions of magnetic field x volume) with a material dielectric$(\varepsilon)$ modified speed of light$(v_\varepsilon)$ (justifiable in the context of resonant condition) and establishes the spin flip/flop as Cooper CPT Charge conjugated pairs resonate from one cavity to the next (see figure 2.3.1). The 1/2 spin factor and electron gyromagnetic factor $(g_e)$ and superconducting gyro factor $(g_s)$ are multipliers of the Bohr magneton where

$$g_s = \frac{3}{4}\frac{m_e}{m_t}\left(\frac{cavity}{(\Delta x\ \Delta y\ \Delta z)} - \frac{m_p}{m_e}\right) = \frac{K_{Dn}}{K_B}\cos(\theta) \qquad (2.1.18a)$$

as related to the Dirac Number by

$(3/4)(g_s\text{-}1)=2\pi(g_d\text{-}1)$.

Also because particles must resonate in pairs, a symmetry is created for superconducting current to advance as bosons.

$$\frac{1}{2}g_e g_s\frac{m_t}{m_e}\frac{e_\pm\hbar}{2m_e v_\varepsilon} = cavity\ H_c \qquad (2.1.19)$$

The proton to electron mass ratio $(m_p/m_e)$ is verified through the Werner Heisenberg Uncertainty Principle volume $(\Delta x\cdot\Delta y\cdot\Delta z)$ as follows in equations 2.1.20a,b,c):

$$\Delta x = \frac{\hbar}{2}\frac{1}{\Delta p_x} = \frac{\hbar}{2}\frac{1}{\Delta(\hbar K_B)} = \Delta\left(\frac{B}{2}\frac{1}{\pi}\right) \qquad (2.1.20a)$$

$$\Delta y = \frac{\hbar}{2}\frac{1}{\Delta p_y} = \frac{\hbar}{2}\frac{1}{\Delta(\hbar K_P)} = \Delta\left(\frac{3B}{2\sqrt{3}}\frac{1}{2\pi}\right) \qquad (2.1.20b)$$

$$\Delta z = \frac{\hbar}{2}\frac{1}{\Delta p_z} = \frac{\hbar}{2}\frac{1}{\Delta(\hbar K_A)} = \Delta\left(\frac{A}{2}\frac{1}{2\pi}\right) \qquad (2.1.20c)$$

$$(\Delta x \cdot \Delta y \cdot \Delta z) = \frac{1}{(2K_B)(2K_P)(2K_A)} \qquad (2.1.21)$$

$$\frac{m_p}{m_e} = \frac{3}{4}\frac{1}{g_s}\frac{cavity}{(\Delta x\ \Delta y\ \Delta z)} - \frac{m_t}{m_e} \qquad (2.1.22)$$

Equations 2.1.20 - 2.1.22 provides us with the confidence ( $\Delta x \cdot \Delta y \cdot \Delta z$ is well contained within the *cavity*) that we can proceed in a semi-classical manner outside the envelop of the uncertainty principle with the resonant transformed mass $(m_t)$ and de Broglie velocities $(v_d)$ used in this trisine superconductor model or in other words equation 2.1.23 holds.

$$\mathbb{C}\left(\frac{\hbar^2(KK)}{2m_t}\right) = \mathbb{C}\left(\frac{1}{2}m_t v_d^2\right) = \frac{(m_t v_d)^2}{m_t} \qquad (2.1.23)$$

The Meissner condition as defined by total magnetic field exclusion from trisine chain as indicated by diamagnetic susceptibility $(\mathrm{X}) = -1/4\pi$ , as per the following equation:

$$\mathrm{X} = -\frac{\mathbb{C}\ k_b T_c}{chain\ H_c^2} = -\frac{1}{4\pi} \qquad (2.1.24)$$

Equation 2.1.25 provides a representation of the resonant transformed mass $(m_t)$

$$m_r c^2 = \frac{1}{2} m_t v_d^2 \qquad where \quad |v_d| > 0 \qquad (2.1.25)$$

**Table 2.1.2 with $(m_t)$, $(m_r)$. Frequency $(2\mathrm{k_b}T_c/\hbar)$ $(\omega)$ & Wave Length $(2\pi c/\,frequency)$ or $(c \cdot time)$ table & plot.**

| | | | | | | |
|---|---|---|---|---|---|---|
| $T_c\ (^\circ K)$ | 8.95E+13 | 2,135 | 1,190 | 447 | 93 | 8.11E-16 |
| $T_b\ (^\circ K)$ | 9.07E+14 | 4.43E+09 | 3.31E+09 | 2.03E+09 | 9.24E+08 | 2.729 |
| $m_t$ | 1.00E-25 | 1.00E-25 | 1.00E-25 | 1.00E-25 | 1.00E-25 | 1.00E-25 |
| $m_r$ | 1.37E-23 | 3.28E-34 | 1.83E-34 | 6.87E-35 | 1.43E-35 | 1.25E-52 |
| $\omega$ | 2.34E+25 | 5.59E+14 | 3.12E+14 | 1.17E+14 | 2.44E+13 | 2.12E-04 |
| Wave Length | 8.04E-15 | 3.37E-04 | 6.05E-04 | 1.61E-03 | 7.74E-03 | 8.87E+14 |

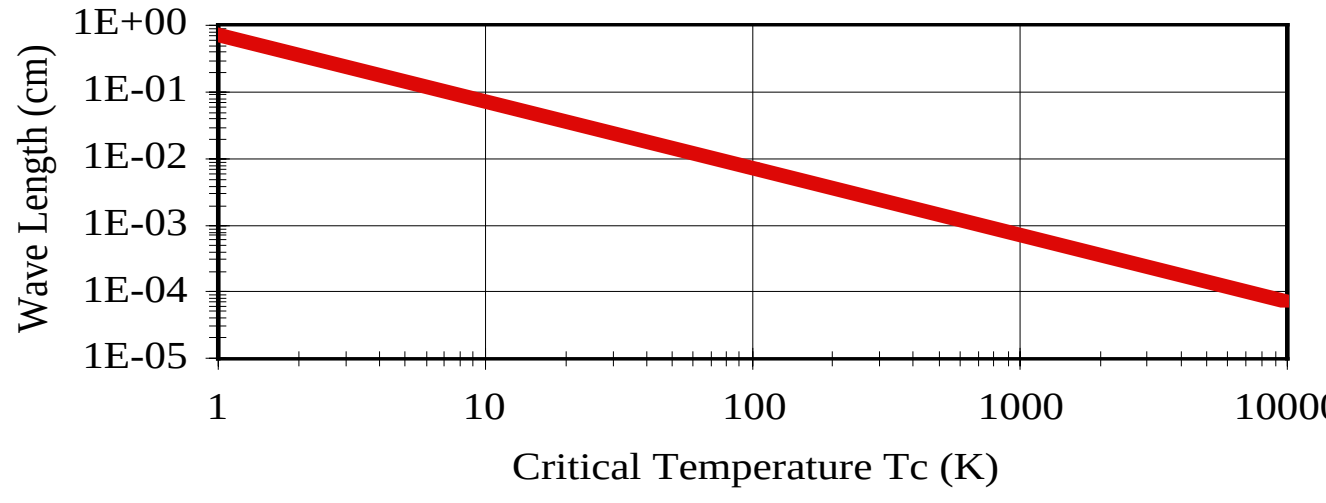

Where $(v_d)$ is the de Broglie velocity as in equation 2.1.26 and also related to resonant CPT $time_\pm$ by equation 2.1.27.

$$v_d = \frac{\hbar}{m_t}\frac{\pi}{B} \qquad (2.1.26)$$

$$time = \frac{2 \cdot B}{v_d} \qquad (2.1.27)$$

The relationship between $(v_{dx})$ and $(v_{\varepsilon x},\ v_{\varepsilon y}\ \&\ v_{\varepsilon z})$ is per the group/phase velocity relationship presented in equation 2.1.30.

$$(group\ velocity)\cdot(phase\ velocity) = c^2$$

$$\left\{\begin{array}{l} (v_{dx})\cdot\left(\frac{1}{\pi}\frac{v_{\varepsilon x}^2}{v_{dx}^2}\cdot c\right) \\ (v_{dx})\cdot\left(\frac{1}{\pi}\frac{v_{\varepsilon y}^2}{v_{dy}^2}\cdot c\right) \\ (v_{dx})\cdot\left(\frac{1}{\pi}\frac{v_{\varepsilon z}^2}{v_{dz}^2}\cdot c\right) \end{array}\right\} = \left\{\begin{array}{l} (v_{dx})\cdot\left(\frac{A}{B}\frac{1}{3^{\frac{1}{4}}}\frac{v_{\varepsilon x}^2}{v_{dx}^2}\cdot c\right) \\ (v_{dx})\cdot\left(\frac{A}{B}\frac{1}{3^{\frac{1}{4}}}\frac{v_{\varepsilon y}^2}{v_{dy}^2}\cdot c\right) \\ (v_{dx})\cdot\left(\frac{A}{B}\frac{1}{3^{\frac{1}{4}}}\frac{v_{\varepsilon z}^2}{v_{dz}^2}\cdot c\right) \end{array}\right\} = c^2 \qquad (2.1.30)$$

## 2.2 Trisine Geometry

The characteristic trisine resonant dimensions A and B are indicated as a function of temperature in table 2.2.1 as computed from the equation 2.2.1

$$k_b T_c = \frac{\hbar^2 K_B^2}{2m_t} = \frac{\hbar^2 K_A^2}{2m_t}\left(\frac{B}{A}\right)^2 \qquad (2.2.1)$$

Trisine Geometry is subject to Charge conjugation Parity change Time reversal (CPT) symmetry which is a fundamental symmetry of physical laws under transformations that involve the inversions of charge, parity and time simultaneously. The CPT symmetry provides the basis for creating particular resonant structures with this trisine geometrical character.

**Table 2.2.1 Dimensions B and A with B vs. Tc plot**

| | | | | | | |
|---|---|---|---|---|---|---|
| $T_c\ (^\circ K)$ | 8.95E+13 | 2,135 | 1,190 | 447 | 93 | 8.11E-16 |
| $T_b\ (^\circ K)$ | 9.07E+14 | 3.38E+09 | 3.31E+09 | 2.03E+09 | 9.24E+08 | 2.729 |
| $A\ (cm)$ | 2.80E-14 | 5.73E-09 | 7.67E-09 | 1.25E-08 | 2.74E-08 | 9.29 |
| $B\ (cm)$ | 6.65E-14 | 1.36E-08 | 1.82E-08 | 2.98E-08 | 6.53E-08 | 22.11 |

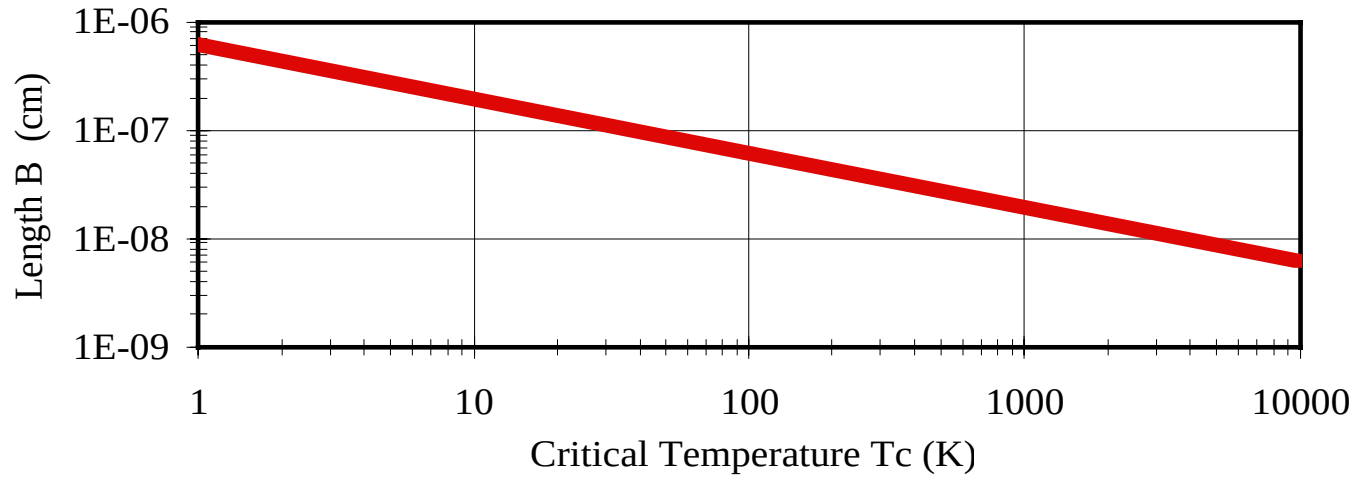

These structures have an analog in various resonant structures postulated in the field of chemistry to explain properties of delocalized $\pi$ electrons in benzene, ozone and a myriad of other molecules and incorporate the Noether theorem concepts of energy – time and momentum – length symmetry.

**Figure 2.2.1** Trisine Steric (Mirror Image) Parity Forms

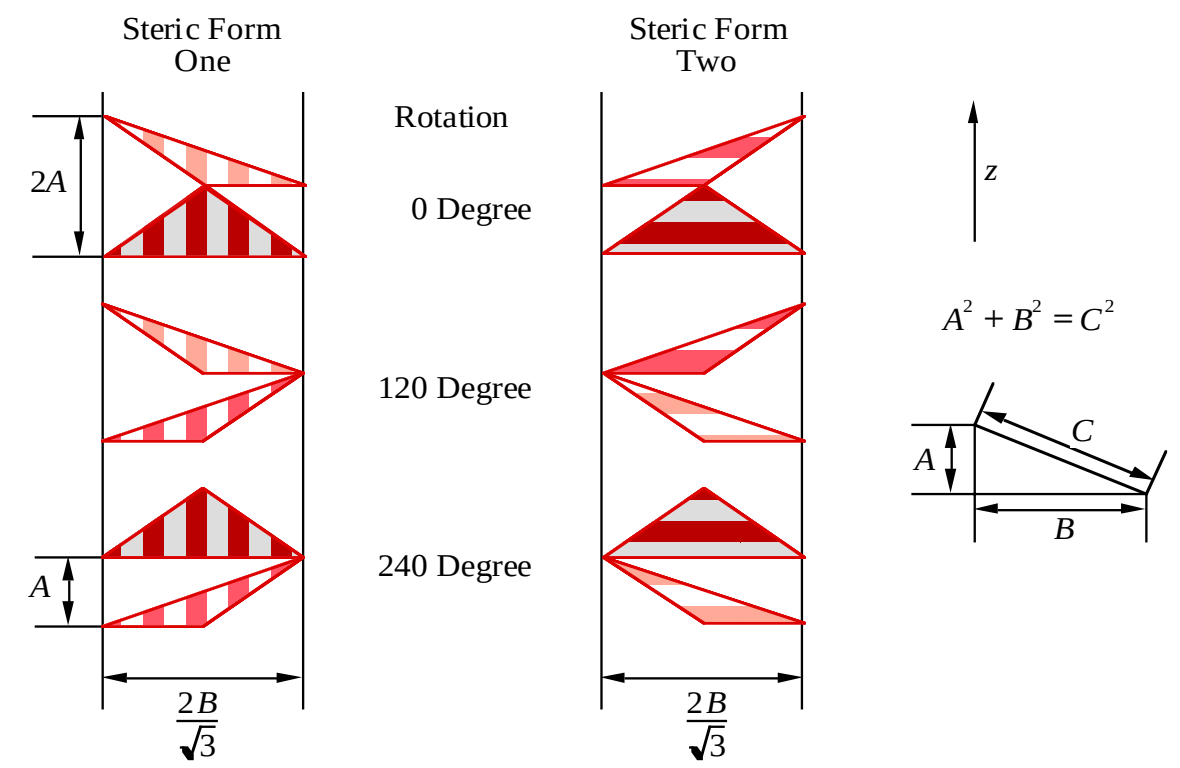

Trisine characteristic volumes with variable names *cavity* and *chain* as well as characteristic areas with variable names *section*, *approach* and *side* are defined in equations 2.2.3 - 2.2.7. See figures 2.2.1 - 2.2.4 for a visual description of these parameters. The mirror image forms in figure 2.2.1 are in conformance with parity requirement in Charge conjugation Parity change Time reversal (CPT) theorem as established by determinant identity in equation 2.1.6b and trivially obvious in cross – product expression in equation 2.2.2 in which case an orthogonal coordinate system is right-handed, or left-handed for result 2AB which is later defined as trisine *cavity side* in equation 2.2.7. Essentially, there are two mirror image *sides* to each *cavity*. At the nuclear scale, such chiral geometry is consistent with the standard model (quarks, gluons etc).

$$(2B)\times\left(\sqrt{A^2+B^2}\right) = 2B\sqrt{A^2+B^2}\sin(\theta) = 2AB \quad (2.2.2)$$

$$cavity = 2\sqrt{3}AB^2 \quad (2.2.3)$$

$$chain = \frac{2}{3}cavity \quad (2.2.4)$$

**Figure 2.2.2** Trisine Cavity and Chain Geometry from Steric (Mirror Image) Forms

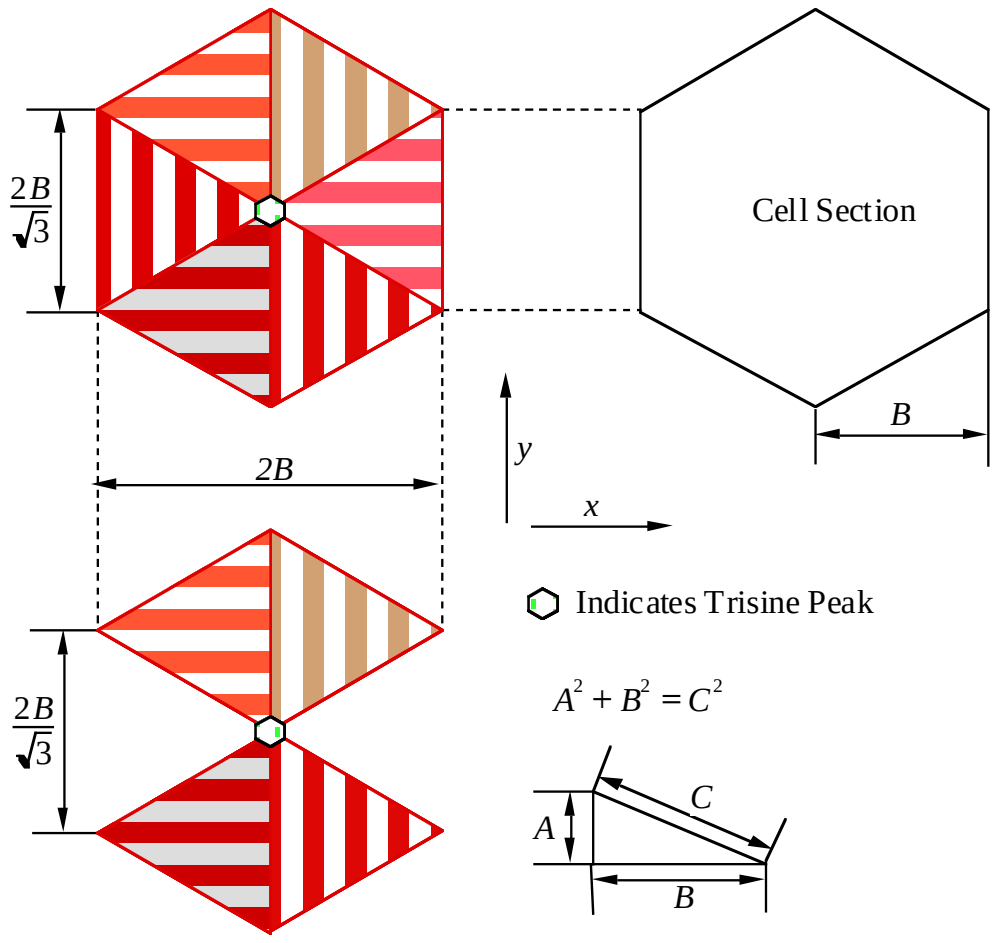

The *section* is the trisine cell *cavity* projection on to the *x*, *y* plane as indicated in figure 2.2.2.

$$section = 2\sqrt{3}B^2 \quad (2.2.5)$$

The *approach* is the trisine cell *cavity* projection on to the *y*, *z* plane as indicated in figure 2.2.3.

$$approach = \frac{1}{2}\frac{3B}{\sqrt{3}}2A = \sqrt{3}AB \quad (2.2.6)$$

**Figure 2.2.3** Trisine *approach* from both *cavity* approaches

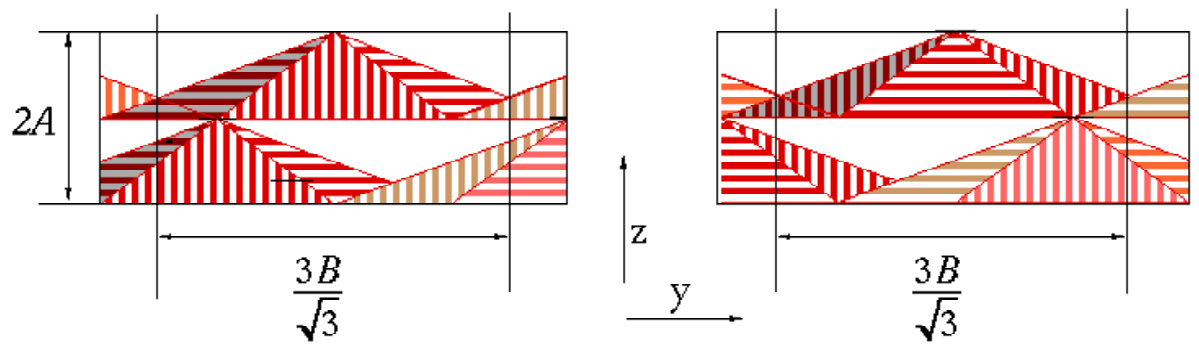

The *side* is the trisine cell *cavity* projection on to the *x*, *z* plane as indicated in figure 2.2.4.

$$side = \frac{1}{2}2A\ 2B = 2AB \quad (2.2.7)$$

**Figure 2.2.4** Trisine *side* view from both *cavity* sides

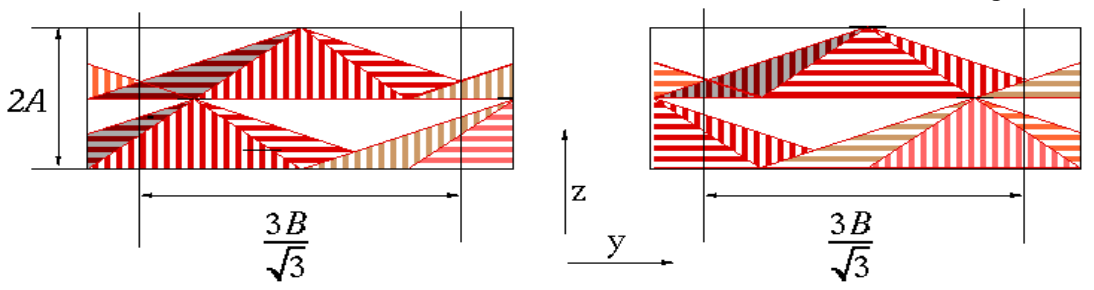

**Figure 2.2.5** Trisine Geometry

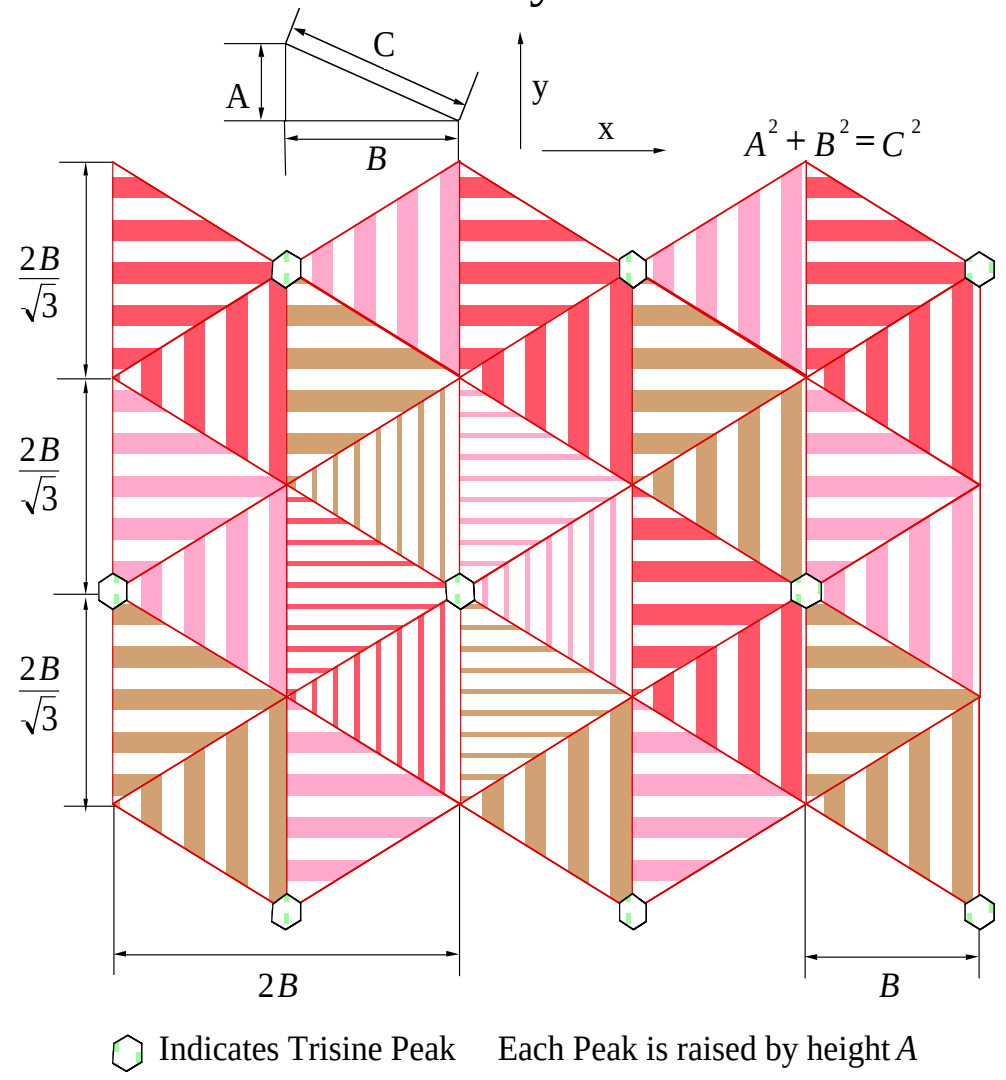

**Table 2.2.2 with cavity and section plots**

| | | | | | | |
|---|---|---|---|---|---|---|
| $T_c\,(^\circ K)$ | 8.95E+13 | 2,135 | 1,190 | 447 | 93 | 8.11E-16 |
| $T_b\,(^\circ K)$ | 9.07E+14 | 3.38E+09 | 3.31E+09 | 2.03E+09 | 9.24E+08 | 2.729 |
| *cavity* ($cm^3$) | 4.29E-40 | 3.68E-24 | 8.85E-24 | 3.84E-23 | 4.05E-22 | 15733 |
| *chain* ($cm^3$) | 2.86E-40 | 2.45E-24 | 5.90E-24 | 2.56E-23 | 2.70E-22 | 10489 |
| *section* ($cm^2$) | 1.53E-26 | 6.43E-16 | 1.15E-15 | 3.07E-15 | 1.48E-14 | 1693 |
| *approach* ($cm^2$) | 3.22E-27 | 1.35E-16 | 2.42E-16 | 6.45E-16 | 3.10E-15 | 356 |
| *side* ($cm^2$) | 3.72E-27 | 1.56E-16 | 2.80E-16 | 7.45E-16 | 3.58E-15 | 411 |

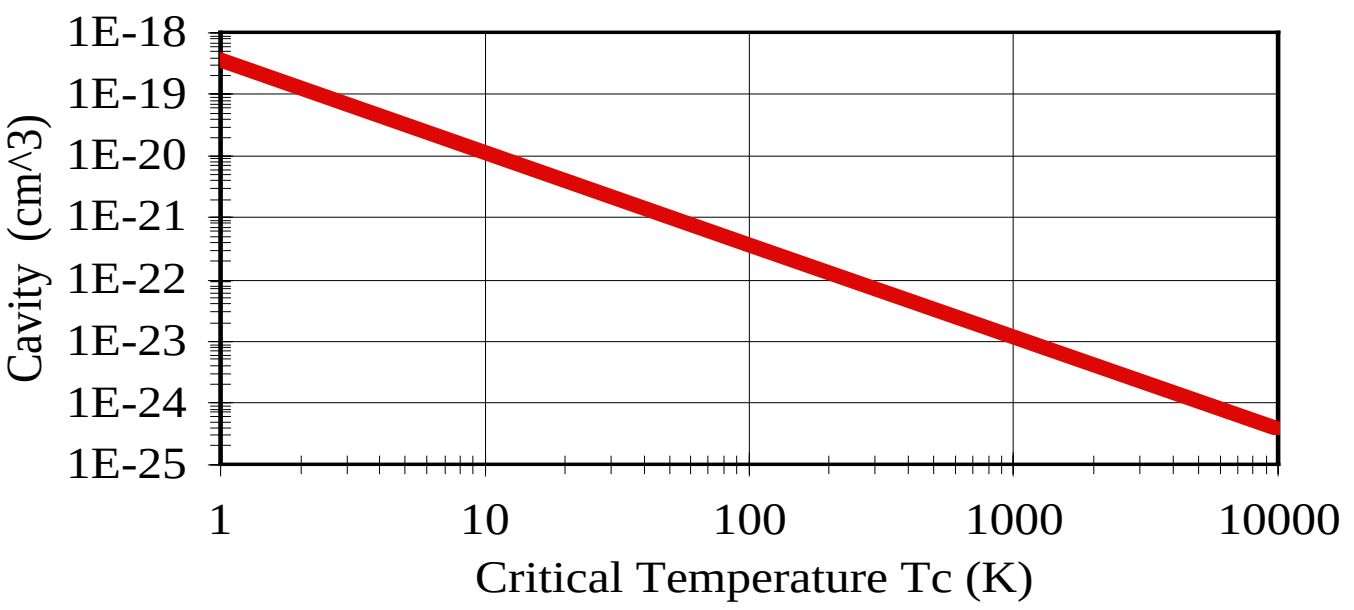

## 2.3 Trisine Characteristic Wave Vectors

Superconducting resonant model trisine characteristic wave vectors $K_{Dn}, K_{Ds}$ and $K_C$ are defined in equations 2.3.1 - 2.3.3 below.

$$K_{Dn} = \left(\frac{6\pi^2}{cavity}\right)^{\frac{1}{3}} \tag{2.3.1}$$

The relationship in equation 2.3.1 represents the normal Debye wave vector $(K_{Dn})$ assuming the *cavity* volume conforming to a sphere in $K$ space and also defined in terms of $K_A, K_P$ and $K_B$ as defined in equations 2.3.4, 2.3.5 and 2.3.6.

See Appendix B for the derivation of equation 2.3.1.

$$K_{Ds} = \left(\frac{8\pi^3}{cavity}\right)^{\frac{1}{3}} = \left(K_A K_B K_P\right)^{\frac{1}{3}} \tag{2.3.2}$$

The relationship in equation 2.3.2 represents the normal Debye wave vector $(K_{Ds})$ assuming the *cavity* volume conforming to a characteristic trisine cell in K space.

See Appendix B for the derivation of equation 2.3.2.

$$K_C = \frac{4\pi}{3\sqrt{3}A} \tag{2.3.3}$$

The relationship in equation 2.3.3 represents the result of equating one-dimensional and trisine density of states. See Appendix C for the derivation of equation 2.3.3.

Equations 2.3.4, 2.3.5, and 2.3.6 translate the trisine wave vectors $K_C$, $K_{Ds}$, and $K_{Dn}$ into x, y, z Cartesian coordinates as represented by $K_B$, $K_P$, and $K_A$ respectively. The superconducting current is in the x direction.

$$K_B = \frac{2\pi}{2B} \tag{2.3.4}$$

$$K_P = \frac{2\pi\sqrt{3}}{3B} \tag{2.3.5}$$

$$K_A = \frac{2\pi}{A} \tag{2.3.6}$$

Note that the sorted energy relationships of wave vectors are as follows:

$$K_A^2 > K_C^2 > K_{Ds}^2 > K_P^2 > K_B^2 \tag{2.3.7}$$

Equation 2.3.8 relates addition of wave function amplitudes (B/√3), (P/2), (B/2) in terms of superconducting resonate energy $(k_b T_c)$.

$$k_b T_c = \frac{cavity}{chain}\frac{\hbar}{time_\pm}\begin{pmatrix} +\frac{K_B}{2B}\left(\frac{B}{\sqrt{3}}\right)^2 \\ +\frac{K_P}{2P}\left(\frac{P}{2}\right)^2 \\ +\frac{K_A}{2A}\left(\frac{A}{2}\right)^2 \end{pmatrix} \tag{2.3.8}$$

The wave vector $K_B$, being the lowest energy, is the carrier of the superconducting energy and in accordance with derivation in appendix A. All of the other wave vectors are contained within the cell *cavity* with many relationships with one indicated in 2.3.9 with the additional wave vector $(K_C)$ defined in 2.3.10

$$\frac{m_e m_t}{m_e + m_t}\frac{1}{K_B^2} + \frac{m_t m_p}{m_t + m_p}\frac{1}{K_A^2} = \frac{1}{6} g_s^2 m_t \frac{1}{K_C^2} \tag{2.3.9}$$

$$K_C = \frac{1}{g_s}\frac{2\pi}{\sqrt{A^2 + B^2}} \tag{2.3.10}$$

The wave vector $K_C$ is considered the hypotenuse vector because of its relationship to $K_A$ and $K_B$.

The conservation of momentum and energy relationships in 2.3.11, 2.3.12 and 2.1.22 result in a convergence to a trisine lattice B/A ratio of 2.379760996 and trisine mass $(m_t)$ of 110.12275343 x electron mass $(m_e)$. Essentially a reversible process or reaction is defined recognizing that momentum is a vector and energy is a scalar.

Conservation Of Momentum

$$\sum_{n=B,C,Ds,Dn} \Delta p_n \Delta x_n = 0$$

$$g_s\left(\pm K_B \pm K_C\right) \rightleftarrows \pm K_{Ds} \pm K_{Dn} \tag{2.3.11}$$

Conservation Of Momentum

$$\sum_{n=B,C,Ds,Dn} \Delta E_n \Delta t_n = 0$$

$$K_B^2 + K_C^2 = g_s\left(K_{Ds}^2 + K_{Dn}^2\right) \tag{2.3.12}$$

The momentum and energy are in equilibrium and define an elastic state in 1 part in ~100,000 after trisine model iteration of several thousand steps which results in a convergence of trisine mass ($m_t$)and (B/A) values and concluding in the following:

The momentum ratio calculates to:

$\pm K_{Ds} \pm K_{Dn} / \left(g_s\left(\pm K_B \pm K_C\right)\right) =$ 0.999865427

The energy ratio calculates to:

$g_s\left(K_{Ds}^2 + K_{Dn}^2\right) / \left(K_B^2 + K_C^2\right) =$ 1.000160309

for an average = 1.000012868

Another perspective of this convergence is contained in Appendix A.

Numerical values of wave vectors are listed in Table 2.3.1.

**Table 2.3.1 with Trisine wave vectors**

| | | | | | | |
|---|---|---|---|---|---|---|
| $T_c\,(^\circ K)$ | 8.95E+13 | 2,135 | 1,190 | 447 | 93 | 8.11E-16 |
| $T_b\,(^\circ K)$ | 9.07E+14 | 3.38E+09 | 3.31E+09 | 2.03E+09 | 9.24E+08 | 2.729 |
| $K_B$ (/cm) | 4.72E+13 | 2.31E+08 | 1.72E+08 | 1.06E+08 | 4.81E+07 | 1.42E-01 |
| $K_{Dn}$ (/cm) | 5.17E+13 | 2.52E+08 | 1.88E+08 | 1.16E+08 | 5.27E+07 | 1.56E-01 |
| $K_{Ds}$ (/cm) | 8.33E+13 | 4.07E+08 | 3.04E+08 | 1.86E+08 | 8.49E+07 | 2.51E-01 |
| $K_C$ (/cm) | 8.65E+13 | 4.22E+08 | 3.15E+08 | 1.93E+08 | 8.82E+07 | 2.60E-01 |
| $K_P$ (/cm) | 5.45E+13 | 2.66E+08 | 1.99E+08 | 1.22E+08 | 5.56E+07 | 1.64E-01 |
| $K_A$ (/cm) | 2.25E+14 | 1.10E+09 | 8.19E+08 | 5.02E+08 | 2.29E+08 | 6.76E-01 |

Figure 2.3.1 in conjunction with mirror image parity images in Figure 2.2.1 clearly depict the trisine symmetry congruent with the Charge conjugation Parity change Time reversal (CPT) theorem. Negative and positive charge reversal as well as $time_\pm$ reversal takes place in mirror image parity pairs as is more visually seen in perspective presented in Figure 2.3.2.

**Figure 2.3.1** Trisine Steric Charge Conjugate Pair Change Time Reversal CPT Geometry.

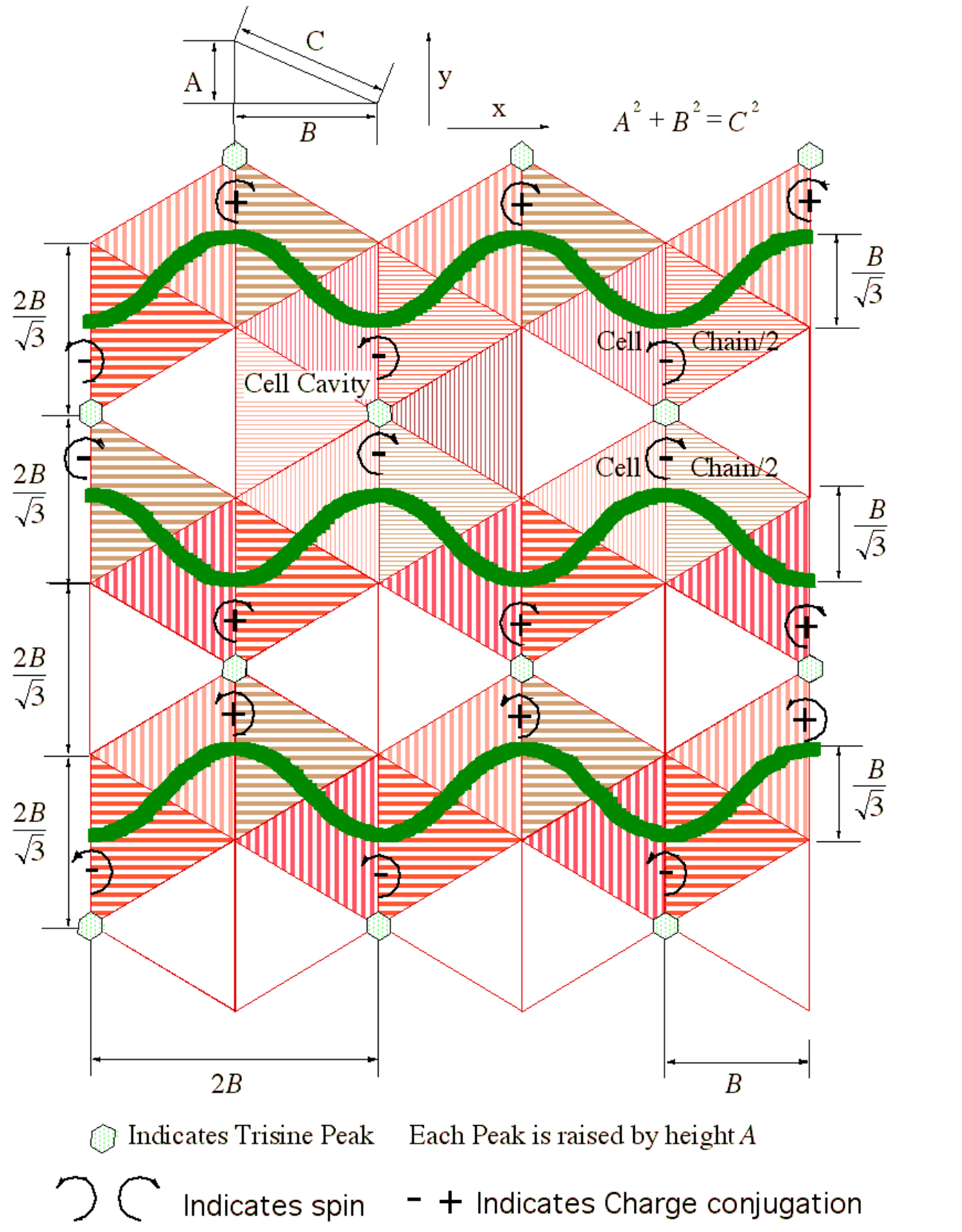

It is important to reiterate that these triangular lattice representations are virtual and equivalent to a virtual reciprocal lattice but as geometric entities, do describe and correlate the physical data quite well.

**Figure 2.3.2** Trisine Steric CPT Geometry in the Superconducting Mode showing the relationship between layered superconducting planes and wave vectors $K_B$. Layers are constructed in *x, y* and *z* directions making up an scaled lattice or Gaussian surface volume *xyz*.

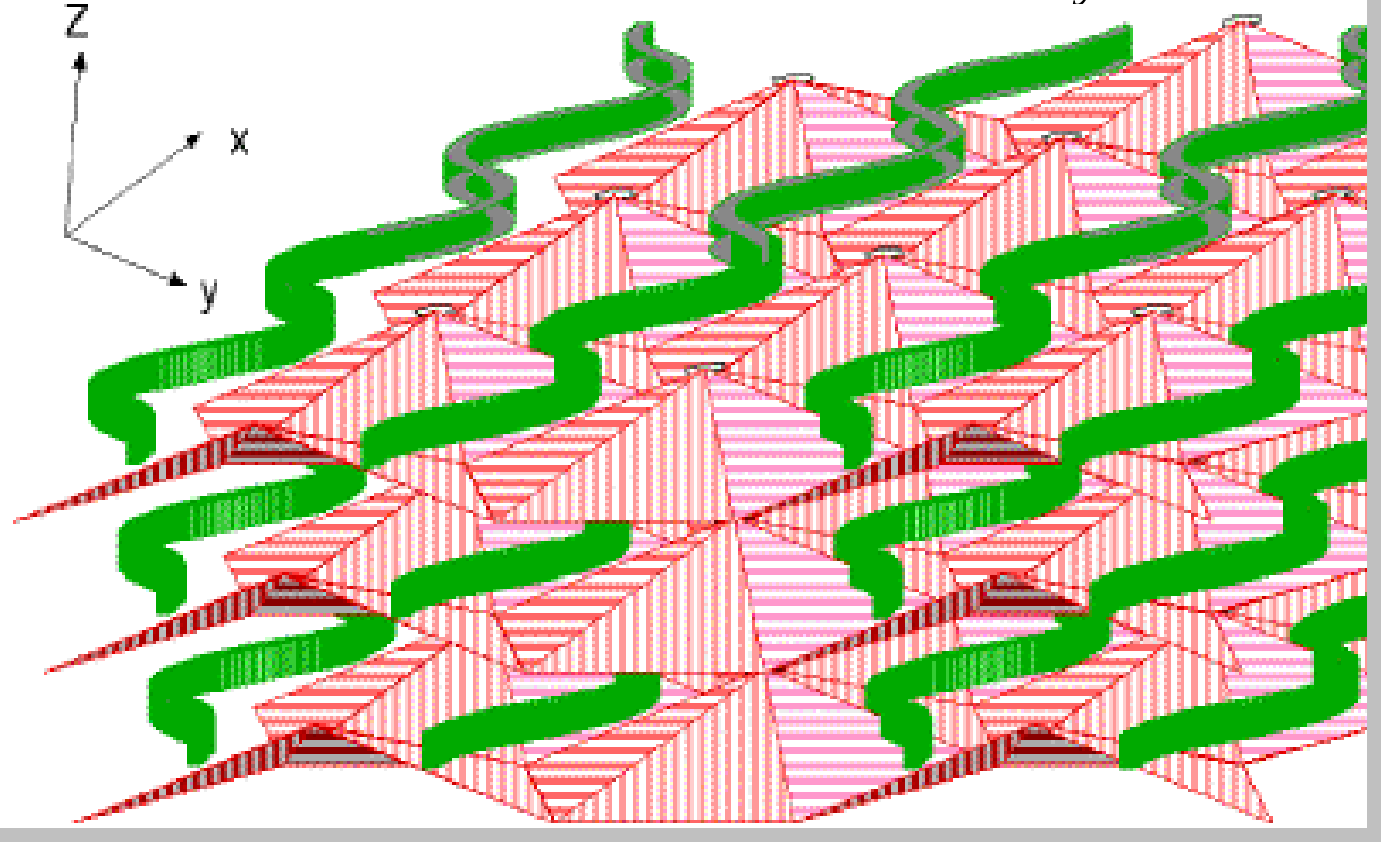

## 2.4 Trisine Characteristic De Broglie Velocities

The Cartesian de Broglie velocities $(v_{dx})$, $(v_{dy})$, and $(v_{dz})$ as well as $(v_{dC})$ are computed with the trisine Cooper CPT Charge conjugated pair residence *resonant CPT time*$_\pm$ and characteristic frequency $(\omega)$ in phase with trisine superconducting dimension $(2B)$ in equations 2.4.1-2.4.5.

$$v_{dx} = \frac{\hbar K_B}{m_t} = \frac{2B}{time_\pm} = \frac{\omega}{2\pi}2B = \frac{eH_c}{m_e v_\varepsilon} g_s^6 B \qquad (2.4.1)$$

In equation 2.4.1, note that $m_e v_\varepsilon / eH_c$ is the electron spin axis precession rate in the critical magnetic field $(H_c)$. This electron spin rate is in tune with the electron moving with de Broglie velocity $(v_{dx})$ with a CPT residence $time_\pm$ in each *cavity*. In other words the Cooper CPT Charge conjugated pair flip spin twice per cavity and because each Cooper CPT Charge conjugated pair flips simultaneously, the quantum is an $2(1/2)$ or integer which corresponds to a boson.

$$v_{dy} = \frac{\hbar K_P}{m_t} \qquad (2.4.2)$$

$$v_{dz} = \frac{\hbar K_A}{m_t} \qquad (2.4.3)$$

$$v_{dC} = \frac{\hbar K_C}{m_t} \qquad (2.4.4)$$

The vector sum of the *x*, *y*, z and *C* de Broglie velocity components are used to compute a three dimensional helical or tangential de Broglie velocity $(v_{dT})$ as follows:

$$\begin{aligned} v_{dT} &= \sqrt{v_{dx}^2 + v_{dy}^2 + v_{dz}^2} \\ &= \frac{\hbar}{m_t}\sqrt{K_A^2 + K_B^2 + K_P^2} \\ &= \frac{3}{g_s}\cos(\theta)\; v_{dC} \end{aligned} \qquad (2.4.5)$$

It is important to note that the one degree of freedom $time_\pm$ (2.96E+04 sec or (164 x 3) minutes) in table 2.4.1 corresponds to a ubiquitous 160 minute resonant signal in the Universe.[68,69] This would be an indication that the momentum and energy conserving (elastic) space lattice existing in universe space is dynamically in tune with stellar, galactic objects. This appears logical, based on the fact the major part of the universe mass consists of this space lattice which can be defined in other terms as dark energy matter.

**Table 2.4.1** Listing of de Broglie velocities, trisine cell pair residence resonant CPT $time_{\pm}$ and characteristic frequency $(\omega)$ as a function of selected critical temperature ($T_c$) and Debye black body temperature ($T_b$) along with a $time_{\pm}$ plot.

| | | | | | | |
|---|---|---|---|---|---|---|
| $T_c\,({}^{\circ}K)$ | 8.95E+13 | 2,135 | 1,190 | 447 | 93 | 8.11E-16 |
| $T_b\,({}^{\circ}K)$ | 9.07E+14 | 3.38E+09 | 3.31E+09 | 2.03E+09 | 9.24E+08 | 2.729 |
| $v_{dx}$ (cm/sec) | 4.96E+11 | 2.42E+06 | 1.81E+06 | 1.11E+06 | 5.06E+05 | 1.49E-03 |
| $v_{dy}$ (cm/sec) | 5.73E+11 | 2.80E+06 | 2.09E+06 | 1.28E+06 | 5.84E+05 | 1.72E-03 |
| $v_{dz}$ (cm/sec) | 2.36E+12 | 1.15E+07 | 8.61E+06 | 5.28E+06 | 2.41E+06 | 7.11E-03 |
| $v_{dC}$ (cm/sec) | 9.09E+11 | 4.44E+06 | 3.31E+06 | 2.03E+06 | 9.27E+05 | 2.74E-03 |
| $v_{dT}$ (cm/sec) | 2.48E+12 | 1.21E+07 | 9.05E+06 | 5.54E+06 | 2.53E+06 | 7.47E-03 |
| $time_{\pm}$ (sec) | 2.68E-25 | 1.12E-14 | 2.02E-14 | 5.37E-14 | 2.58E-13 | 2.96E+04 |
| $\omega$ (/sec) | 2.34E+25 | 5.59E+14 | 3.12E+14 | 1.17E+14 | 2.44E+13 | 2.12E-04 |

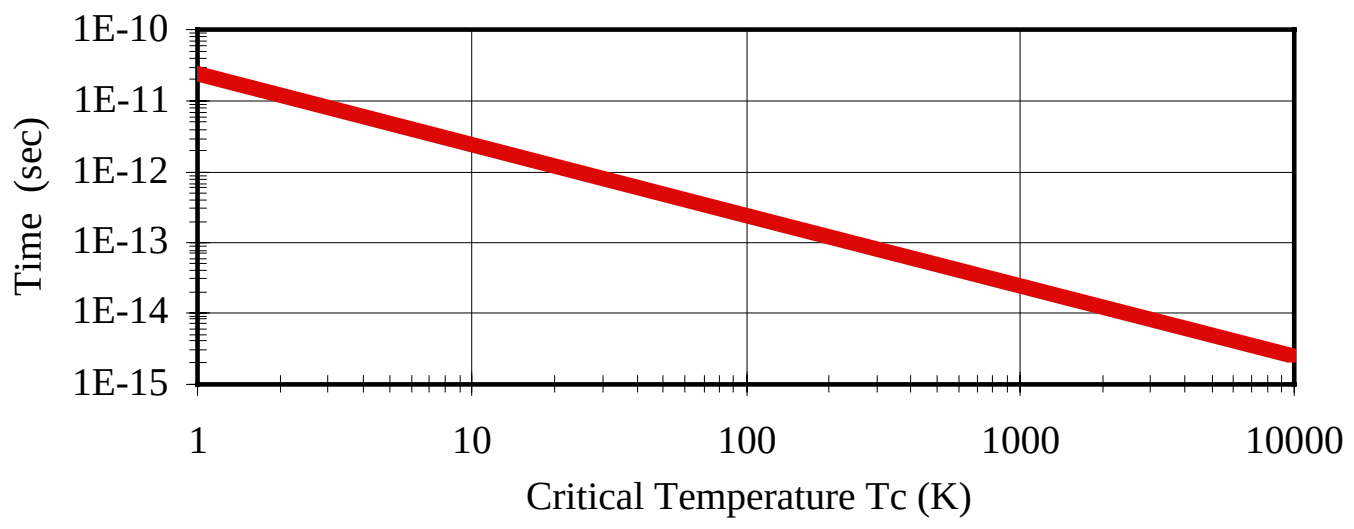

Equation 2.4.6 represents a check on the trisine cell residence *resonant CPT* $time_{\pm}$ as computed from the precession of each electron in a Cooper CPT Charge conjugated pair under the influence of the perpendicular critical field$(H_c)$. Note that the precession is based on the electron mass$(m_e)$ and not trisine resonant transformed mass$(m_t)$.

$$time_{\pm} = \frac{1}{g_s^6}\frac{m_e v_{\varepsilon}}{e_{\pm} H_c}\frac{1}{B} \qquad (2.4.6)$$

Another check on the trisine cell residence resonant CPT $time_{\pm}$ is computed on the position of the Cooper CPT Charge conjugated pair$(\mathbb{C}\,e)$ particles under the charge influence of each other within the dielectric$(\varepsilon)$ as they travel in the $y$ direction as expressed in equation 2.4.7.

$$m_t \frac{d^2 y}{dt^2} = \frac{\mathrm{M}_a e_{\pm}^2}{\varepsilon\cos(\theta)}\frac{1}{y^2} \qquad (2.4.7)$$

From figure 2.3.1 it is seen that the Cooper CPT Charge conjugated pairs travel over distance

$$\left(3B/\sqrt{3} - 2B/\sqrt{3}\right)$$

in each quarter cycle or *resonant CPT* $time_{\pm}/4$.

$$time_{\pm} = 4\left(\frac{8}{81}\frac{m_t\varepsilon\cos(\theta)}{\mathrm{M}_a e_{\pm}^2}\right)^{\frac{1}{2}}\left(\frac{3B}{\sqrt{3}} - \frac{2B}{\sqrt{3}}\right)^{\frac{3}{2}} \qquad (2.4.8)$$

where the Madelung constant$(M_y)$ in the $y$ direction is calculated as:

$$M_y = \sqrt{3}\sum_{n=0}^{n=\infty}\left(\frac{1}{3(1+2n)-1} - \frac{1}{3(1+2n)+1}\right) \qquad (2.4.9)$$
$$= 0.523598679$$

And the Thomas scattering formula[43] holds in terms of the following equation:

$$side = \left(\frac{K_B}{K_{Ds}}\right)\left(\frac{8\pi}{3}\right)\left(\frac{e_{\pm}^2}{M_y \varepsilon m_t v_{dy}^2}\right)^2 \qquad (2.4.10)$$

A sense of the rotational character of the de Broglie $(v_{dT})$ velocity can be attained by the following energy equation:

$$\frac{1}{2}m_t v_{dT}^2 = \frac{1}{g_s}\cos(\theta)\frac{chain}{cavity}\frac{1}{2}\left(m_t(2B)^2\right)\omega^2 + \frac{1}{g_s}\cos(\theta)\frac{chain}{cavity}\frac{1}{2}\left(m_t(A)^2\right)\omega^2 \qquad (2.4.11)$$

Equation 2.4.12 provides an expression for the Sagnac time.

$$time_{\pm} = \frac{\sqrt{A^2+B^2}}{B}\frac{2\pi}{time_{\pm}}\frac{section}{6}\frac{1}{v_{dx}^2} \qquad (2.4.12)$$

## 2.5 Superconductor Resonant Dielectric Constant And Magnetic Permeability

The material dielectric is computed by determining a displacement$(D)$ (equation 2.5.1) and electric field$(E)$ (equation 2.5.2) and which establishes a dielectric$(\varepsilon$ or $D/E)$ (equation 2.5.3) and modified speed of light $(v_{\varepsilon})$ (equations 2.5.4, 2.5.5). Then the superconducting fluxoid $(\Phi_{\varepsilon})$ (equation 2.5.6) can be calculated.

The electric field(E) is calculated by taking the translational energy ($m_t v_{dx} v_{dx}/2$) (which is equivalent to $k_b T_c$) over a distance, which is the trisine cell volume to surface ratio. This concept of the surface to volume ratio is used extensively in fluid mechanics for computing energies of fluids passing through various geometric forms and is called the hydraulic ratio.

$$E = \left(\frac{1}{g_s^2}\right)\left(\mathbb{C}\frac{m_t v_{dx}^2}{2}\right)\left(\frac{2\ section}{\cos(\theta)}\right)\left(\frac{1}{\mathbb{C}\ e_{\pm}\ cavity}\right)$$
$$= \left(\frac{1}{g_s^2}\right)(2\mathbb{C})\left(1+\left(\frac{B}{A}\right)^2\right)^{\frac{1}{2}}\left(\frac{m_t v_{dx}}{time_{\pm}}\right)\left(\frac{1}{\mathbb{C}\ e_{\pm}}\right) \qquad (2.5.1)$$
$$= 4.25\ \frac{1}{2}\frac{m_t v_{dx}^2}{\mathbb{C}\ e_{\pm}\ A}$$

Secondarily, the electric field(E) is calculated in terms of the forces$(m_t v_d / time)$ exerted by the wave vectors$(K)$.

$$E=\begin{Bmatrix} 10\left(g_s^2\frac{m_t v_{dx}}{time_\pm}\cos(\theta)\right)\left(\frac{1}{\mathbb{C}\ e_\pm}\right) \\ 9\left(\frac{1}{g_s^2}\frac{m_t v_{dy}}{time_\pm}\right)\left(\frac{1}{\mathbb{C}\ e_\pm}\right) \\ 2\left(\frac{1}{g_s}\frac{1}{\cos(\theta)}\frac{m_t v_{dz}}{time_\pm}\right)\left(\frac{1}{\mathbb{C}\ e_\pm}\right) \\ 6\left(\frac{m_t v_{dC}}{time_\pm}\cos(\theta)\right)\left(\frac{1}{\mathbb{C}\ e_\pm}\right) \end{Bmatrix} \tag{2.5.2}$$

The trisine cell surface is ($section/cos(\theta)$) and the volume is expressed as *cavity*. Note that the trisine *cavity*, although bounded by ($2section/cos(\theta)$), still has the passageways for Cooper CPT Charge conjugated pairs ($\mathbb{C}e$) to move from cell *cavity* to cell *cavity*.

The displacement$(D)$ as a measure of Gaussian surface containing free charges$(\mathbb{C}\ e_\pm)$ is computed by taking the $(4\pi)$ solid angle divided by the characteristic trisine area $(section/\cos(\theta))$ with the two(2) charges $(\mathbb{C}\ e_\pm)$ contained therein and in accordance with Gauss's law as expressed in equation 2.1.1.

$$D=4\pi\ \mathbb{C}\ e_\pm\frac{\cos(\theta)}{2\ section} \tag{2.5.3}$$

Now a dielectric coefficient $(\varepsilon)$ can be calculated from the electric field$(E)$ and displacement field$(D)$ [61]. Note that $\varepsilon<1$ for superluminal velocities as justified by resonant condition as numerically indicated in table 2.5.1.

$$\varepsilon=\frac{D}{E} \tag{2.5.4}$$

Assuming the trisine geometry has the relative magnetic permeability of a vacuum$(k_m=1)$ then a modified velocity of light$(v_\varepsilon)$ can be computed from the dielectric coefficient$(\varepsilon)$ and the speed of light$(c)$ where $(\varepsilon_0)$ = 1.

$$v_\varepsilon=\frac{c}{\sqrt{k_m\frac{\varepsilon}{\varepsilon_0}}}=\frac{c}{\sqrt{\varepsilon}} \tag{2.5.5}$$

Trisine incident/reflective angle (Figure 2.3.1) of 30 degrees is less than Brewster angle of $\tan^{-1}(\varepsilon/\varepsilon)^{1/2}$=$45^0$ assuring total reflectivity as particles travel from trisine lattice cell to trisine lattice cell.

Now the fluxoid $(\Phi_\varepsilon)$ can be computed quantized according to $(\mathbb{C}\ e)$ as experimentally observed in superconductors.

$$\Phi_\varepsilon=\frac{2\pi\hbar v_\varepsilon}{\mathbb{C}\ e_\pm} \tag{2.5.6}$$

What is generally called the Tao effect [65,66,67], wherein it is found that superconducting micro particles in the presence of an electrostatic field aggregate into balls of macroscopic dimensions, is modeled by the calculated values for E in table 2.5.1. using equation 2.5.2. When the electrostatic field exceeds a critical value coincident with the critical temperature $(T_b)$, the microscopic balls dissipate. This experimental evidence can be taken as another measure of superconductivity equivalent to the Meissner effect.

**Table 2.5.1 with dielectric plot**

| | | | | | | |
|---|---|---|---|---|---|---|
| $T_c$ ($^\circ K$) | 8.95E+13 | 2,135 | 1,190 | 447 | 93 | 8.11E-16 |
| $T_b$ ($^\circ K$) | 9.07E+14 | 3.38E+09 | 3.31E+09 | 2.03E+09 | 9.24E+08 | 2.729 |
| $D$ (erg/e/cm) | 3.63E+17 | 8.65E+06 | 4.82E+06 | 1.81E+06 | 3.77E+05 | 3.29E-12 |
| $E$ (erg/e/cm) | 1.96E+21 | 2.28E+05 | 9.49E+04 | 2.18E+04 | 2.07E+03 | 5.34E-23 |
| $E$ (volt/cm) | 5.87E+23 | 6.84E+07 | 2.85E+07 | 6.55E+06 | 6.22E+05 | 1.60E-20 |
| $\varepsilon$ | 1.85E-04 | 37.95 | 50.83 | 82.93 | 181.82 | 6.16E+10 |
| $v_\varepsilon$ (cm/sec) | 2.20E+12 | 4.87E+09 | 4.21E+09 | 3.29E+09 | 2.22E+09 | 1.21E+05 |
| $\Phi_\varepsilon$ (gauss cm$^2$) | 1.52E-05 | 3.36E-08 | 2.90E-08 | 2.27E-08 | 1.53E-08 | 8.33E-13 |

Using the computed dielectric, the energy associated with superconductivity can be calculated in terms of the standard Coulomb's law electrostatic relationship $e^2/(\varepsilon\ B)$ as presented in equation 2.5.7.

$$k_bT_c=\begin{Bmatrix} \frac{chain}{cavity}\frac{e_\pm^2}{\varepsilon\ B} \\ g_s\frac{1}{\cos(\theta)}\mathrm{M}_y\frac{chain}{cavity}\frac{e_\pm^2\sqrt{3}}{\varepsilon\ B} \\ \frac{1}{g_s}\tan(\theta)\frac{chain}{cavity}\frac{e_\pm^2}{\varepsilon\ A} \end{Bmatrix} \tag{2.5.7}$$

where $\mathrm{M}_a$ is the Madelung constant in *y* direction as computed in equation 2.4.9.

Significantly, equation 2.5.7 is dimensionally congruent with equation 2.5.7a. This implies that mass $(m_t)$ is created and destroyed in each $time_\pm$ cycle.

$$k_bT_c=\frac{1}{\sqrt{3}}\left(\frac{m_t}{time_\pm}\right)^2\frac{cavity}{m_t}\frac{1}{A} \tag{2.5.7a}$$

A conversion between superconducting temperature and black body temperature is calculated in equation 2.5.9. Two primary energy related factors are involved, the first being the superconducting velocity$(v_{dx}^\pm)$ to light velocity $(v_\varepsilon^\pm)$ and the second is normal Debye wave vector$(K_{Dn}^\pm)$ to superconducting Debye wave vector$(K_{Ds}^\pm)$ all of this followed by a minor rotational factor for each factor involving $(m_e)$ and$(m_p)$. For reference, a value of 2.729 degrees Kelvin is used for the universe black body temperature$(T_b)$ as indicated by experimentally observed microwave radiation by the Cosmic Background Explorer (COBE) and later satellites. Although the observed minor fluctuations (1 part in 100,000) in this universe background radiation indicative of clumps of matter forming shortly after the big bang, for the purposes of this report we will assume that the experimentally observed uniform radiation is indicative of present universe that is isotropic and homogeneous.

Verification of equation 2.5.6 is indicated by the calculation that superconducting density $(m_t/cavity)$ (table 2.9.1) and present universe density (equation 2.11.4) are equal at this Debye black body temperature $(T_b)$ of $2.729\,^oK$.

$$T_b = \frac{1}{\mathbb{C}}\left(\frac{v_\varepsilon}{v_{dx}}\right)^2 g_s^3\ T_c \tag{2.5.8}$$

$$k_b T_b = 2\pi\hbar\frac{c}{\lambda_b} = \left(\frac{18}{g_s^2}\right)\hbar K_A c \tag{2.5.9}$$

The black body temperatures $(T_b)$ appear to be high relative to superconducting $(T_c)$, but when the corresponding wave length $(\lambda_b)$ is calculated in accordance with equation 2.5.7, it is nearly the same and just within the Heisenberg $(\Delta z)$ parameter for all $T_c$ and $T_b$ as calculated from equation 2.1.18 and presented in table 2.5.2. It is suggested that a black body oscillator exists within such a volume as defined by $(\Delta x \Delta y \Delta z)$ and is the source of the microwave radiation at $2.729\,^oK$. Further, it is suggested that the trisine geometry when scaled to nuclear dimensions is congruent with the experimentally observed Standard Model energy and mass (quark, gluon, weak and strong force) parameters.

**Table 2.5.2 based on equations 2.1.20 and 2.5.9**

| | | | | | | |
|---|---|---|---|---|---|---|
| $T_c\ (^oK)$ | 8.95E+13 | 2,135 | 1,190 | 447 | 93 | 8.11E-16 |
| $T_b\ (^oK)$ | 9.07E+14 | 3.38E+09 | 3.31E+09 | 2.03E+09 | 9.24E+08 | 2.729 |
| $\Delta x$ (cm) | 1.06E-14 | 2.17E-09 | 2.90E-09 | 4.74E-09 | 1.04E-08 | 3.52 |
| $\Delta y$ (cm) | 9.17E-15 | 1.88E-09 | 2.52E-09 | 4.10E-09 | 9.00E-09 | 3.05 |
| $\Delta z$ (cm) | 2.23E-15 | 4.56E-10 | 6.10E-10 | 9.96E-10 | 2.18E-09 | 0.74 |
| $\lambda_b$ (cm) | 1.59E-15 | 3.25E-10 | 4.35E-10 | 7.10E-10 | 1.56E-09 | 0.53 |

Based on the same approach as presented in equations 2.5.1, 2.5.2 and 2.5.3, Cartesian x, y, and z values for electric and displacement fields are presented in equations 2.5.10 and 2.5.11.

$$\begin{Bmatrix} E_x \\ E_y \\ E_z \end{Bmatrix} = \begin{Bmatrix} \dfrac{F_x}{\mathbb{C}\ e_\pm} \\ \dfrac{F_y}{\mathbb{C}\ e_\pm} \\ \dfrac{F_z}{\mathbb{C}\ e_\pm} \end{Bmatrix} = \begin{Bmatrix} \dfrac{m_t v_{dx}}{time_\pm}\dfrac{1}{\mathbb{C}\ e_\pm} \\ \dfrac{m_t v_{dy}}{time_\pm}\dfrac{1}{\mathbb{C}\ e_\pm} \\ \dfrac{m_t v_{dz}}{time_\pm}\dfrac{1}{\mathbb{C}\ e_\pm} \end{Bmatrix} \tag{2.5.10}$$

$$\begin{Bmatrix} D_x \\ D_y \\ D_z \end{Bmatrix} = \begin{Bmatrix} \dfrac{4\pi}{approach}\mathbb{C}\ e_\pm \\ \dfrac{4\pi}{side}\mathbb{C}\ e_\pm \\ \dfrac{4\pi}{section}\mathbb{C}\ e_\pm \end{Bmatrix} \tag{2.5.11}$$

Now assume that the superconductor material magnetic permeability $(k_m)$ is defined as per equation 2.5.12 noting that $k_m = k_{mx} = k_{my} = k_{mz}$.

$$k_m = \begin{Bmatrix} k_{mx} \\ k_{my} \\ k_{mz} \end{Bmatrix} = \begin{Bmatrix} \dfrac{D_x v_{\varepsilon x}^2}{E_x}\dfrac{E_x}{D_x v_{dx}^2} \\ \dfrac{D_y v_{\varepsilon y}^2}{E_y}\dfrac{E_y}{D_y v_{dy}^2} \\ \dfrac{D_z v_{\varepsilon z}^2}{E_z}\dfrac{E_z}{D_z v_{dz}^2} \end{Bmatrix} = \begin{Bmatrix} \dfrac{v_{\varepsilon x}^2}{v_{dx}^2} \\ \dfrac{v_{\varepsilon y}^2}{v_{dy}^2} \\ \dfrac{v_{\varepsilon z}^2}{v_{dz}^2} \end{Bmatrix} \tag{2.5.12}$$

Also note that the dielectric $(\varepsilon)$ and permeability $(k_m)$ are related as follows:

$$k_{m\pm} = \varepsilon_\pm\left(\frac{m_t}{m_e}\right)^{\frac{3}{2}}\cos^2(\theta) \tag{2.5.12a}$$

Then the de Broglie velocities ($v_{dx}$, $v_{dy}$ and $v_{dz}$) as per equations 2.4.1, 2.4.2, 2.4.3 can be considered a sub- and super- luminal speeds of light internal to the superconductor resonant medium as per equation 2.5.13. These sub- and super- luminal resonant de Broglie velocities as presented in equation 2.5.13 logically relate to wave vectors $K_A$, $K_B$ and $K_P$.

$$\begin{Bmatrix} v_{dx} \\ v_{dy} \\ v_{dz} \end{Bmatrix} = \begin{Bmatrix} \dfrac{c}{\sqrt{k_m \varepsilon_x}} \\ \dfrac{c}{\sqrt{k_m \varepsilon_y}} \\ \dfrac{c}{\sqrt{k_m \varepsilon_z}} \end{Bmatrix} = \begin{Bmatrix} \dfrac{1}{k_m}\dfrac{E_x}{D_x}c^2 \\ \dfrac{1}{k_m}\dfrac{E_y}{D_y}c^2 \\ \dfrac{1}{k_m}\dfrac{E_z}{D_z}c^2 \end{Bmatrix} \tag{2.5.13}$$

Combining equations 2.5.12 and 2.5.13, the Cartesian dielectric sub- and super- luminal velocities ($v_{\varepsilon x}$, $v_{\varepsilon y}$ & $v_{\varepsilon z}$) can be computed and are presented in Table 2.5.3.

**Table 2.5.3 Dielectric Velocities** ($v_{\varepsilon x}$, $v_{\varepsilon y}$ & $v_{\varepsilon z}$)

| | | | | | | |
|---|---|---|---|---|---|---|
| $T_c\ (^oK)$ | 8.95E+13 | 2,135 | 1,190 | 447 | 93 | 8.11E-16 |
| $T_b\ (^oK)$ | 9.07E+14 | 4.43E+09 | 3.31E+09 | 2.03E+09 | 9.24E+08 | 2.729 |
| $k_m$ | 1.88E-01 | 3.86E+04 | 5.16E+04 | 8.43E+04 | 1.85E+05 | 6.26E+13 |
| $v_{\varepsilon x}$ (cm/sec) | 2.15E+11 | 4.76E+08 | 4.11E+08 | 3.22E+08 | 2.17E+08 | 1.18E+04 |
| $v_{\varepsilon v}$ (cm/sec) | 2.49E+11 | 5.50E+08 | 4.75E+08 | 3.72E+08 | 2.51E+08 | 1.36E+04 |
| $v_{\varepsilon z}$ (cm/sec) | 1.02E+12 | 2.27E+09 | 1.96E+09 | 1.53E+09 | 1.03E+09 | 5.62E+04 |

We note with special interest that the Lorentz-Einstein resonant relationship expressed in equation 2.5.14 equals 2, which we define as *Cooper* $(\mathbb{C})$ for all $T_c$.

$$\frac{1}{\sqrt{1-\dfrac{v_{dx}^2}{v_{dy}^2}}} = \mathbb{C} = 2 \tag{2.5.14}$$

As a check on these dielectric Cartesian velocities, note that they are vectorially related to $v_\varepsilon$ from equation 2.5.5 as indicated in equation 2.5.15a. As indicated, the factor '2' is related to the ratio of Cartesian surfaces *approach*, *section* and *side* to trisine area $\cos(\theta)/section$ and *cavity/chain*.

$$(approach)\left(\frac{\cos(\theta)}{section}\right)\left(\frac{chain}{cavity}\right) + (section)\left(\frac{\cos(\theta)}{section}\right)\left(\frac{chain}{cavity}\right) + (side)\left(\frac{\cos(\theta)}{section}\right)\left(\frac{chain}{cavity}\right) = 2 = \mathbb{C} \quad (2.5.15)$$

$$v_\varepsilon = 2\sqrt{v_{\varepsilon x}^2 + v_{\varepsilon y}^2 + v_{\varepsilon z}^2} = \mathbb{C}\sqrt{v_{\varepsilon x}^2 + v_{\varepsilon y}^2 + v_{\varepsilon z}^2} \quad (2.5.15a)$$

Based on the Cartesian dielectric sub- and super- luminal resonant justified velocities ($v_{\varepsilon x}$, $v_{\varepsilon y}$ & $v_{\varepsilon z}$), corresponding Cartesian fluxoids can be computed as per equation 2.5.16.

$$\begin{Bmatrix} \Phi_{\varepsilon x} \\ \Phi_{\varepsilon y} \\ \Phi_{\varepsilon z} \end{Bmatrix} = \begin{Bmatrix} \dfrac{2\pi\hbar v_{\varepsilon x}}{\mathbb{C}\ e_\pm} \\ \dfrac{2\pi\hbar v_{\varepsilon y}}{\mathbb{C}\ e_\pm} \\ \dfrac{2\pi\hbar v_{\varepsilon z}}{\mathbb{C}\ e_\pm} \end{Bmatrix} \quad (2.5.16)$$

The trisine residence *resonant CPT time*$_\pm$ is confirmed in terms of conventional capacitance $(C)$ and inductance $(L)$ resonant circuit relationships in *x*, *y* and *z* as well as *trisine* dimensions.

$$time_\pm = \begin{Bmatrix} \sqrt{L_x}\sqrt{C_x} \\ \sqrt{L_y}\sqrt{C_y} \\ \sqrt{L_z}\sqrt{C_z} \end{Bmatrix} = \sqrt{L}\sqrt{C} \quad (2.5.17a)$$

Where relationships between Cartesian and trisine capacitance and inductance is as follows:

$$2\ C = C_x = C_y = C_z \quad (2.5.18)$$

$$L = 2\ L_x = 2\ L_y = 2\ L_z \quad (2.5.19)$$

$$L_x = L_y = L_z \quad (2.5.20)$$

$$C_x = C_y = C_z \quad (2.5.21)$$

**Table 2.5.4 Capacitance and Inductive Density**

| | | | | | | |
|---|---|---|---|---|---|---|
| $T_c\,(^\circ K)$ | 8.95E+13 | 2,135 | 1,190 | 447 | 93 | 8.11E-16 |
| $T_b\,(^\circ K)$ | 9.07E+14 | 4.43E+09 | 3.31E+09 | 2.03E+09 | 9.24E+08 | 2.729 |
| $C$/v (fd/cm$^3$) | 7.70E+09 | 3.76E+04 | 2.81E+04 | 1.72E+04 | 7.85E+03 | 2.32E-05 |
| $L$/v (h/cm$^3$) | 4.85E+19 | 2.37E+14 | 1.77E+14 | 1.08E+14 | 4.94E+13 | 1.46E+05 |

In terms of an extended Thompson cross-section $(\sigma_T)$, it is noted that the *(1/R²)* factor [62] is analogous to a dielectric $(\varepsilon)$ equation 2.5.4. Dimensionally the dielectric $(\varepsilon)$ is proportional to trisine cell length dimension.

The extended Thompson scattering cross section $(\sigma_T)$ [43 equation 78.5, 62 equation 33] then becomes as in equation 2.5.22.

$$\sigma_T = \left(\frac{K_B}{K_{Ds}}\right)^2 \left(\frac{8\pi}{3}\right)\left(\frac{e_\pm^2}{\varepsilon m_t v_{dy}^2}\right)^2 \approx side \quad (2.5.22)$$

## 2.6 Fluxoid And Critical Fields

Based on the material fluxoid $(\Phi_\varepsilon)$ (equation 2.5.5), the critical fields $(H_{c1})$ (equation 2.6.1), $(H_{c2})$ (equation 2.6.2) & $(H_c)$ (equation 2.6.3) as well as penetration depth $(\lambda)$ (equation 2.6.5) and Ginzburg-Landau coherence length $(\xi)$ (equation 2.6.6) are computed. Also the critical field $(H_c)$ is alternately computed from a variation on the Biot-Savart law.

$$H_{c1} = \frac{\Phi_\varepsilon}{\pi\lambda_t^2} \quad (2.6.1)$$

$$H_{c2} = \frac{4\pi\ \mathbb{C}}{section}\Phi_\varepsilon \quad (2.6.2)$$

$$H_c = \sqrt{H_{c1}H_{c2}} = \frac{e_\pm}{v_{\varepsilon x}\cdot A\cdot time}\frac{m_t}{m_e} = \frac{g_s}{6}\frac{\mathbb{C}}{cavity}\frac{v_{ey}}{c}e_\pm\lambda \quad (2.6.3)$$

$$n_c = \frac{\mathbb{C}}{cavity} \quad (2.6.4)$$

It was observed by Huxley [70] that an increasing magnetic field quenched the URhGe superconductor at 2 tesla (20,000 gauss) initiated the URhGe superconductor at 8 tesla (20,000 gauss) quenched the URhGe superconductor at 13 tesla (130,000 gauss)

The trisine model $(H_{c2})$ correlates well with the reported [70] 2 (20,000 gauss) with a calculated modeled superconductor magnetic critical field value $(H_{c2})$ of 1.8 tesla (18,000 gauss) for critical temperature $(T_c)$ of .28 K.

At critical temperature $(T_c)$ of .4 Kelvin, the trisine model predicts superconductor magnetic critical field $(H_{c2})$ of 2.9 tesla (29,000 gauss), which is substantially lower than the observed 13 tesla (130,000 gauss). This could be explained in terms of the ferromagnetic properties of URhGe, which essentially mask the superconductor critical magnetic field. But it still remarkable how well the model correlates to the observed superconducting magnetic critical field $(H_{c2})$ in this case.

**Figure 2.6.1** Experimental (Harshman Data) and Trisine Model Cooper CPT Charge conjugated Pair Three Dimensional Concentration as per equation 2.6.4

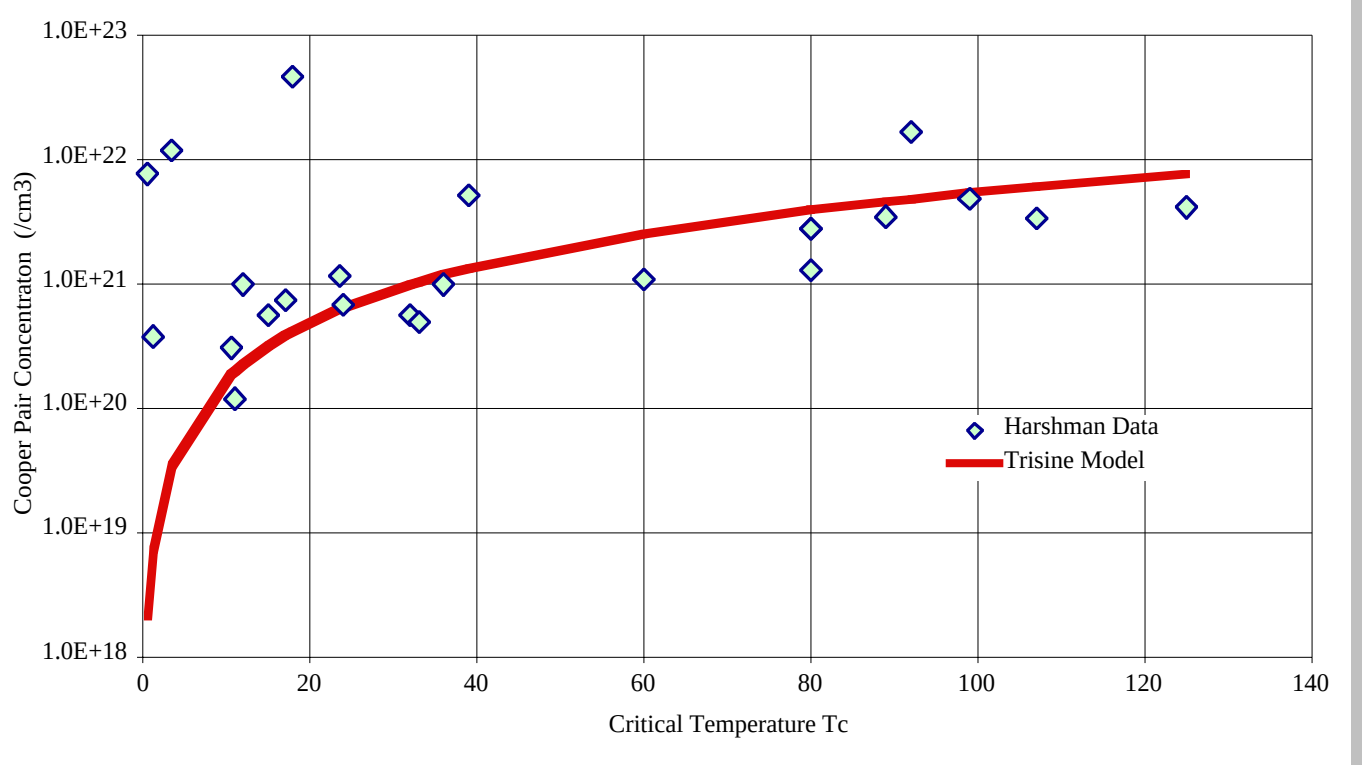

Also it interesting to note that the following relationship holds

$$2\pi \frac{e_\pm^2}{A}\frac{e_\pm^2}{B} = m_e v_{dx}^2 m_e c^2 \qquad (2.6.4a)$$

**Figure 2.6.2** Experimental (Harshman Data) and Trisine Model Cooper CPT Charge conjugated Pair Two Dimensional Area Concentration or ℂ/section

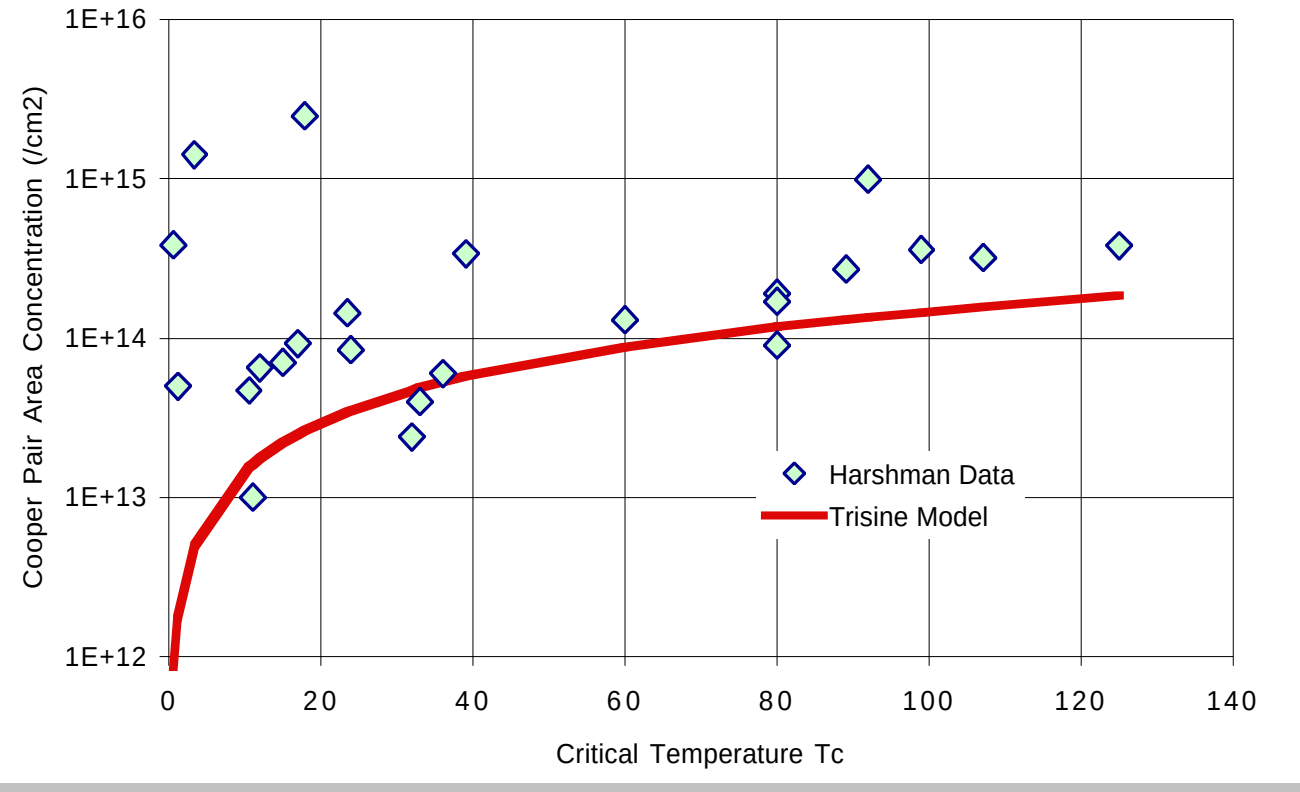

$$\lambda^2 = \left(\frac{1}{g_s^2}\right) \mathbb{C} \left(\frac{m_t v_\varepsilon^2}{2}\right) \frac{1}{n_c} \frac{1}{(\mathbb{C}\, e)^2}$$
$$= \left(\frac{1}{g_s^2}\right) \mathbb{C} \left(\frac{1}{\varepsilon}\right) \left(\frac{m_t c^2}{2}\right) \frac{1}{n_c} \frac{1}{(\mathbb{C}\, e_\pm)^2} \qquad (2.6.5)$$

From equation 2.6.5, the trisine penetration depth $\lambda$ is calculated. This is plotted in the above figure along with Harshman[17] and Homes[51, 59] penetration data. The Homes data is multiplied by a factor of $(n_e\ chain/\mathbb{C})^{1/2}$ to get $n_c$ in equation 2.6.5. This is consistent with $\lambda^2 \propto 1/(\varepsilon n_c)$ in equation 2.6.5 and consistent with Uemura Plot $\lambda^2 \propto 1/n_c$ over short increments of penetration depth $\lambda$. It appears to be a good fit and better than the fit with data compiled by Harshman[17].

Penetration depth and gap data for magnesium diboride $MgB_2$ [55], which has a critical temperature $T_c$ of 39 K indicates a fit to the trisine model when a critical temperature of 39/3 or 13 K is assumed (see Table 2.6.1). This is interpreted as an indication that $MgB_2$ is a three dimensional superconductor.

**Table 2.6.1** Experimental ($MgB_2$ Data [55]) and the related Trisine Model Prediction

| | Trisine Data at $T_c$ 39/3 K or 13 K | Observed Data [55] |
|---|---|---|
| Penetration Depth (nm) | 276 | 260 ±20 |
| Gap (meV) | 1.12 | 3.3/3 or 1.1 ±.3 |

As indicated in equation 2.5.5, the material medium dielectric sub- and super- luminal speed of light $(v_\varepsilon)$ is a function of medium dielectric $(\varepsilon)$ constant, which is calculated in general, and specifically for trisine by $\varepsilon = D/E$ (equation 2.5.4). The fact that chain rather than cavity (related by 2/3 factor) must be used in obtaining a trisine model fit to Homes' data indicates that the stripe concept may be valid in as much as chains in the trisine lattice visually form a striped pattern as in Figure 2.3.1.

**Figure 2.6.3** Experimental (Homes, Pratt & Harshman) and Trisine Model Penetration Depth

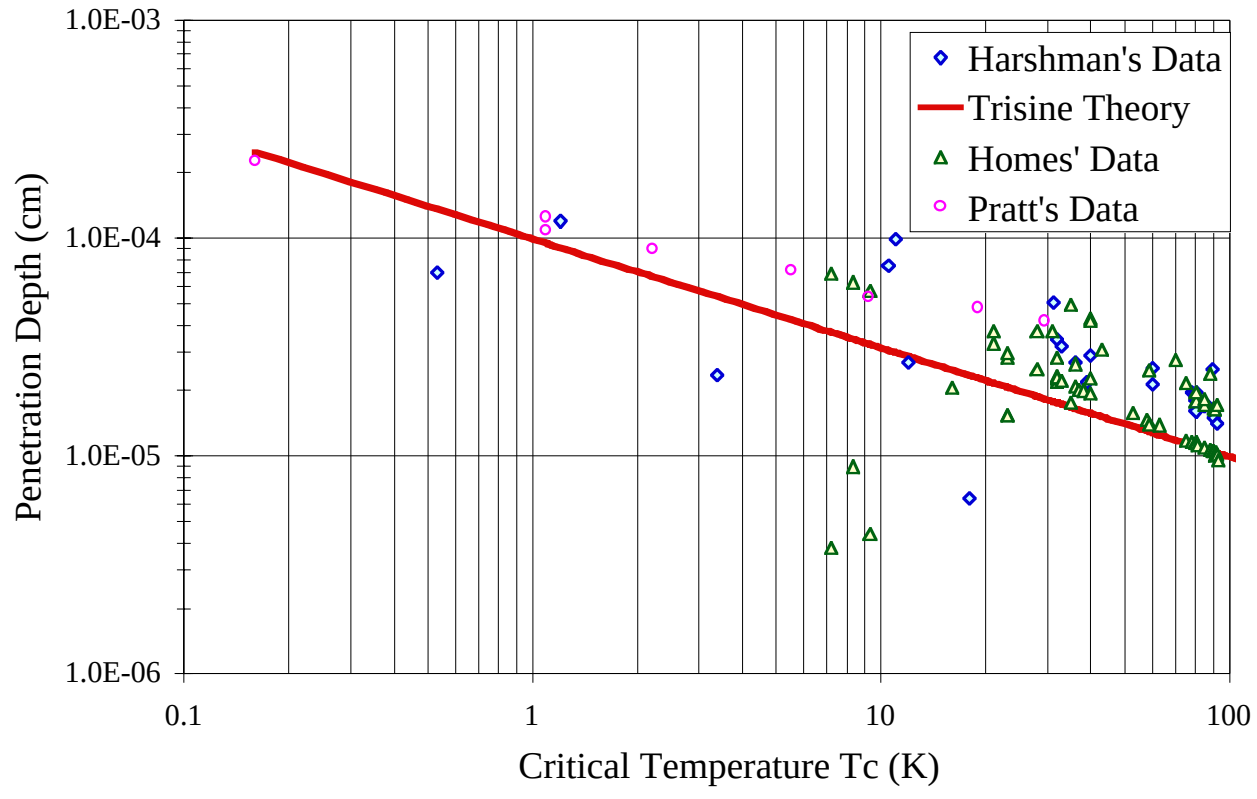

See derivation of Ginzberg-Landau coherence length $\xi$ in equation 2.6.6 in Appendix D. It is interesting to note that $B/\xi = 2.73 \approx \mathrm{e}$ for all $T_c$, which makes the trisine superconductor mode fit the conventional description of operating in the "dirty" limit. [54]

$$\xi = \sqrt{\frac{m_t}{m_e}\frac{1}{K_B^2}\frac{chain}{cavity}} \qquad (2.6.6)$$

**Figure 2.6.4** Trisine Geometry in the Fluxoid Mode (Interference Pattern $\Psi^2$)

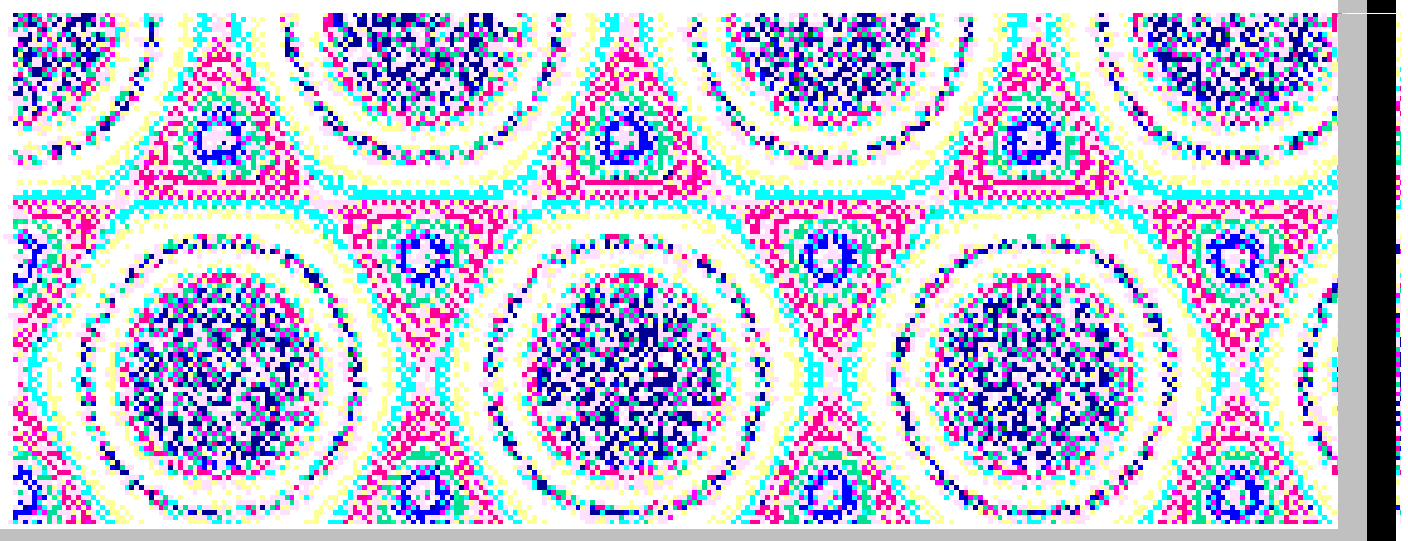

The geometry presented in Figure 2.6.4 is produced by three intersecting coherent polarized standing waves, which can be called the trisine wave function. This can be compared to figure 2.6.5, which presents the dimensional trisine geometry.

Equation 2.6.7 represents the trisine wave function $(\Psi)$. This wave function is the addition of three sin functions that are 120 degrees from each other in the x y plane and form an angle 22.8 degrees with this same x y plane. Also note that 22.8 degrees is related to the $(B/A)$ ratio by $\tan(90^0 - 22.8^0) = B/A$.

$$\Psi = \exp\left(\sin\left(2\left(a_{x1}x_0 + b_{x1}y_0 + c_{x1}z_0\right)\right)\right)$$
$$+ \exp\left(\sin\left(2\left(a_{x2}x_0 + b_{x2}y_0 + c_{x2}z_0\right)\right)\right) \qquad (2.6.7)$$
$$+ \exp\left(\sin\left(2\left(a_{x3}x_0 + b_{x3}y_0 + c_{x3}z_0\right)\right)\right)$$

Where:

$a_{x1} = \cos(030^0)\cos(22.8^0) \quad b_{x1} = \sin(030^0)\cos(22.8^0)$

$c_{x1} = -\sin(030^0)$

$$a_{x2} = \cos\left(150^0\right)\cos\left(22.8^0\right) \quad b_{x2} = \sin\left(150^0\right)\cos\left(22.8^0\right)$$

$$c_{x2} = -\sin\left(150^0\right)$$

$$a_{x3} = \cos\left(270^0\right)\cos\left(22.8^0\right) \quad b_{x3} = \sin\left(270^0\right)\cos\left(22.8^0\right)$$

$$c_{x3} = -\sin\left(270^0\right)$$

**Figure 2.6.5** Trisine Geometry in the Fluxoid Mode

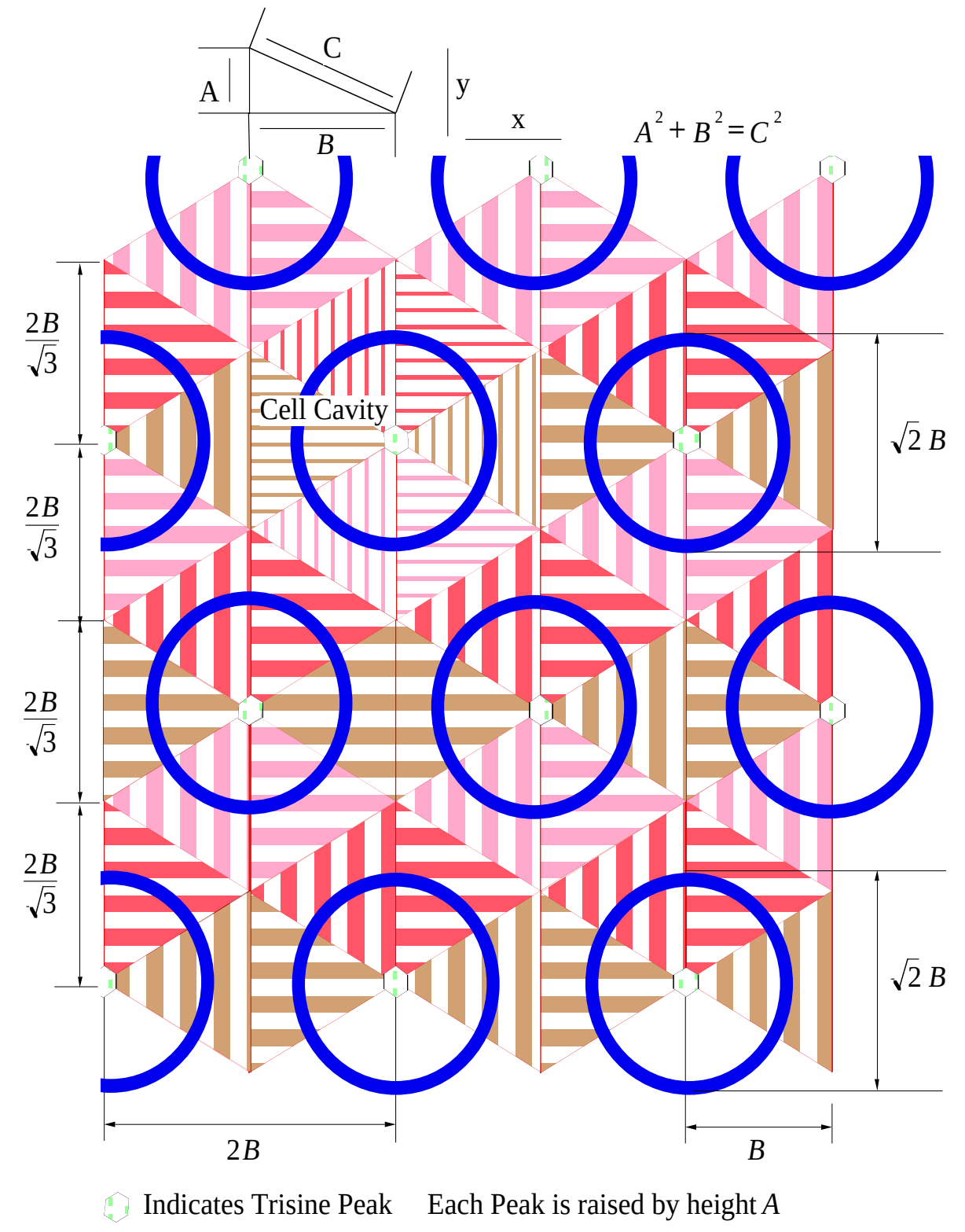

In table 2.6.2, note that the ratio of London penetration length $(\lambda)$ to Ginzburg-Landau correlation distance $(\xi)$ is the constant $(\kappa)$ of 58. The trisine values compares favorably with London penetration length $(\lambda)$ and constant $(\kappa)$ of 1155(3) Å and 68 respectively as reported in reference [11, 17] based on experimental data.

**Table 2.6.2 Critical Fields and Lengths**

| | | | | | | |
|---|---|---|---|---|---|---|
| $T_c\,(^0K)$ | 8.95E+13 | 2,135 | 1,190 | 447 | 93 | 8.11E-16 |
| $T_b\,(^0K)$ | 9.07E+14 | 4.43E+09 | 3.31E+09 | 2.03E+09 | 9.24E+08 | 2.729 |
| $H_c$ (gauss) | 3.29E+19 | 1.74E+06 | 8.36E+05 | 2.46E+05 | 3.46E+04 | 1.64E-17 |
| $H_{c1}$ (gauss) | 4.36E+16 | 2.30E+03 | 1.11E+03 | 3.26E+02 | 4.57E+01 | 2.17E-20 |
| $H_c \frac{c}{v_{dx}} \frac{m_e}{m_t} \frac{2}{3}$ | 1.20E+16 | 1.30E+08 | 8.39E+07 | 4.02E+07 | 1.24E+07 | 1.99E-06 |
| $H_{c2}$ (gauss) | 2.49E+22 | 1.31E+09 | 6.32E+08 | 1.86E+08 | 2.61E+07 | 1.24E-14 |
| $n_c$ (/cm$^3$) | 4.66E+39 | 5.43E+23 | 2.26E+23 | 5.20E+22 | 4.94E+21 | 1.27E-04 |
| $\lambda$ (cm) | 1.05E-11 | 2.16E-06 | 2.89E-06 | 4.71E-06 | 1.03E-05 | 3.50E+03 |
| $\xi$ (cm) | 1.81E-13 | 3.72E-08 | 4.98E-08 | 8.12E-08 | 1.78E-07 | 60.30 |

Figure 2.6.5 is graph of generally available critical field data as compile in Weast [16].

**Figure 2.6.5** Critical Magnetic Field Data (gauss) compared to Trisine Model as a function of $T_c$

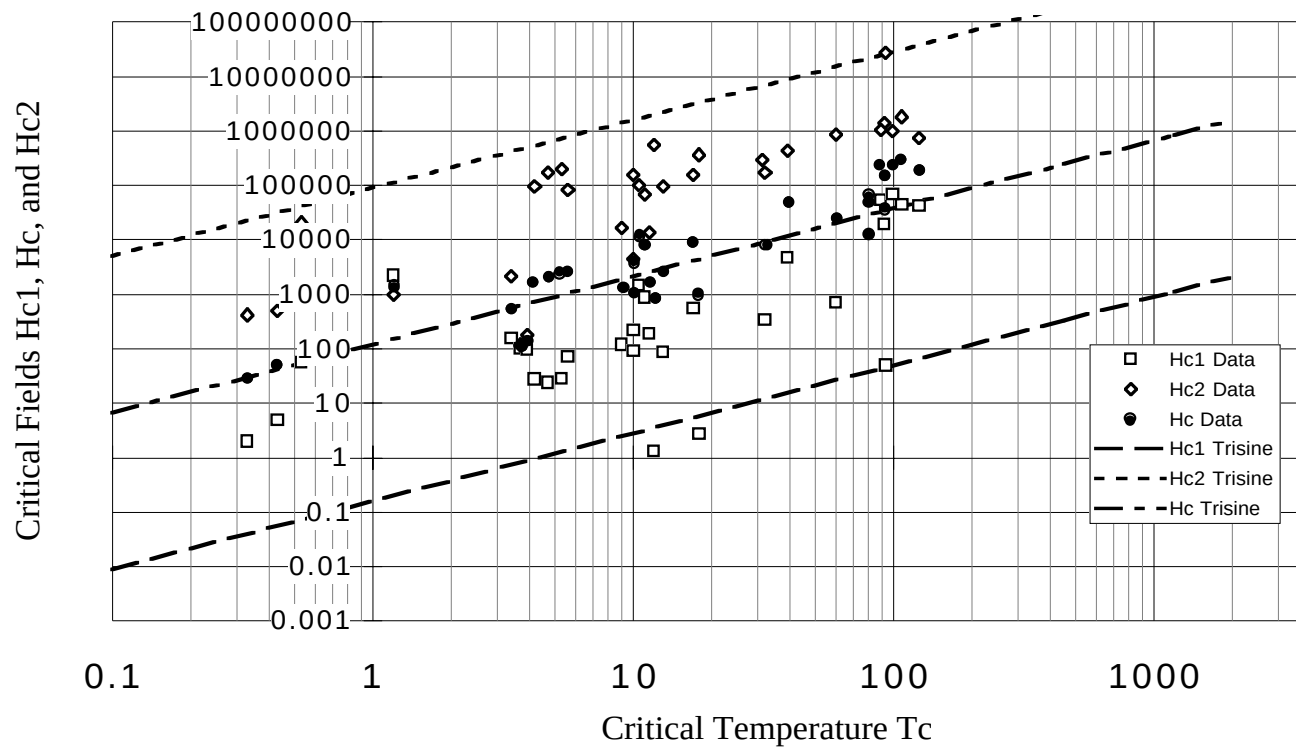

## 2.7 Superconductor Internal Pressure, Casimir Force And Deep Space Thrust Force (Pioneer 10 & 11)

The following pressure (P) values in Table 2.7.1 as based on equations 2.7.1 indicate rather high values as superconducting $(T_c)$ increases. This is an indication of the forces that must be contained in these anticipated materials or lattices. These forces are rather large for conventional materials especially at nuclear dimensions, but range to very small to the virtual lattice existing in outer space.

$$P = \left\{ \begin{array}{c} \dfrac{\hbar K_B}{time_\pm \ approach} \\ \dfrac{\hbar K_P}{time_\pm \ side} \\ \dfrac{\hbar K_A}{time_\pm \ section} \end{array} \right\} = \left\{ \begin{array}{c} \dfrac{F_x}{approach} \\ \dfrac{F_y}{side} \\ \dfrac{F_z}{section} \end{array} \right\} = \frac{\mathbb{C} k_b T_c}{cavity} \qquad (2.7.1a)$$

$$P = \frac{\sqrt{A^2 + B^2}}{B} \frac{cavity}{chain} \frac{\pi^2 \hbar \ v_{dx}}{240 \ A^4} \left( \begin{array}{c} \text{Casimir} \\ \text{pressure} \end{array} \right) \qquad (2.7.1b)$$

$$P = \left\{ \begin{array}{c} \dfrac{m_t v_{dx}}{time_\pm \ approach} \\ \dfrac{m_t v_{dy}}{time_\pm \ side} \\ \dfrac{m_t v_{dz}}{time_\pm \ section} \end{array} \right\} = \frac{\mathbb{C} \ k_b T_c}{cavity} \qquad (2.7.1c)$$

To put the calculated pressures in perspective, the C-C bond has a reported energy of 88 kcal/mole and a bond length of 1.54 Å [24]. Given these parameters, the internal chemical bond pressure $(CBP)$ for this bond is estimated to be:

$$88 \frac{kcal}{mole} 4.18x10^{10} \frac{erg}{kcal} \frac{1}{6.02x10^{23}} \frac{mole}{bond} \frac{1}{\left(1.54x10^{-8}\right)^3} \frac{bond}{cm^3}$$

$$= 1.67x10^{12} \frac{erg}{cm^3}$$

Within the context of the superconductor internal pressure numbers in table 2.7.1, the internal pressure requirement for the superconductor is less than the C-C chemical bond pressure at the order of design $T_c$ of a few thousand degrees Kelvin indicating that there is the possibility of using materials with the bond strength of carbon to chemically engineer high performance resonant superconducting materials.

In the context of interstellar space vacuum, the total pressure $m_t c^2/cavity$ may be available (potential energy) and will be of such a magnitude as to provide for universe expansion as astronomically observed. (see equation 2.11.21). Indeed, there is evidence in the Cosmic Gamma Ray Background Radiation for energies on the order of $m_t c^2$ or 56 Mev or ($\log_{10}$(Hz) = 22.13) or ( 2.20E-12 cm) [73]. In the available spectrum [73], there is a hint of a black body peak at 56 Mev or ($\log_{10}$(Hz) = 22.13) or ( 2.20E-12 cm). This radiation has not had an identifiable source until now.

**Table 2.7.1 with pressure (P) plot**

| | | | | | | |
|---|---|---|---|---|---|---|
| $T_c\,(^\circ K)$ | 8.95E+13 | 2,135 | 1,190 | 447 | 93 | 8.11E-16 |
| $T_b\,(^\circ K)$ | 9.07E+14 | 4.43E+09 | 3.31E+09 | 2.03E+09 | 9.24E+08 | 2.729 |
| *pressure* erg/cm$^3$ | 5.76E+37 | 1.60E+11 | 3.71E+10 | 3.21E+09 | 6.34E+07 | 1.42E-35 |
| *pressure* (*psi*) | 8.35E+32 | 2.32E+06 | 5.39E+05 | 4.66E+04 | 9.20E+02 | 2.06E-40 |
| *pressure* (*pascal*) | 5.76E+36 | 1.60E+10 | 3.71E+09 | 3.21E+08 | 6.34E+06 | 1.42E-36 |
| Total pressure erg/cm$^3$ | 1.36E+24 | 2.45E+19 | 1.02E+19 | 2.35E+18 | 2.23E+17 | 5.73E-09 |

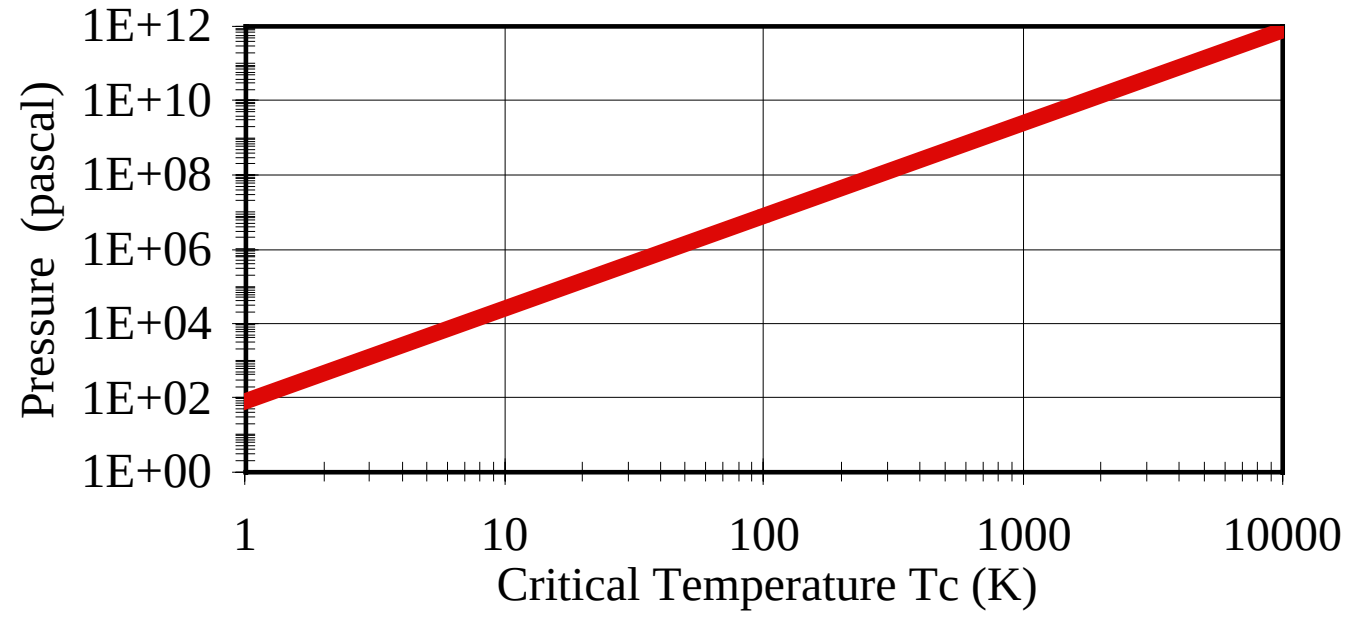

Assuming the superconducting current moves as a longitudinal wave with an adiabatic character, then equation 2.7.2 for current or de Broglie velocity ($v_{dx}$) holds where the ratio of heat capacity at constant pressure to the heat capacity at constant volume $(\delta)$ equals 1.

$$v_{dx} = \sqrt{\delta\ pressure \frac{cavity}{m_t}} \qquad (2.7.2)$$

The Casimir force is related to the trisine geometry by equation 2.7.3.

$$Casimir\ Force = g_s^3 \frac{section}{3} \frac{\pi^2 \hbar v_{dz}}{240 A^4} \qquad (2.7.3)$$

In a recent report[53] by Jet Propulsion Laboratory, an unmodeled deceleration was observed in regards to Pioneer 10 and 11 spacecraft as they exited the solar system. This deceleration can be related empirically to the following dimensional correct force expression (2.7.4) where $(m_t/cavity)$ is the mass equivalent energy density (pressure) $(m_t c^2/cavity)$ value at $T_c =\ 8.11E-16\ ^0K$ representing conditions in space, which is 100 percent, transferred to an object passing through it.

$$Force = -AREA\frac{m_t}{cavity}c^2 = -AREA(\rho_U)c^2 \qquad (2.7.4)$$

Note that equation 2.7.4 is similar to the conventional drag equation as used in design of aircraft with $v^2$ replaced with $c^2$. The use of $c^2$ vs. $v^2$ means that the entire resonant energy content *($m_t c^2$/cavity)* is swept out of volume as the spacecraft passes through it. Of course $Force = Ma$ , so the Pioneer spacecraft $M\,/\,AREA$ , is an important factor in establishing deceleration, which was observed to be on the order of 8.74E-8 cm$^2$/sec.

The force equation is repeated below as the conventional drag equation in terms of Pioneer spacecraft and the assumption that space density $(\rho_U)$ (2.11.16) is essentially what has been determined to be a potential candidate for dark matter:

$$F = Ma = C_d A_c \rho_U \frac{v^2}{2}$$

$$where:\ C_d = \frac{24}{R_e} + \frac{6}{1+\sqrt{R_e}} + .4 \quad [60]\ \text{empirical} \qquad (2.7.4)$$

and where:

$F$ = force

$M$ = Pioneer 10 & 11 Mass

$a$ = acceleration

$A_c$ = Pioneer 10 & 11 cross section

When the Reynolds' number ($R_e$) is low (laminar flow condition), then $C_d = (24/R_e)$ and the drag equation reduces to the Stokes' equation $F = 3\pi\mu v D$. [60]. When the Pioneer spacecraft data is fitted to this drag equation, the computed drag force is 6 orders of magnitude lower than observed values. $\mu$ (fluid absolute viscosity)

In other parts of this report, a model is developed as to what this dark matter actually is. The model dark matter is keyed to a 6.38E-30 g/cc value $(\rho_U)$ (2.11.16) in trisine model (NASA observed value of 6E-30 g/cc). The proposed dark matter consists of mass units of $m_t$ (110.12275343 x electron mass) per volume (Table 2.2.2 *cavity*- 15733 cm$^3$). These mass units are virtual particles that exist in a coordinated lattice at base energy $\hbar\omega/2$ under the condition that momentum and energy are conserved - in other words complete elastic character. This dark matter lattice would have internal pressure (table 2.7.1) to withstand collapse into gravitational clumps.

Under these circumstances, the traditional drag equation form is valid but velocity (v) should be replaced by speed of light (c). Conceptually, the equation 2.7.4 becomes a thrust equation rather than a drag equation. Correspondingly, the space viscosity is computed as momentum/area or $(m_t c/(2\Delta x^2))$ where $\Delta x$ is uncertainty dimension in table 2.5.1 at $T_c$ = 8.1E-16 K and as per equation 2.1.20. This is in general agreement with gaseous kinetic theory [19].

The general derivation is in one linear dimensional path (s) and in accordance with Newton's Second Law, Resonant Mass and Work-Energy Principles in conjunction with the concept of Momentum Space, which defines the trisine elastic space lattice. These considerations are developed and presented in Appendix F with linear path 's' used here.

$$\int_B^0 F ds = m_t \int_{v_{dx}+vo}^{c} \frac{v}{\sqrt{1-\frac{v^2}{c^2}}} dv = m_t \int_{k_m\varepsilon+\frac{v_o^2}{c^2}}^{1} \frac{-1}{\alpha^2} \frac{c^2}{\sqrt{1-\frac{1}{\alpha}}} d\alpha \qquad (2.7.5)$$

then

$$F\ s\Big|_B^0 = -m_t c^2 \sqrt{1-\frac{v^2}{c^2}}\ \Bigg|_{v_{dx}+vo}^{c} = -m_t c^2 \sqrt{1-\frac{1}{\alpha}}\ \Bigg|_{k_m\varepsilon+\frac{v_o^2}{c^2}}^{1}$$

$$\approx -m_t c^2 = -\frac{(m_t c)^2}{m_t} \text{ at } v_{dx} << c \qquad (2.7.5a)$$

$$\text{and } \left(k_m\varepsilon + \frac{v_o^2}{c^2}\right) >> 1$$

$$= 0 \text{ at } v = c \text{ (photon has zero mass)}$$

Clearly, an adiabatic process is described wherein the change in internal energy of the system (universe) is equal in absolute magnitude to the work (related to objects moving through the universe) or Work = Energy. Momentum is transferred from the trisine elastic space lattice to the object moving through it.

The hypothesis can be stated as:

**An object moving through momentum space will slow down.**

It is understood that the above development assumes that $dv/dt \neq 0$, an approach which is not covered in standard texts[61], but is assumed to be valid here because of it describes in part the actual physical phenomenon taking place when an object impacts or collides with an trisine elastic space lattice cell. The elastic space lattice cell essentially collapses $(B \to 0)$ and the entire cell momentum $(m_t c)$ is transferred to the object with cell permittivity $(\varepsilon)$ and permeability $(k_m)$ proceeds to a value of one(1).

In terms of an object passing through the trisine elastic space CPT lattice at some velocity $(v_o)$, the factor $(Fs/m_t)$ is no longer dependent of object velocity $(v_o)$ as defined in $(v_{dx} < v_o << c)$ but is a constant relative to $c^2$ which reflects the fact that the speed of light is a constant in the universe. Inspection would indicate that as $(v_o)$ is in range $(v_{dx} < v_o << c)$ then the observed object deceleration "a" is a constant (not tied to any particular reference frame either translational or rotational):

$$a = \frac{F}{M}$$

and specifically for object and trisine elastic space lattice relative velocities $(v_o)$ as follows:

The sun referenced Pioneer velocity of about 12 km/s
or earth (around sun) orbital velocity of 29.8 km/sec
or sun (circa milky way center) rotation velocity of 400 km/sec
or the CMBR dipole of 620 km/s
or any other velocity 'v' magnitude (0<v<<c)

do not contribute to the observed Pioneer deceleration
(as observed)
(noting that c ~ 300,000 km/sec).
and in consideration of the following established relationship:

$$F \approx -C_t \frac{m_t c^2}{s} = C_t A_c \rho_U c^2 \qquad (2.7.6)$$

Where $C_t$ is in the form of the standard fluid mechanical drag coefficient but is defined as a thrust coefficient.

This model was used to analyze the Pioneer Unmodeled Deceleration data 8.74E-08 cm/sec$^2$ and there appeared to be a good fit with the observed space density $(\rho)$ 6E-30 g/cm$^3$ (trisine model $(\rho)$ 6.38E-30 g/cm$^3$). A thrust coefficient of 59.67 indicates a laminar flow condition. It is interesting to note that this value correlates well with $m_t/(2m_e)$ or 55.06. An absolute space viscosity $(\mu)$ of 1.21E-16 g/(cm sec) is established within trisine model and used.

Pioneer (P) Translational Calculations

| | | |
|---|---|---|
| P mass | 241,000 | gram |
| P diameter | 274 | cm (effective) |
| P cross section | 58,965 | cm$^2$ (effective) |
| P area/mass | 0.24 | cm$^2$/g |
| P velocity | 1,117,600 | cm/sec |
| space kinematic viscosity | 1.90E+13 | cm$^2$/sec |
| space density | 6.38E-30 | g/cm$^3$ |
| Pioneer Reynold's number | 4.31E-01 | unitless |
| thrust coefficient | 59.67 | |
| P deceleration | 8.37E-08 | cm/sec$^2$ |
| one year thrust distance | 258.51 | mile |
| laminar thrust force | 9.40E-03 | dyne |
| Time of object to stop | 1.34E+13 | sec |

JPL, NASA raised a question concerning the universal application of the observed deceleration on Pioneer 10 & 11. In other words, why do not the planets and their satellites experience such an deceleration? The answer is in the $A_c/M$ ratio in the modified thrust relationship.

$$a \approx -C_t \frac{m_t c^2}{s} = -C_t \frac{A_c}{M} \rho_U c^2 = -C_t \frac{A_c}{M} \frac{m_t}{cavity} c^2 \qquad (2.7.6)$$

Using the earth as an example, the earth differential movement per year due to modeled deceleration of 4.94E-19 $cm/\sec^2$ would be 0.000246 centimeters (calculated as $at^2/2$, an unobservable distance amount). Also assuming the trisine superconductor model, the reported 6 nanoTesla (nT) (6E-5 gauss) interplanetary magnetic field (IMF) in the vicinity of earth, (as compared to .2 nanoTesla (nT) (2E-6 gauss [77]) interstellar magnetic field congruent with CMBR) will not allow the formation of a superconductor CPT lattice due to the well known properties of a superconductor in that magnetic fields above critical fields will destroy it. In this case, the space superconductor at $T_c$ = 8.1E-16 K will be destroyed by a magnetic field above

$H_c(c/v_{dx})(m_e/m_t)(chain/cavity)$ = 2.E-6 gauss

as in table 2.6.2. It is conceivable that the IMF decreases by some power law with distance from the sun, and perhaps at some distance the IMF diminishes to an extent wherein the space superconductor CPT lattice is allowed to form. This may be an explanation for the observed "kick in" of the Pioneer spacecraft deceleration phenomenon at 10 Astronomical Units (AU).

Earth Translational Calculations

| | | |
|---|---|---|
| Earth mass | 5.98E+27 | gram |
| Earth diameter | 1.27E+09 | cm |
| Earth cross section | 1.28E+18 | $cm^2$ |
| Earth area/mass | 2.13E-10 | $cm^2/g$ |
| Earth velocity | 2,980,010 | cm/sec |
| space kinematic viscosity | 1.90E+13 | $cm^2/sec$ |
| space density | 6.38E-30 | $g/cm^3$ |
| Earth Reynold's number | 2.01E+06 | unitless |
| thrust coefficient | 0.40 | unitless |
| thrust force (with $C_t$) | 2.954E+09 | dyne |
| Earth deceleration | 4.94E-19 | $cm/sec^2$ |
| One year thrust distance | 0.000246 | cm |
| laminar thrust force | 4.61E-01 | dyne |
| Time for Earth to stop rotating | 6.03E+24 | sec |

Now to test the universal applicability of the unmodeled Pioneer 10 & 11 decelerations for other man made satellites, one can use the dimensions and mass of the Hubble space telescope. One arrives at a smaller deceleration of 6.79E-09 $cm/sec^2$ because of its smaller area/mass ratio. This low deceleration is swamped by other decelerations in the vicinity of earth, which would be multiples of those well detailed in reference 1, Table II Pioneer Deceleration Budget. Also the trisine elastic space CPT lattice probably does not exist in the immediate vicinity of the earth because of the earth's milligauss magnetic field, which would destroy the coherence of the trisine elastic space CPT lattice.

Hubble Satellite Translational Calculations

| | | |
|---|---|---|
| Hubble mass | 1.11E+07 | gram |
| Hubble diameter | 7.45E+02 | centimeter |
| Hubble cross section | 5.54E+05 | $cm^2$ |
| Hubble area/mass | 0.0499 | $cm^2/g$ |
| Hubble velocity | 790,183 | cm/sec |
| space kinematic viscosity | 1.90E+13 | $cm^2/sec$ |
| space density | 6.38E-30 | $g/cm^3$ |
| Hubble Reynolds' number | 1.17E+00 | unitless |
| thrust coefficient | 23.76 | |
| thrust force (with $C_t$) | 7.549E-02 | dyne |
| Hubble deceleration | 6.79E-09 | $cm/sec^2$ |
| One year thrust distance | 3,378,634 | cm |
| laminar thrust force | 7.15E-08 | dyne |
| Time for Hubble to stop rotating | 1.16E+14 | sec |

The fact that the unmodeled Pioneer 10 & 11 deceleration data are statistically equal to each other and the two space craft exited the solar system essentially 190 degrees from each other implies that the supporting fluid space density through which they are traveling is co-moving with the solar system. But this may not be the case. Assuming that the supporting fluid density is actually related to the cosmic background microwave radiation (CMBR), it is known that the solar system is moving at 500 km/sec relative to CMBR. Equivalent decelerations in irrespective to spacecraft heading relative to the CMBR would further support the hypothesis that deceleration is independent of spacecraft velocity.

**Figure 2.7.1 Pioneer Trajectories (NASA JPL [53]) with delineated solar satellite radial component r.**

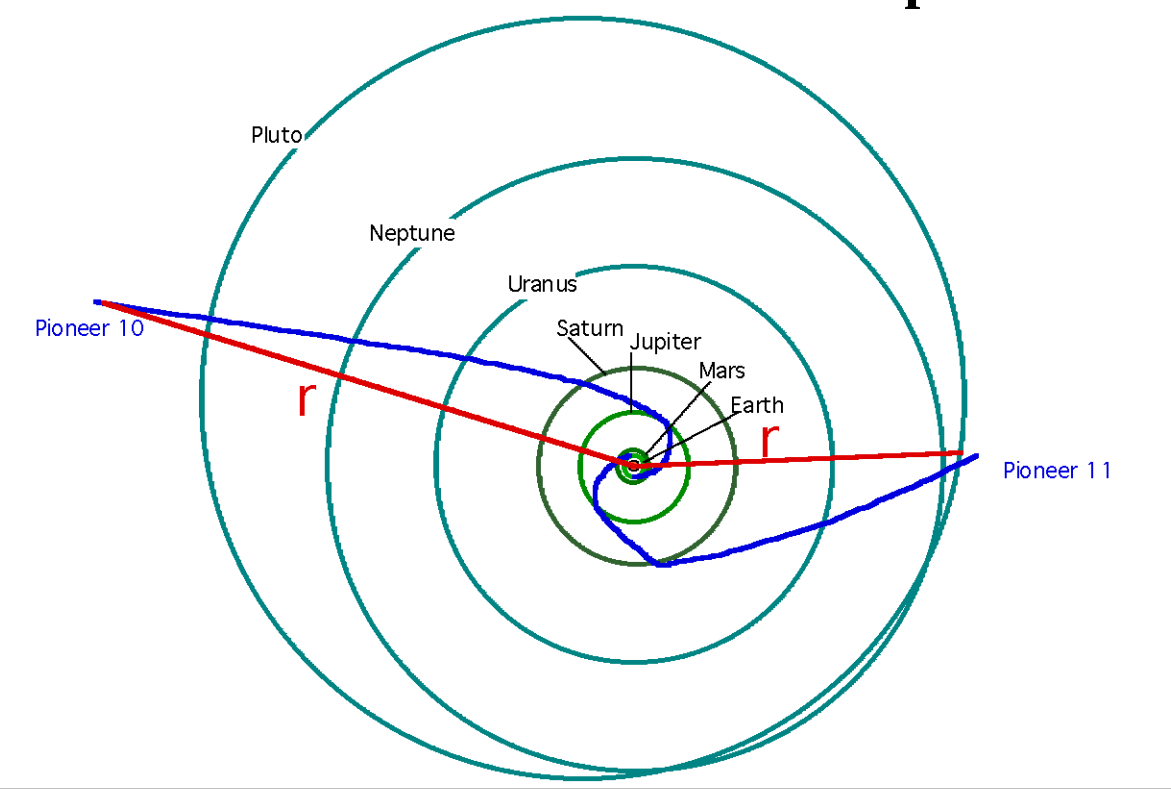

Executing a web based [58] computerized radial rate dr/dt (where r is distance of spacecraft to sun) of Pioneer 10 & 11 Figure 2.7.1), it is apparent that the solar systems exiting velocities are not equal. Based on this graphical method, the spacecraft's velocities were in a range of about 9% (28,600 – 27,500 mph for Pioneer 10 "3Jan87 - 22July88" and 26,145 – 26,339 mph for Pioneer 11 "5Jan87 - 01Oct90"). This reaffirms the basic thrust equation used for computation in that, velocity magnitude of the spacecraft is not a factor, only its direction with the thrust force opposite to that direction. The observed equal deceleration at different velocities would appear to rule out deceleration due to Kuiper belt dust as a source of the deceleration for a variation at $1.09^2$ or 1.19 would be expected.

Also it is observed that the Pioneer spacecraft spin is slowing down from 7.32 rpm in 1987 to 7.23 rpm in 1991. This equates to an angular deceleration of .0225 rpm/year or 1.19E-11 $rotation/sec^2$. JPL[57] has acknowledged some systematic forces that may contribute to this deceleration rate and suggests an unmodeled deceleration rate of .0067 rpm/year or 3.54E-12 rotations/sec. [57]

Now using the standard torque formula:

$$\Gamma = I\dot{\omega} \tag{2.7.7}$$

The deceleration value of 3.54E-12 $rotation/sec^2$ can be replicated assuming a Pioneer Moment of Inertia about spin axis(I) of 5.88E+09 g $cm^2$ and a 'paddle' cross section area of 3,874 $cm^2$ which is 6.5 % spacecraft 'frontal" cross section which seems reasonable. The gross spin deceleration rate of .0225 rpm/year or 1.19E-11 rotation/sec$^2$ results in a pseudo paddle area of 13,000 cm$^2$ or 22.1% of spacecraft 'frontal" cross section which again seems reasonable.

Also it is important to note that the angular deceleration rate for each spacecraft (Pioneer 10, 11) is the same even though they are spinning at the two angular rates 4 and 7 rpm respectfully. This would further confirm that velocity is not a factor in measuring the space mass or energy density. The calculations below are based on NASA JPL problem set values[57]. Also, the sun is moving through the CMBR at 600 km/sec. According to this theory, this velocity would not be a factor in the spacecraft deceleration.

Pioneer (P) Rotational Calculations

| | | |
|---|---|---|
| P mass | 241,000 | g |
| P moment of inertia | 5.88E+09 | g cm$^2$ |
| P diameter | 274 | cm |
| P translational cross section | 58,965 | cm$^2$ |
| P radius of gyration k | 99 | cm |
| P radius r | 137 | cm |
| paddle cross section | 3,874 | cm$^2$ |
| paddle area/mass | 0.02 | cm$^2$/g |
| P rotation speed at k | 4,517 | cm/sec |
| P rotation rate change | 0.0067 | rpm/year |
| P rotation rate change | 3.54E-12 | rotation/sec$^2$ |
| P rotation deceleration at k | 2.20E-09 | cm/sec$^2$ |
| P force slowing it down | 1.32E-03 | dyne |
| space kinematic viscosity | 1.90E+13 | cm$^2$/sec |
| space density | 6.38E-30 | g/cm$^3$ |
| P Reynolds' number | 4.31E-01 | unitless |
| thrust coefficient | 59.67 | |
| thrust force | 1.32E-03 | dyne |
| One year rotational thrust distance | 1,093,344 | cm |
| P rotational laminar thrust force | 7.32E-23 | dyne |
| Time for P to stop rotating | 2.05E+12 | sec |

These calculations suggest a future spacecraft with general design below to test this hypothesis. This design incorporates general features to measure translational and rotational thrust. A wide spectrum of variations on this theme is envisioned. General Space Craft dimensions should be on the order of Trisine geometry B at $T_b = 2.711\,K \;\; T_c = 8.12E-16\,K$ or 22 cm (Table 2.2.1) or larger. The Space Craft Reynold's number would be nearly 1 at this dimension and the resulting thrust the greatest. The coefficient of thrust would decrease towards one with larger Space Craft dimensions. It is difficult to foresee what will happen with Space Craft dimensions below 22 cm.

The spacecraft (Figure 2.7.2) would rotate on green bar axis while traveling in the direction of the green bar. The red paddles would interact with the space matter as mc$^2$ slowing the rotation with time in conjunction with the overall spacecraft slowing decelerating.

A basic experiment cries out here to be executed and that is to launch a series of spacecraft of varying $A_c/M$ ratios and exiting the solar system at various angles while measuring their rotational and translational decelerations. This would give a more precise measure of the spatial orientation of dark matter in the immediate vicinity of our solar system. An equal calculation of space energy density by translational and rotational observation would add credibility to accumulated data.

**Figure 2.7.2 Space craft general design to quantify translational and rotary thrust effect. Brown is parabolic antenna, green is translational capture element and blue is rotational capture element.**

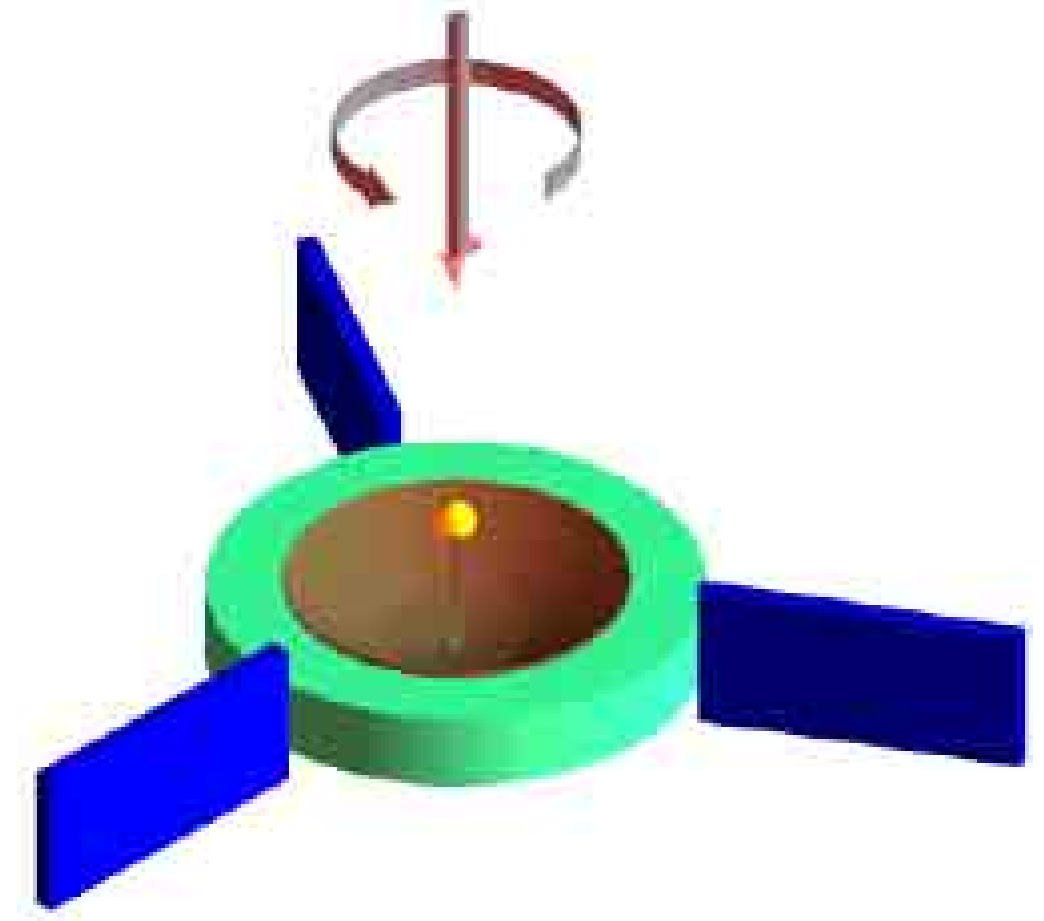

The assumption that the Pioneer spacecraft will continue on forever and eventually reach the next star constellation may be wrong. Calculations indicate the Pioneers will come to a stop relative to space fluid in about 400,000 years. We may be more isolated in our celestial position than previously thought.

When this thrust model is applied to a generalized design for a Solar Sail [63], a deceleration of 3.49E-05 cm/sec$^2$ is calculated which is 3 orders higher than experienced by Pioneer spacecraft. Consideration should be given to spin paddles for these solar sails in order to accumulate spin deceleration data also. In general, a solar sail with its characteristic high $A_c/M$ ratio such as this example should be very sensitive to the trisine elastic space CPT lattice. The integrity of the sail should be maintained beyond its projected solar wind usefulness as the space craft travels into space beyond Jupiter.

Sail Deceleration Calculations

| | | |
|---|---|---|
| Sail mass | 300,000 | gram |
| Sail diameter | 40,000 | centimeter |
| Sail cross section | 1,256,637,000 | cm$^2$ |
| Sail area/mass | 4,189 | cm$^2$/g |
| Sail velocity | 1,117,600 | cm/sec |
| space kinematic viscosity | 1.90E+13 | cm$^2$/sec |
| space density | 6.38E-30 | g/cm$^3$ |
| Sail Reynold's number | 6.30E+01 | unitless |
| thrust coefficient | 1.45 | |
| thrust force | 10 | dyne |
| Sail acceleration | 3.49E-05 | $cm/sec^2$ |
| one year thrust distance | 1.73E+10 | cm |
| laminar thrust force | 1.37E+00 | dyne |
| Time of Sail to stop | 1015 | year |

The following calculation establishes a distance from the sun at which the elastic space CPT lattice kicks in. There is a energy density competition between the solar wind and the elastic space CPT lattice. Assuming the charged particle density at earth is 5 / $cm^3$ [57] and varying as $1/R^2$ and comparing with elastic space density of kTc/cavity, then an equivalent energy density is achieved at 3.64 AU. This of course is a dynamic situation with solar wind changing with time, which would result in the boundary of the elastic space CPT lattice changing with time. This is monitored by a space craft (Advanced Composition Explorer (ACE)) at the L1 position (1% distance to sun) and can be viewed at http://space.rice.edu/ISTP/dials.html. A more descriptive term would be making the elastic space CPT lattice decoherent or coherent. It is interesting to note that the calculated boundary is close to that occupied by the Astroids and is considered the Astroid Belt.

| Elastic Space CPT lattice intersecting with solar wind | AU | |
|---|---|---|
| Mercury | 0.39 | |
| Venus | 0.72 | |
| Earth | 1.00 | |
| Mars | 1.53 | |
| Jupiter | 5.22 | |
| Saturn | 9.57 | |
| Uranus | 19.26 | |
| Neptune | 30.17 | |
| Pluto | 39.60 | |
| AU | 3.64 | |
| n at AU | 0.38 | $cm^{-3}$ |
| $r$ $1/n^{1/3}$ | 1.38 | cm |
| coordination # | 6 | unitless |
| *volume* | 0.44 | $cm^3$ |
| $\frac{e^2}{r}\ \frac{1}{volume}\ \frac{T_c}{T_b}\ \frac{energy}{volume}$ | 1.13E-34 | erg / $cm^3$ |
| Trisine $\frac{k_b T_c}{cavity}\ \frac{energy}{volume}$ | 1.13E-34 | erg / $cm^3$ |

Now assuming the basic conservation of total kinetic and potential energy for an object in circular orbit of the sun with velocity $(v_0)$ and at distance $(R_0)$ equation 2.7.8 provides an indication of the time required to perturb the object out of the circular orbit in the direction of the sun.

$$v_0^2 + \frac{GM_\odot}{R_0} = v^2 + (v_0 - at)^2 \qquad (2.7.8a)$$

$$v_0^2 + \frac{GM_\odot}{R_0} = v^2 + \left(v_0 - C_t \frac{Area}{M} \rho c^2 t\right)^2 \qquad (2.7.8b)$$

$$where \quad v = \delta v_0$$

Assuming the sun orbital velocity perturbing factor $\delta$ to be .9 and the time (t) at the estimated age of the solar system (5,000,000,000 years), then the smallest size objects that could remain in sun orbit are listed below as a function of distance (AU) from the sun.

| | AU | Space object upper size diameter (meter) |
|---|---|---|
| Mercury | 0.39 | 9.04 |
| Venus | 0.72 | 10.67 |
| Earth | 1.00 | 11.63 |
| Mars | 1.53 | 13.04 |
| Jupiter | 5.22 | 18.20 |
| Saturn | 9.57 | 21.50 |
| Uranus | 19.26 | 26.10 |
| Neptune | 30.17 | 29.60 |
| Pluto | 39.60 | 31.96 |
| Kuiper Belt | 50 | 35.95 |
| | 60 | 37.57 |
| | 70 | 39.03 |
| Oort Cloud? | 80 | 40.36 |
| | 100 | 41.60 |
| | 150 | 46.75 |
| | 200 | 50.82 |
| | 300 | 57.20 |
| | 400 | 62.24 |

This calculation would indicate that there is not small particle dust remaining in the solar system, which would include the Kuiper Belt and hypothesized Oort Cloud. Also, the calculated small object size would seem to be consistent with Asteroid Belt objects, which exist at the coherence/decoherence boundary for the elastic space CPT lattice.

This analysis has implications for large-scale galactic structures as well. Indeed, there is evidence in the Cosmic Gamma Ray Background Radiation for energies on the order of $m_t c^2$ or 56 Mev or ($\log_{10}$(Hz) = 22.13) or (2.20E-12 cm) [73]. In the available spectrum [73], there is a hint of a black body peak at 56 Mev or ($\log_{10}$(Hz) = 22.13) or ( 2.20E-12 cm). This radiation has not had an identifiable source until now. It is conceivable that energy densities may vary spatially within a galaxy with generally increased energy density towards the center. This would imply that a generally radial boundary condition would exist within a galaxy and may be responsible for observed anomalous dark matter galactic rotation. Apparent clumping of galactic matter may be due to localities where elastic space CPT lattice does not exist due to decoherent effects of various energy sources existing at that particular locality. Dark energy and Dark matter may be one and the same. Reference [76] indicates the very existence of dark matter and just such clumping thereof as well as a thrust effect on the dark matter on galactic material. Reference [79] substantiates the independence of dark and visible matter distribution. According to this correlation, the observed dark matter distributions are due to discrete quantum mechanical states of the ubiquitous dark matter which is uniform and currently unobservable in most of space but clumps in the vicinity of massive objects such as galaxies or residual effects of historical massive collision events.

Reference 64 presents an experimental protocol in which a comoving optical and sodium atom lattice is studied at nanoKelvin a temperature. When the moving lattice is stopped (generating laser beams extinguished), the sodium atoms stop rather than proceeding inertially. This unexplained anomaly may be explained in a similar manner as Pioneer deceleration. The following calculation indicates a sodium atom deceleration large enough to indicate essentially an instantaneous stop.

| | | |
|---|---|---|
| sodium temperature | 1.00E-06 | Kelvin |
| sodium mass | 3.82E-23 | gram |
| sodium van der Waals diameter | 4.54E-08 | centimeter |
| sodium cross-section | 1.62E-15 | $cm^2$ |
| sodium area/mass | 4.24E+07 | $cm^2/g$ |
| optical lattice velocity | 3.0 | cm/sec |
| space kinematic viscosity | 1.90E+13 | $cm^2/sec$ |
| space density | 6.38E-30 | g/ $cm^2$ |
| sodium Reynold's number | 7.15E-11 | unitless |
| thrust coefficient | 3.36E+11 | unitless |
| thrust force (with thrust coefficient) | 3.116E-12 | dyne |
| sodium deceleration | 8.16E+10 | $cm/sec^2$ |
| time to stop sodium atom | 3.68E-11 | sec |
| distance to stop sodium atom | 5.51E-11 | cm |

An experiment is envisioned whereby larger atomic clusters are decelerated in a more observable manner because of their increased mass. It assumed that such an anomalous deceleration will occur only at nanoKelvin temperatures where the kT energies are essentially zero.

It is understood that the New Horizon spacecraft launched on January 19, 2006 presents an opportunity to replicate the Pioneer 10 & 11 deceleration. The tables below provide a predictive estimate of this deceleration. Because of the New Horizon smaller area to mass ratio than Pioneer, it is anticipated that the New Horizon anomalous deceleration will be smaller than the Pioneer anomalous deceleration by a factor of about .4. Also, because the New Horizon Moment of Inertia about the spin axis is smaller than Pioneer, it is predicted that the rate of spin deceleration will be greater for the New Horizon than observed for the Pioneer space crafts.

New Horizon (NH) Predictive Translation Calculations

| | |
|---|---|
| New Horizon mass | 470,000 gram |
| New Horizon diameter | 210 cm (effective) |
| New Horizon cross section | 34,636 $cm^2$ (effective) |
| New Horizon area/mass | 0.07 $cm^2/g$ |
| trisine area/mass | 5.58E+30 $cm^2/g$ |
| New Horizon velocity | 1,622,755 cm/sec |
| space kinematic viscosity | 1.90E+13 $cm^2/sec$ |
| space density | 6.38E-30 $g/cm^3$ |
| New Horizon Reynold's number | 3.31E-01 unitless |
| drag coefficient | 76.82 |
| drag force (with drag coefficient) | 1.525E-02 dyne |
| New Horizon deceleration | 3.24E-08 $cm/sec^2$ |
| one year drag distance | 16,132,184 cm |
| laminar drag force | 7.21E-03 dyne |
| Time of New Horizon to stop | 5.00E+13 sec |

New Horizon (NH) Predictive Rotational Calculations

| | | |
|---|---|---|
| NH mass | 470,000 | g |
| NH moment of inertia NASA JPL | 4.01E+09 | g $cm^2$ |
| NH diameter | 315 | cm |
| NH translational cross-section | 77,931 | $cm^2$ |
| NH radius of gyration k | 131 | cm |
| NH radius r | 158 | cm |
| paddle cross-section | 60,210 | $cm^2$ |
| paddle area/mass | 1.28E-01 | $cm^2/g$ |
| NH rotation speed at k | 5,971 | cm/sec |
| NH rotation rate change | 0.2000 | rpm/year |
| NH rotation rate change | 1.06E-10 | $rotation/sec^2$ |
| NH rotation deceleration at k | 8.68E-08 | $cm/sec^2$ |
| NH force slowing it down | 2.04E-02 | dyne |
| space kinematic viscosity | 1.90E+13 | $cm^2/sec$ |
| space density | 6.38E-30 | $g/cm^3$ |
| NH Reynold's number | 4.36E-01 | unitless |
| thrust coefficient | 59.09 | |
| thrust force (with thrust coefficient) | 2.04E-02 | dyne |
| One year rotational thrust distance | 43,139,338 | cm |
| NH rotational laminar thrust force | 1.90E-02 | dyne |
| Time for NH to stop rotating | 6.88E+10 | sec |

This New Horizon deceleration characteristics will be monitored in the vicinity of Jupiter passage on February 28, 2007 and thereafter.

## 2.8 Superconducting Resonant Current, Voltage And Conductance

The superconducting electrical current $(I_e)$ is expressed as in terms of a volume *chain* containing a Cooper CPT Charge conjugated pair moving through the trisine CPT lattice at a de Broglie velocity $(v_{dx})$.

$$\frac{I_e}{area} = \frac{Cooper\ e_{\pm}\ v_{dx}}{chain} \tag{2.8.1}$$

The superconducting mass current $(I_m)$ is expressed as in terms of a trisine resonant transformed mass $(m_t)$ per *cavity* containing a Cooper CPT Charge conjugated pair and the *cavity* Cooper CPT Charge conjugated pair velocity *($v_{dx}$)*.

$$I_m = \left(\frac{m_t}{cavity}\right)v_{dx} \tag{2.8.2}$$

It is recognized that under strict CPT symmetry, there would be zero current flow. It must be assumed that pinning of one side of CPT will result in observed current flow. Pinning (beyond that which naturally occurs) of course has been observed in real superconductors to enhance actual supercurrents. This is usually performed by inserting unlike atoms into the superconducting molecular lattice or mechanically distorting the lattice at localized points [72]. Also, there are reports that superconductor surface roughness enhances super currents by 30% [72]. This could be due to the inherent "roughness" of the CPT symmetry better expressing itself at a rough or corrugated surface rather than a smooth one.

Secondarily, this reasoning would provide reasoning why a net charge or current is not observed in a vacuum such as trisine model extended to critical space density in section 2.11.

**Table 2.8.1 with super current $\left(amp/cm^2\right)$ plot.**

| | | | | | | |
|---|---|---|---|---|---|---|
| $T_c\left({}^\circ K\right)$ | 8.95E+13 | 2,135 | 1,190 | 447 | 93 | 8.11E-16 |
| $T_b\left({}^\circ K\right)$ | 9.07E+14 | 4.43E+09 | 3.31E+09 | 2.03E+09 | 9.24E+08 | 2.729 |
| $I_e\left(amp/cm^2\right)$ | 5.56E+32 | 3.17E+11 | 9.83E+10 | 1.39E+10 | 6.01E+08 | 4.56E-26 |
| $I_m\left(g/cm^2/\sec\right)$ | 1.16E+26 | 6.61E+04 | 2.05E+04 | 2.90E+03 | 1.25E+02 | 9.52E-33 |

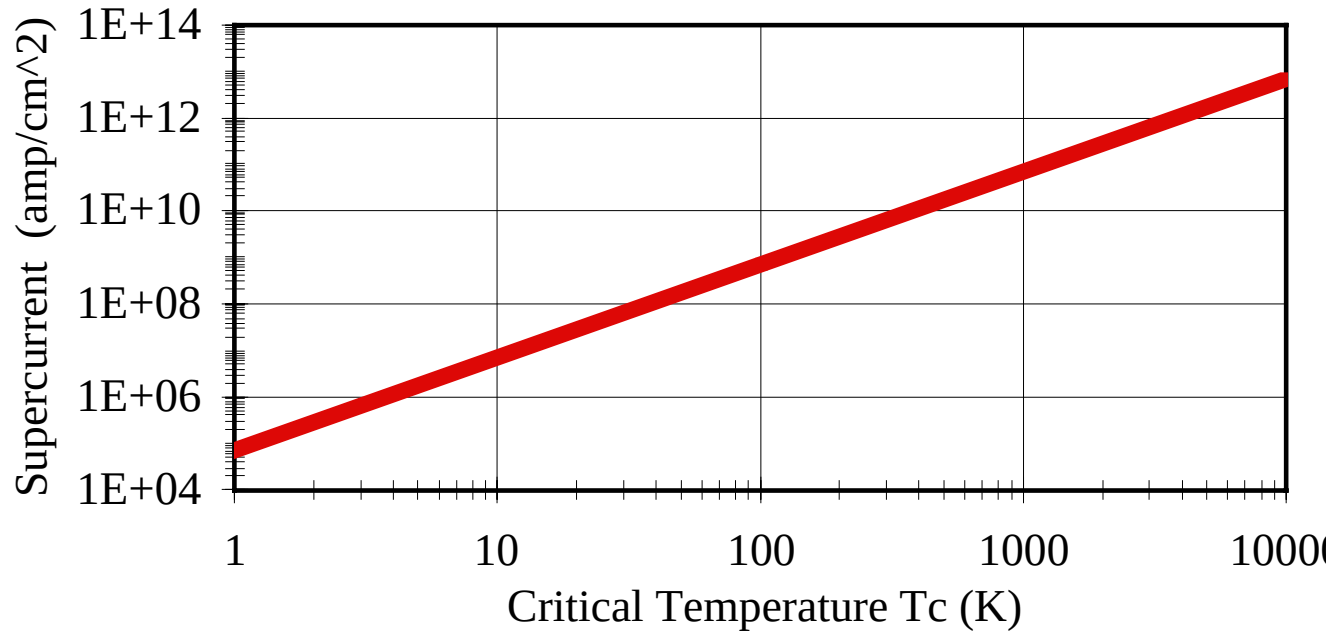

A standard Ohm's law can be expressed as follows with the resistance expressed as the Hall resistance with a value of 25,815.62 ohms (von Klitzing constant).

$$voltage = (\ current\ )\ (\ resistance)$$

$$\frac{k_b T_c}{\mathbb{C}\ e_\pm} = \left(\frac{\mathbb{C}\ e_\pm}{chain}\ approach\ v_{dx}\right)\left(\frac{2\pi\hbar}{\left(\mathbb{C}\ e_\pm\right)^2}\right) \qquad (2.8.3)$$

Equation 2.8.4 represents the Poynting vector relationship or in other words the power per area per resonant CPT time$_\pm$ being transmitted from each superconducting *cavity*:

$$k_b T_c = \frac{m_e}{m_t}\frac{1}{2}\ E_x\ H_c(2\ \text{side})\ time_\pm\ v_\varepsilon \qquad (2.8.4)$$

Equation 2.8.4 can be rearranged in terms of a relationship between the de Broglie velocity $(v_{dx})$ and modified speed of light velocity $(v_{\varepsilon x})$ as expressed in equation 2.8.5.

$$v_{\varepsilon x}^2 = \frac{\sqrt{3}}{48\pi^2}\left(\frac{m_t}{m_e}\right)^2\frac{\left(\mathbb{C}\ e_\pm\right)^2}{\hbar}\frac{B}{A}v_{dx} = 3^{\frac{1}{4}}\frac{B}{A}\cdot c\cdot v_{dx} = \pi\cdot c\cdot v_{dx} \qquad (2.8.5)$$

This leads to the group/phase velocity relationship (as previously stated in 2.1.30) in equation 2.8.6, which holds for the sub- and super- luminal nuclear conditions $\left(v_{dx} < c\ \ \&\ v_{\varepsilon x} < c\right)\&\left(v_{dx} > c\ \ \&\ v_{\varepsilon x} > c\right)$.

$$(group\ velocity)(phase\ velocity) = c^2$$

$$(v_{dx})\left(\frac{1}{\pi}\frac{v_{\varepsilon x}^2}{v_{dx}^2}\cdot c\right) \cong \frac{1}{2}(v_{dx})\left(\frac{chain}{cavity}\frac{v_{\varepsilon x}^2}{v_{dx}^2}\cdot c\right) \cong c^2 \qquad (2.8.6)$$

In the Nature reference [51] and in [59], Homes reports experimental data conforming to the empirical relationship Homes' Law relating superconducting density $\rho_s$ at just above $T_c$, superconductor conductivity $\sigma_{dc}$ and $T_c$.

$$\rho_s \propto \sigma_{dc} T_c \qquad \text{Homes' Law}\quad \text{ref: 51} \qquad (2.8.7)$$

The Trisine model is developed for $T_c$, but using the Tanner's law observation that $n_S = n_N/4$ (ref 52), we can assume that $E_{fN} = E_{fS}/4$. This is verified by equation 2.5.1. Now substituting $I_e/(E_{fN} area)$ for superconductor conductivity $\sigma_{dc}$, and using $\mathbb{C}/cavity$ for super current density $\rho_s$, the following relationship is obtained and is given the name Homes' constant.

Note: $\sigma_s(cgs) = 8.99377374\text{E+11 x } \sigma_s(\text{S/cm})$

$$\text{Homes' constant} = \frac{\sigma_{dc} k_b T_c}{\rho_s} = \frac{3\pi}{2}\frac{\mathbb{C}^2 e_\pm^2 \hbar}{m_t^2}\frac{A}{B}$$

$$= 191{,}537\,\frac{cm^5}{\sec^3} \qquad (2.8.8)$$

*or*

$$\rho_s = \frac{\sigma_{dc} k_b T_c}{191{,}537} \quad \text{(cgs units)}$$

Equation 2.8.8 is consistent with equation 2.8.3.

The results for both c axis and a-b plane (ref 51) are graphically presented in the following figure 2.8.1. On average the Homes' constant calculated from data exceeds the Homes' constant derived from Trisine geometry by 1.72.

**Figure 2.8.1** Homes' data compared to Trisine Model Homes' Constant

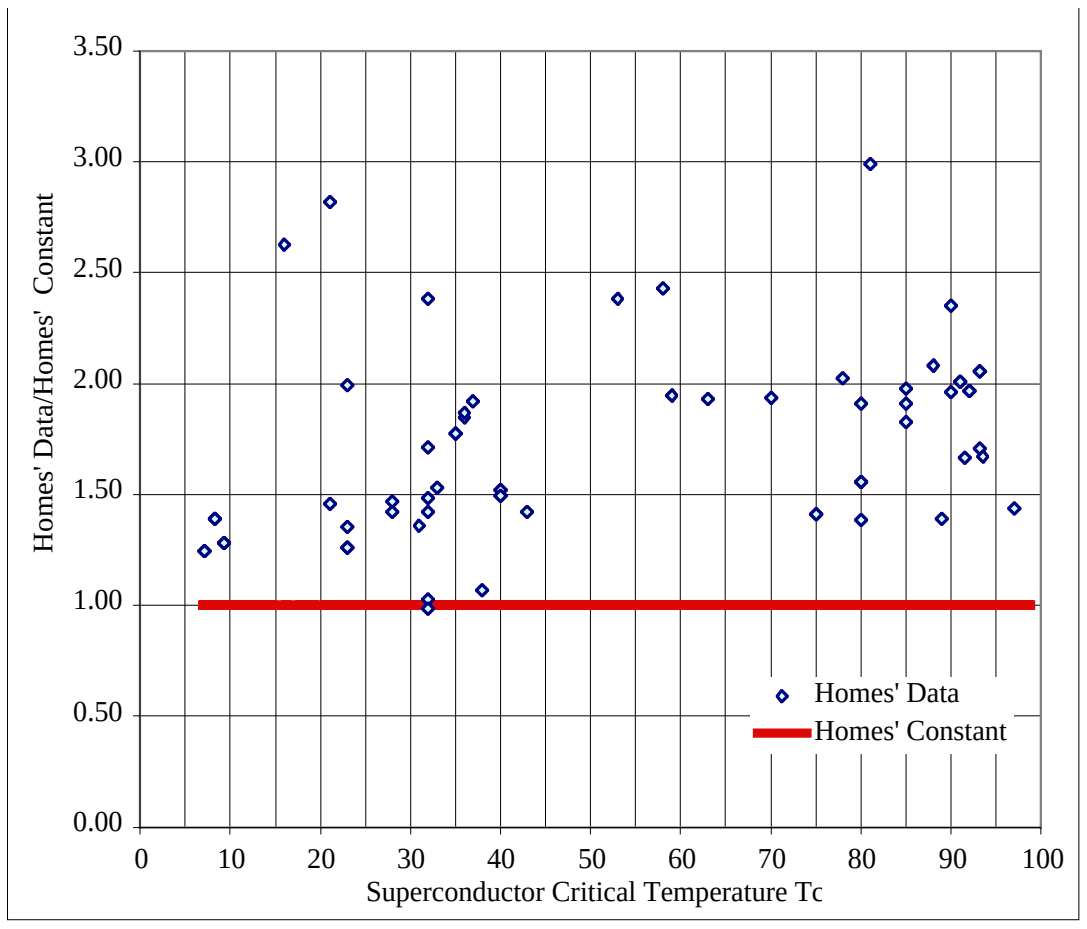

There may be some degrees of freedom considerations $1\cdot k_b T_c/2$, $2\cdot k_b T_c/2$, $3\cdot k_b T_c/2$ that warrant looking into that may account for this divergence although conformity is noteworthy. Also 1.72 is close to sqrt(3), a value that is inherent to Trisine geometry. This correspondence warrants a closer look. In general, the conventional and high temperature superconductor data falls in the same pattern. It is concluded that Homes' Law can be extrapolated to any critical temperature $T_c$.

## 2.9 Superconductor Apparent Weight Reduction In A Gravitational Field

Based on the superconducting helical or tangential velocity $(v_{dT})$ calculated from equation 2.4.5, the superconducting gravitational shielding effect $(\Theta)$ is computed based on Newton's gravitational law and the geometry as presented in Figure 2.9.1.

**Figure 2.9.1** Superconductor Gravitational Model Based on Super current mass flow with Mass Velocities $(v_{dx})$ and $(v_{dT})$.

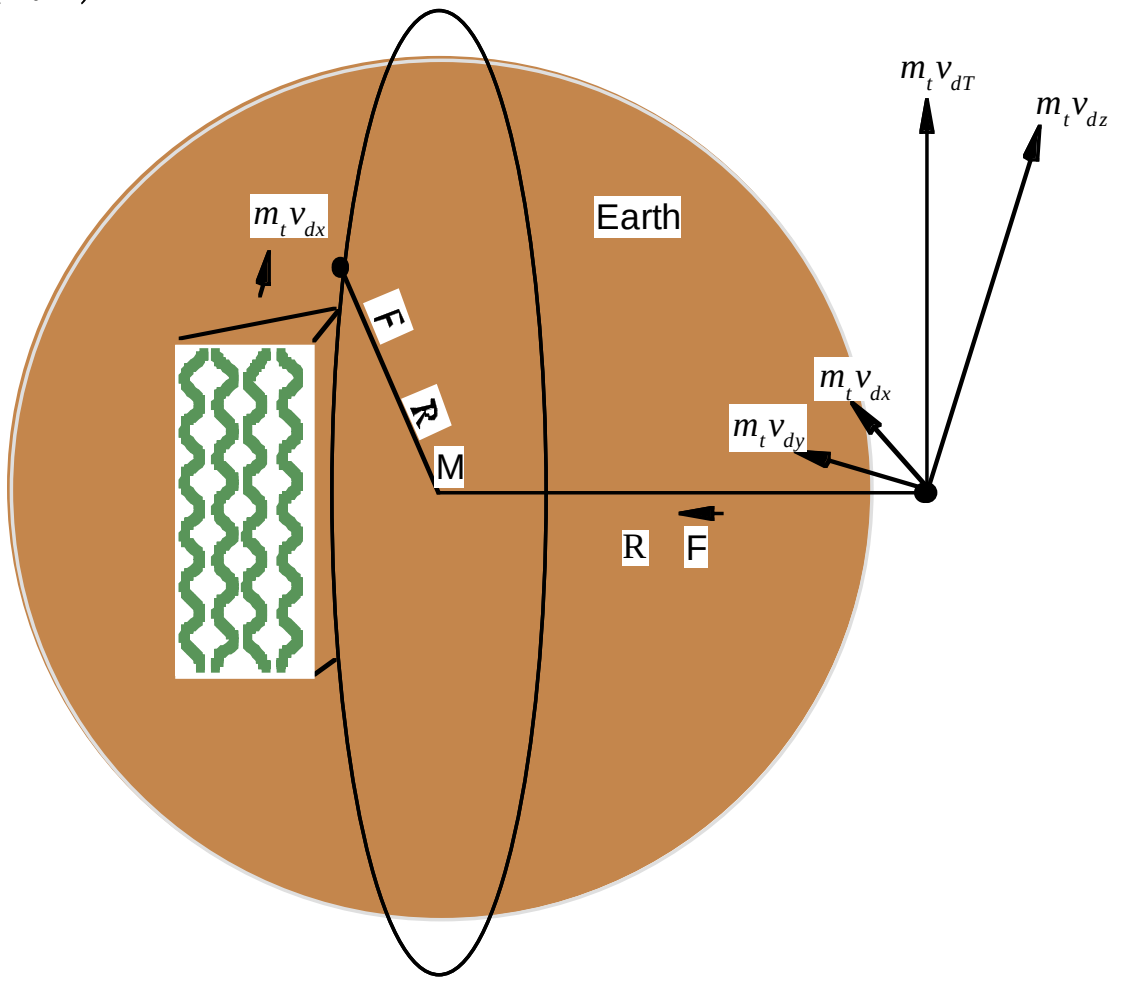

With a material density $(\rho)$ of 6.39 g/cc [17], the gravitational shielding effect $(\Theta)$ is .05% as observed by Podkletnov and Nieminen (see Table 2.9.1).

$$\rho_{\mathbb{C}} = \frac{m_t}{cavity} = \frac{n\,m_t}{2} \tag{2.9.1}$$

$$F_{\uparrow} = m_t \frac{v_d^2}{R} \qquad F_{\downarrow} = \frac{G m_t M}{R^2} \tag{2.9.2}$$

$$\Theta_{dx} = \frac{Superconducting\ Force\ (F_{\uparrow dx})}{Earth's\ Gravitational\ Force\ (F_{\downarrow})} = \frac{\rho_{\mathbb{C}}}{\rho}\frac{Earth\ Radius}{Earth\ Mass}\frac{v_{dx}^2}{G} \tag{2.9.3}$$

$$\Theta_{dT} = \frac{Superconducting\ Force\ (F_{\uparrow dT})}{Earth's\ Gravitational\ Force\ (F_{\downarrow})} = \frac{\rho_{\mathbb{C}}}{\rho}\frac{Earth\ Radius}{Earth\ Mass}\frac{v_{dT}^2}{G} \tag{2.9.4}$$

**Table 2.9.1 Cooper Pair Concentration and Shielding $(\Theta)$ with Plot of Cooper $(\mathbb{C})$ Pair density**

| | | | | | | |
|---|---|---|---|---|---|---|
| $T_c\,({}^\circ K)$ | 8.95E+13 | 2,135 | 1,190 | 447 | 93 | 8.11E-16 |
| $T_b\,({}^\circ K)$ | 9.07E+14 | 4.43E+09 | 3.31E+09 | 2.03E+09 | 9.24E+08 | 2.729 |
| $\rho_C$ (g/cm$^3$) | 2.34E+14 | 2.73E-02 | 1.13E-02 | 2.61E-03 | 2.48E-04 | 6.38E-30 |
| $\Theta_{dx}$ | 1.44E+25 | 4.00E-02 | 9.28E-03 | 8.03E-04 | 1.58E-05 | 3.55E-48 |
| $\Theta_{dT}$ | 3.59E+26 | 1.00 | 2.32E-01 | 2.00E-02 | 3.96E-04 | 8.88E-47 |

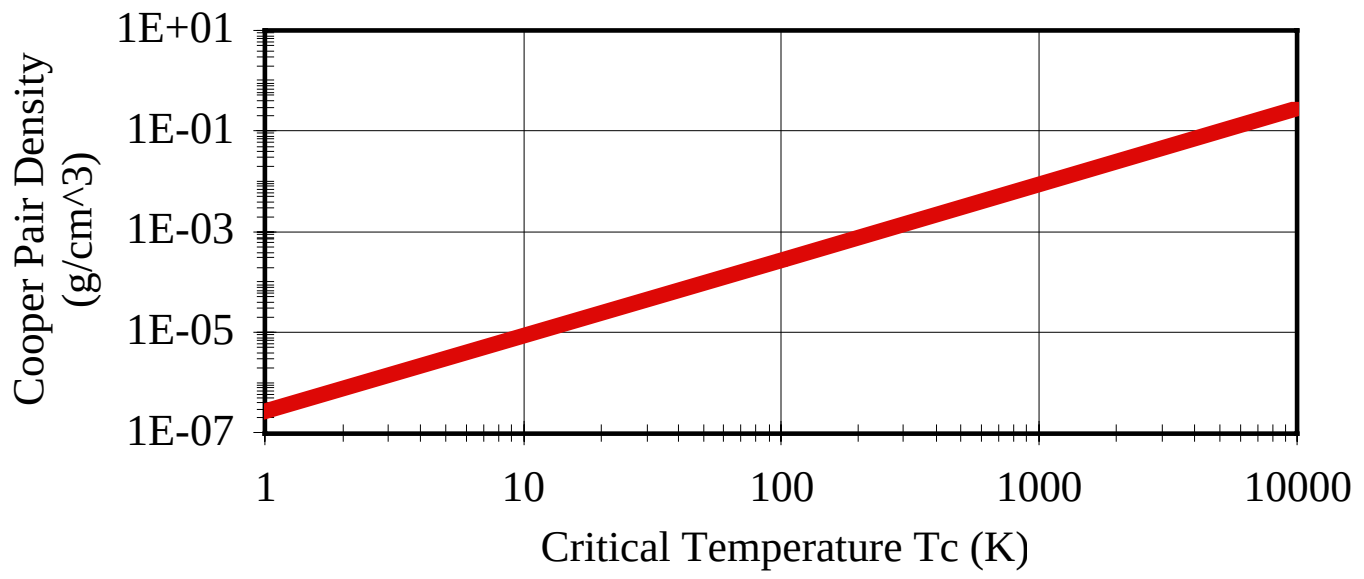

Experiments have been conducted to see if a rotating body's weight is changed from that of a non-rotating body [21, 22, 23]. The results of these experiments were negative as they clearly should be because the center of rotation equaled the center mass. A variety of gyroscopes turned at various angular velocities were tested. Typically a gyroscope with effective radius 1.5 cm was turned at a maximum angular frequency of 22,000 rpm. This equates to a tangential velocity of 3,502 cm/sec. At the surface of the earth, the tangential velocity is calculated to be:

$$\sqrt{gR_E} \text{ or } 791{,}310 \text{ cm/sec.}$$

The expected weight reduction if center of mass was different then center of rotation $(r)$ according to equations 2.9.2 and 2.9.3 would be:

$$(3{,}502/791{,}310)^2 \text{ or } 1.964\text{E-5}$$

of the gyroscope weight. The experiments were done with gyroscopes with weights of about 142 grams. This would indicate a required balance sensitivity of 1.964E-5 x 142 g or 2.8 mg, which approaches the sensitivity of laboratory balances. Both positive[20] and negative results [22, 23] are experimentally observed in this area although it has been generally accepted that the positive results were spurious.

Under the geometric scenario for superconducting materials presented herein, the Cooper CPT Charge conjugated pair centers of mass and centers of rotation $(r)$ are different making the numerical results in table 2.9.1 valid. Under this scenario, it is the superconducting material that looses weight due to transverse Cooper pair movement relative to a center of gravity. The experimentally observed superconducting shielding effect cannot be explained in these terms.

## 2.10 BCS Verifying Constants

Equations 2.9.1 - 2.9.8 represent constants as derived from BCS theory [2] and considered by Little [5] to be primary constraints on any model depicting the superconducting phenomenon. It can be seen that the trisine model constants compare very favorably with BCS constants in brackets {}.

$$C_D = 10.2D\left(\epsilon_T\right)k_b^2 T_c \frac{\mathbb{C}}{cavity} = 10.2k_b \frac{\mathbb{C}}{cavity} \quad (2.10.1)$$

$$\gamma = \frac{2}{3}\pi^2 D\left(\varepsilon_T\right)k_b^2 \frac{\mathbb{C}}{cavity} = \frac{2}{3}\frac{\pi^2 k_b}{T_c}\frac{\mathbb{C}}{cavity} \quad (2.10.2)$$

$$\frac{1}{\mathbb{C}}\frac{\gamma T_c^2}{H_c^2} = .178\left\{\frac{\pi}{18} or .170\right\} \quad (2.10.3)$$

$$\frac{C_D}{\gamma T_c} = \frac{3 \cdot jump}{2\pi^2}\{1.52\} = 1.55\{1.52\} \quad (2.10.4)$$

$$\frac{\hbar^2\left(\frac{K_A^2}{\mathbb{C}} - K_{Dn}^2\right)}{2m_t k_b T_c} = jump\{10.2\} = 10.12\{10.2\} \quad (2.10.5)$$

$$T_c^{\frac{3}{2}} cavity = \frac{3\pi^3\hbar^3}{6^{1/2} m_t^{3/2}\left(\frac{B}{A}\right)_t k_b^{3/2}} \quad (2.10.6)$$

$$= 3.6196E-19 \quad \text{(Saam's Constant)}$$

Equation 2.10.6 basically assumes the ideal gas law for superconducting charge carriers.

**Figure 2.10.1** Homes' Data Compared to Saam's Constant

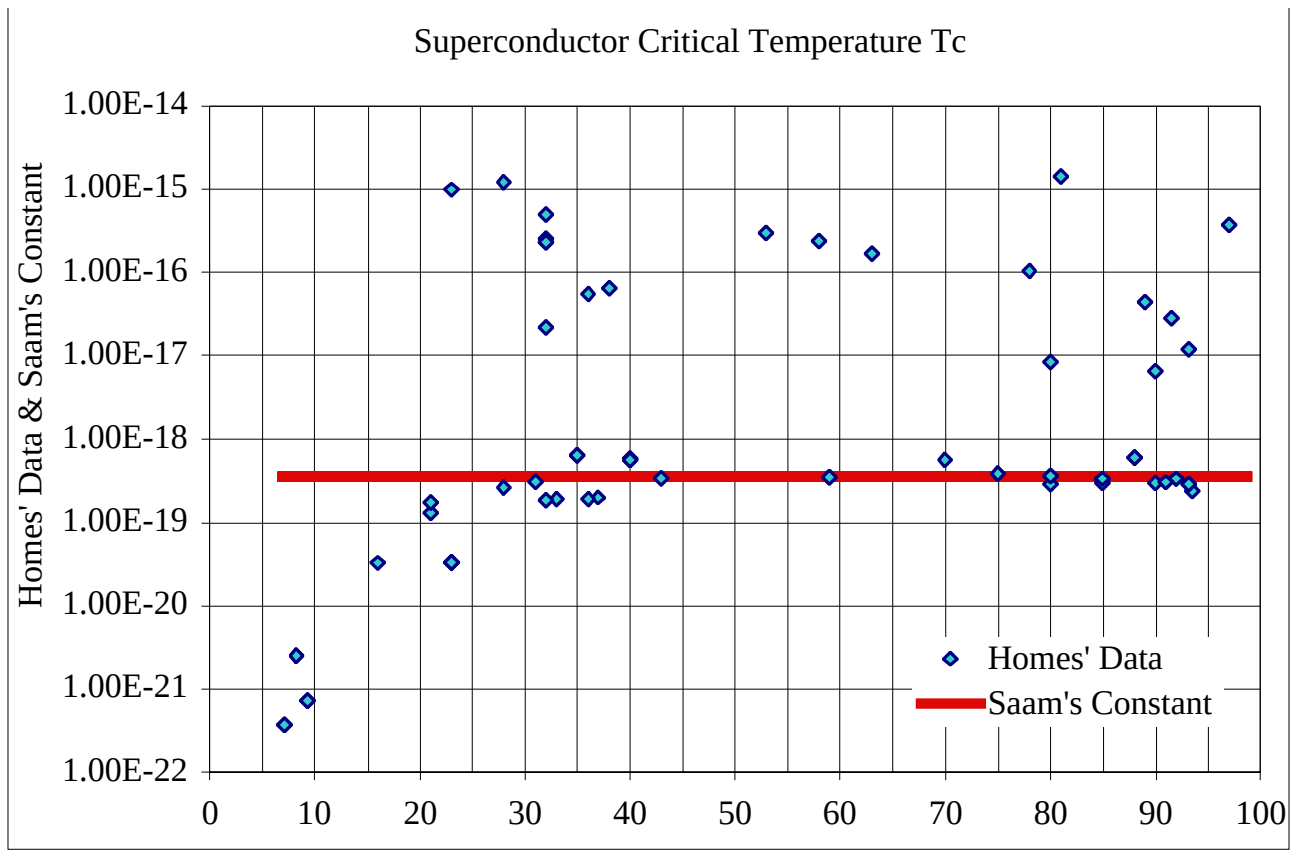

When the $1/\rho_s$ data from reference [51, 59] is used as a measure of cavity volume in the equation 2.10.6 , the following plot is achieved. It is a good fit for some of the data and not for others (figure 2.10.1). One can only conclude that the number of charge carriers in the cavity volume varies considerably from one superconductor to another with perhaps only the Cooper CPT Charge conjugated pair being a small fraction of the total. Also, degrees of freedom may be a factor as in the kinetic theory of gas or other words – is a particular superconductor one-, two- or three- dimensional $(1\cdot k_b T_c/2,\ \ 2\cdot k_b T_c/2,\ \ 3\cdot k_b T_c/2)$ as Table 2.6.1 data for $MgB_2$ would imply? Also it is important to remember that Homes' data is from just above $T_c$ and not at or below $T_c$.

$$\frac{\gamma T_c}{k_b}\frac{cavity}{\mathbb{C}} = \left\{\frac{2\pi^2}{3}\right\} \quad (2.10.7)$$

$$\frac{2\Delta_o^{BCS}}{k_b T_c} = 2\pi e^{-Euler}\{3.527754\} \quad (2.10.8)$$

Within the context of the trisine model, it is important to note that the Bohr radius is related to the proton radius $(B_p)$ by this BCS gap as defined in 2.10.8

$$\frac{cavity}{\Delta x\Delta y\Delta z}\frac{B_p}{Bohr\ radius}\frac{m_t}{m_e}\frac{e^{Euler}}{2\pi} \quad (2.10.9)$$

**Table 2.10.1** Electronic Heat Jump $k_b\mathbb{C}/cavity$ at $T_c$

| | | | | | | |
|---|---|---|---|---|---|---|
| $T_c\ (^\circ K)$ | 8.95E+13 | 2,135 | 1,190 | 447 | 93 | 8.11E-16 |
| $T_b\ (^\circ K)$ | 9.07E+14 | 4.43E+09 | 3.31E+09 | 2.03E+09 | 9.24E+08 | 2.729 |
| $\frac{k_b\mathbb{C}}{cavity}\left(\frac{erg}{cm^3}\right)$ | 6.52E+24 | 7.59E+08 | 3.16E+08 | 7.28E+07 | 6.90E+06 | 1.78E-19 |

Harshman [17] has compiled and reported an extended listing of electronic specific heat jump at $T_c$ data for a number of superconducting materials. This data is reported in units of mJ/mole $K^2$. With the volume per formula weight data for the superconductors also reported by Harshman and after multiplying by $T_c^2$, the specific heat jump at $T_c$ is expressed in terms of erg/cm$^3$ and graphed in figure 2.10.2.

**Figure 2.10.2** Electronic Specific Heat Jump$\left(C_d\right)$ at $T_c$ comparing Harshman compiled experimental data and predicted trisine values based on equation 2.10.1

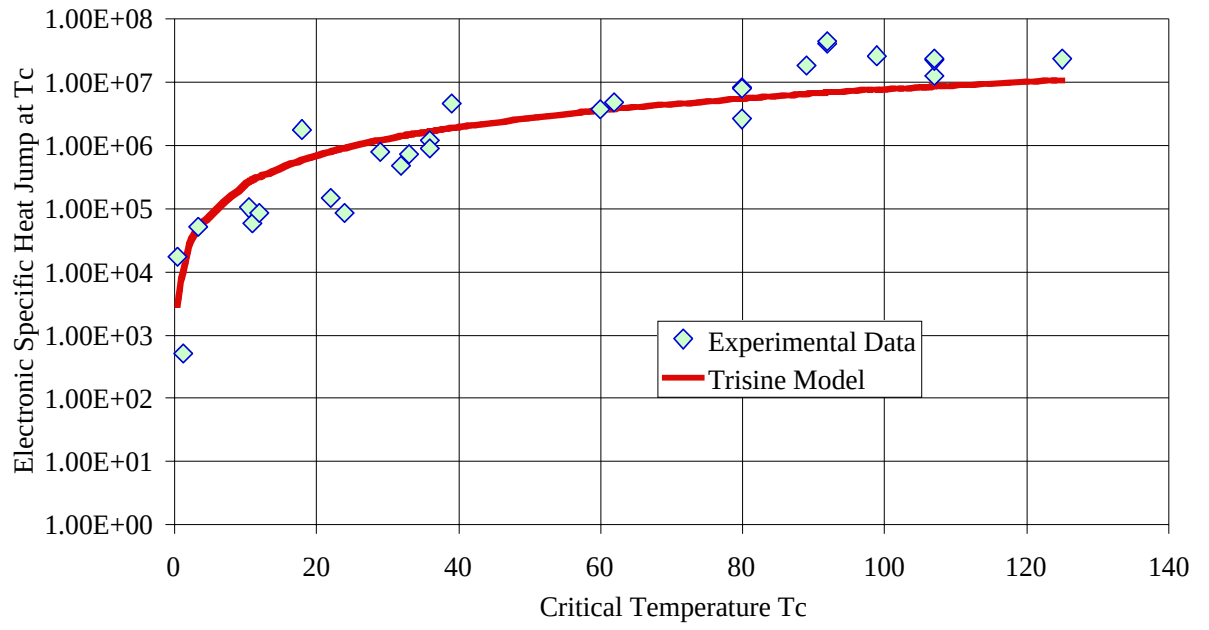

In addition it is important to note a universal resonant superconductor diffusion coefficient $D_c$ based on Einstein's 1905 publication [75] which adds to the universality of the trisine model.

$$D_c = \frac{(2B)^2}{time} = \frac{3^{\frac{1}{2}} kT_c}{3\pi\mu_U(\Delta x)} = 6.60526079\times 10^{-2} \quad (2.10.10)$$

## 2.11 Superconducting Resonant Cosmological Constant

The following relations of $radius(R_U)$, $mass(M_U)$, $density(\rho_U)$ and $time(T_U)$ in terms of the superconducting *cosmological constant* are in general agreement with known data. This could be significant. The cosmological constant as presented in equation 2.11.1 is based on energy/area or *universe surface tension* $(\sigma_U)$, as well as energy/volume or *universe pressure* $(p_U)$. The *universe surface tension* $(\sigma_U)$ is independent of $(T_c)$ while the *universe pressure* $(p_U)$ is the value at current CMBR temperature of $2.729\,^{o}K$ as presented in table 2.7.1 and observed by COBE [26] and later satellites. The *cosmological constant* $(C_U)$ also reduces to an expression relating trisine resonant velocity transformed mass $(m_t)$ the gravitational constant $(G)$ and Planck's constant $(\hbar)$. Also a reasonable value of 71.2 km/sec-million parsec for the Hubble constant $(H_U)$ from equation 2.11.13 is achieved based on a *universe absolute viscosity* $(\mu_U)$ and *energy dissipation rate* $(P_U)$.

$$C_U = \frac{16\pi G}{v_{dx}^4}\sigma_U = \frac{H_U}{\sqrt{3}c} = \frac{4}{\sqrt{3\pi}}\frac{m_t^3 G}{\hbar^2} \qquad (2.11.1)$$

$$\sigma_U = \frac{k_b T_c}{section} \qquad (2.11.2)$$

$$R_U = \frac{1}{C_U} = 2.25E28 \text{ cm (47.5 billion light years)} \qquad (2.11.3)$$

**Figure 2.11.1** Closed Universe with Mass $M_U$ and Radius $R_U$

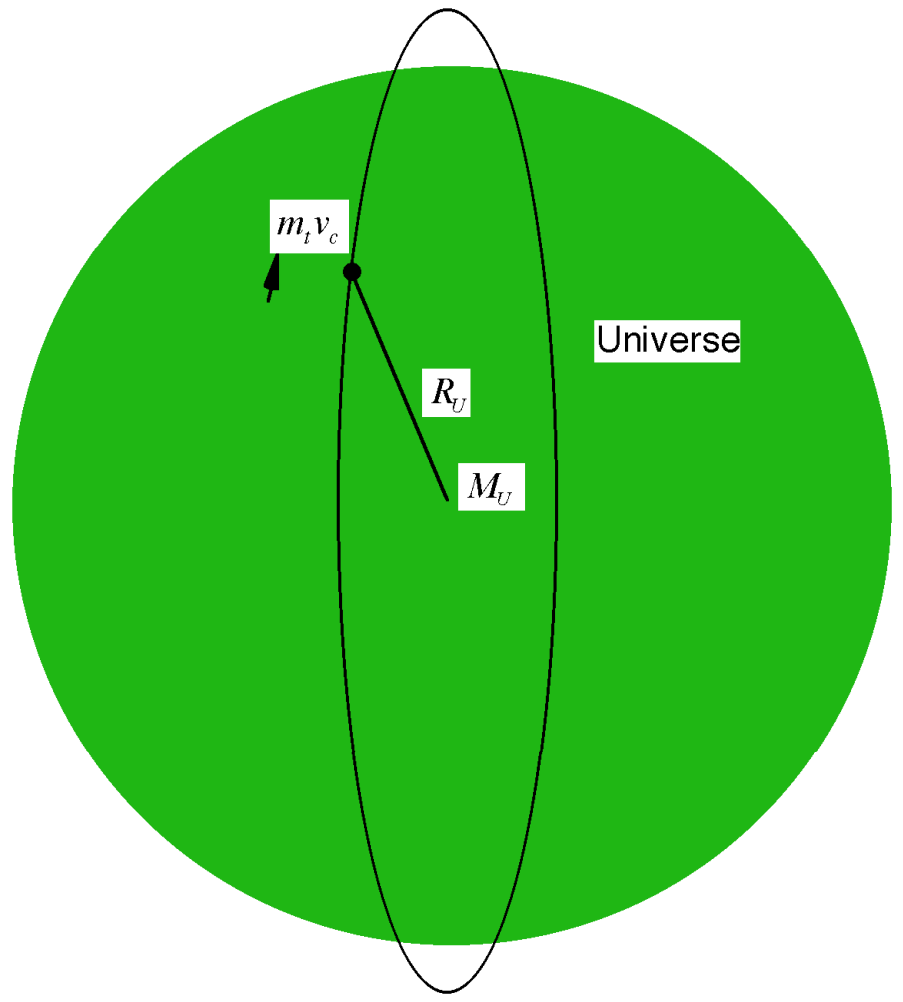

The mass of the universe $(M_U)$ is established by assuming a spherical isotropic universe (center of mass in the universe center) of radius $(R_U)$ such that a particle with the speed of light cannot escape from it as expressed in equation 2.11.4 and graphically described in figure 2.11.1.

$$M_U = R_U \cdot \frac{c^2}{G} = \frac{\sqrt{3\pi}}{4}\frac{\hbar^2 c^2}{m_t^3 G^2} \qquad (2.11.4)$$
$$= 3.02\text{E+56 } g$$

$$\rho_U = M_U \cdot \frac{3}{4\pi}\frac{1}{R_U^3} = \frac{4}{\pi^3}\frac{m_t^6 c^2 G}{\hbar^4} = \frac{m_t}{cavity}\Bigg|_{\substack{T_c = 8.11E-16 \\ T_b = 2.729}} \qquad (2.11.5)$$
$$= 6.38E-30 \text{ g/cm}^3$$

This value for Universe density is much greater than the density of the universe calculated from observed celestial objects. As reported in reference [27], standard cosmology model allows accurate determination of the universe baryon density of between 1.7E-31 to 4.1E-31 g/cm$^3$. These values are 2.7 - 6.4% of the universe density reported herein based the background universe as superconductor.

$$Universe\ time(T_U) = \frac{R_U}{\sqrt{3}c} = \frac{\pi}{4}\frac{\hbar^2}{m_t^3 c\, G} \qquad (2.11.6)$$
$$= 4.32E17 \text{ sec } (1.37E10 \text{ years})$$

**Figure 2.11.2** Shear forces along parallel planes of an elemental volume of fluid

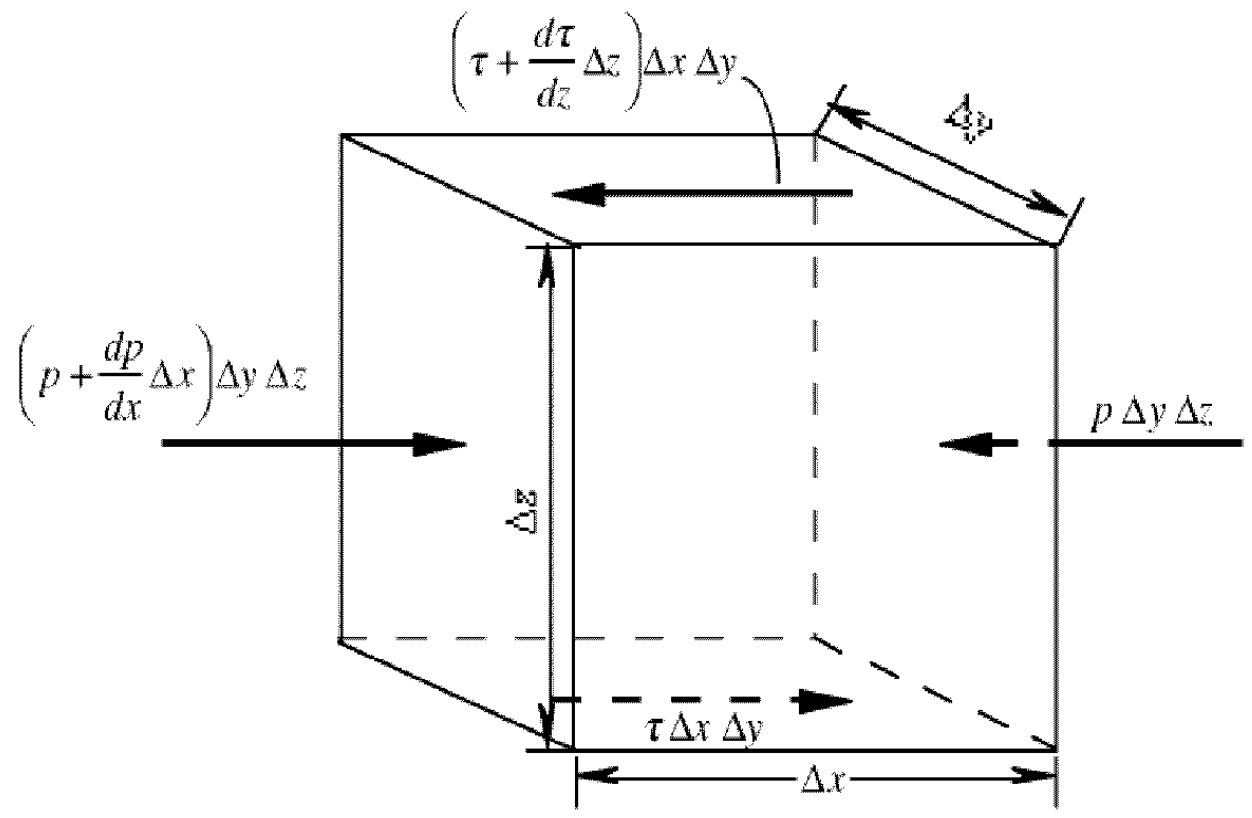

This is an indication that the universe has been expanding at the speed of light ($c$) since the beginning accounting for the early universe inflation and present universe expansion.

As illustrated in Figure 2.11.2 equation 2.11.13 for the Hubble constant $(H_U)$ is derived by equating the forces acting on a cube of fluid in shear in one direction to those in the opposite direction, or

$$p_U\ \Delta y\ \Delta z + \left[\tau + \frac{d\tau}{dz}\Delta z\right]\Delta x\ \Delta y$$
$$= \tau\ \Delta x\ \Delta y + \left[p_U + \frac{dp_U}{dx}\Delta x\right]\Delta y\ \Delta z \qquad (2.11.7)$$

Here $p_U$ is the universe pressure intensity, $\tau$ the shear intensity, and $\Delta x$, $\Delta y$, and $\Delta z$ are the dimensions of the cube. It follows that:

$$\frac{d\tau}{dz} = \frac{dp_U}{dx} \qquad (2.11.8)$$

The power expended, or the rate at which the couple does work $(\tau\ \Delta x\ \Delta y)$, equals the torque $(\tau\ \Delta x\ \Delta y)\Delta z$ times the angular velocity $dv/dz$. Hence the universe power $(P_U)$ consumption per universe volume $(V_U)$ of fluid is

$$P_U/V_U = \left[(\tau\ \Delta x\ \Delta y)\Delta z\ \ dv/dz\right]/(\Delta x\ \Delta y\ \Delta z) = \tau\ (\text{dv/dz}).$$

Defining $\quad \tau = \mu_U\ \ dv/dz$

and $\quad H_U = dv/dz$

then $\quad P_U/V_U = \mu_U\left(dv/dz\right)^2 = \mu_U H_U^2$

or $\quad H_U^2 = P_U/\left(\mu_U V_U\right)$

The *universe absolute viscosity* $(\mu_U)$ is based on momentum/surface area as an outcome of classical kinetic theory of gas viscosity [19]. In terms of our development, the momentum changes over sectional area in each *cavity* such that:

$$\mu_U = \frac{m_t v_{dx}}{section} \qquad (2.11.9)$$

Derivation of the kinematic viscosity $(\upsilon_U)$ in terms of absolute viscosity as typically presented in fluid mechanics [18], a constant universe kinematic viscosity $(\upsilon_U)$ is achieved in terms of Planck's constant$(\hbar)$ and Cooper CPT Charge conjugated pair resonant transformed mass$(m_t)$.

$$\upsilon_U = \frac{cavity}{m_t}\mu_U = \frac{\mu_U}{\rho_U} = \pi\frac{\hbar}{m_t}\frac{A}{B} = 1.39E-02\ \frac{\text{cm}^2}{\text{sec}} \qquad (2.11.10)$$

Derivation of the Universe Energy Dissipation Rate $(P_U)$ assumes that all of the universe mass $(M_U)$ in a universe volume $(V_U)$is in the superconducting state. In other words, our visible universe contributes an insignificant amount of mass to the universe mass. This assumption appears to be valid due to the subsequent calculation of the Hubble constant $(H_U)$ of 71.2km/sec-million parsec (equation 2.11.13) which falls within the experimentally observed value[20] on the order of 70-80 km/sec-million parsec and the recently measured 70 km/sec-million parsec within 10% as measured by a NASA Hubble Space Telescope Key Project Team on May 25, 1999 after analysis of 800 Cepheid stars, in 18 galaxies, as far away as 65 million light years.

$$P_U = \frac{M_U\ v_{dx}^2}{2}\frac{v_{dx}}{R_U} \qquad (2.11.11)$$

$$V_U = \frac{4}{3}\pi R_U^3 = \frac{\sqrt{3}\pi^4\hbar^6}{16 m_t^9 G^3} = 4.74E+85\ cm^3 \qquad (2.11.12)$$

$$H_U = \sqrt{\frac{P_U}{\mu_U V_U}} = \frac{4\ m_t^3 Gc}{\pi\hbar^2} = 2.31E-18\,/\,\text{sec} \qquad (2.11.13)$$

or in more conventionally reported units:

$$H_U = 71.23\frac{\text{km}}{\text{sec}}\frac{1}{\text{million parsec}}$$

Now from equation 2.11.1 and 2.11.13 the *cosmological constant* $(C_U)$is defined in terms of the Hubble constant $(H_U)$ in equation 2.11.14.

$$C_U = \frac{H_U}{\sqrt{3}c} = \frac{1}{R_U} = 4.45\text{E-}29\ \ cm^{-1} \qquad (2.11.14)$$

And the Universe Escape Velocity $(v_U)$

$$v_U = c = \frac{H_U R_U}{\sqrt{3}} = \sqrt{\frac{GM_U}{R_U}} \qquad (2.11.14a)$$

Now from equation 2.11.4 and 2.11.13 the *universe mass* $(M_U)$ is defined in terms of the Hubble constant $(H_U)$ in equation 2.11.15.

$$M_U = \rho_U\frac{4\pi}{3}R_U^3 = \frac{\sqrt{3}c^3}{H_U G} = \frac{\sqrt{3}\pi\hbar^2 c^2}{4m_t^3 G^2} = \frac{R_U c^2}{G} = 3.02E56\ g \qquad (2.11.15)$$

Now from equation 2.11.13 and 2.11.5 the *universe density* $(\rho_U)$ is defined in terms of the Hubble constant $(H_U)$ in equation 2.11.16. This differs from the universe critical density equation in reference [27] by a factor of 2/3, which indicates an inflationary universe.

$$\rho_U = \frac{2}{8\pi}\frac{H_U^2}{G} = \frac{3}{8\pi}\frac{chain}{cavity}\frac{H_U^2}{G} = 6.38E-30\ \frac{g}{cm^3} \qquad (2.11.16)$$

Now from equation 2.11.13 and 2.11.6 the *universe time* $(T_U)$ is defined in terms of the Hubble constant $(H_U)$ in equation 2.11.17.

$$T_U = \frac{1}{H_U} = 4.32\text{E+}17 \text{ or } 1.37E10 \text{ years} \qquad (2.11.17)$$

$$Universe\ mass\ creation\ rate(\dot{M}_U) = \frac{P_U}{c^2} \qquad (2.11.18)$$

**Table 2.11.1** Vacuum Space parameters

| | | | | | | |
|---|---|---|---|---|---|---|
| $T_c\,(^0K)$ | 8.95E+13 | 2,135 | 1,190 | 447 | 93 | 8.11E-16 |
| $T_b\,(^0K)$ | 9.07E+14 | 4.43E+09 | 3.31E+09 | 2.03E+09 | 9.24E+08 | 2.729 |
| $\mu_U$ (g/cm/sec) | 2.09E+01 | 3.78E-04 | 1.57E-04 | 3.62E-05 | 3.44E-06 | 8.85E-32 |
| $\sigma_U$ (erg/cm$^2$) | 8.05E+23 | 4.58E+02 | 1.42E+02 | 2.01E+01 | 8.70E-01 | 6.61E-35 |
| $P_U$ (erg/sec) | 8.23E+62 | 9.59E+46 | 3.99E+46 | 9.19E+45 | 8.72E+44 | 2.24E+19 |
| $\dot{M}_U$ (g/sec) | 9.16E+41 | 1.07E+26 | 4.44E+25 | 1.02E+25 | 9.70E+23 | 2.50E-02 |

This development would allow the galactic collision rate calculation in terms of the Hubble constant$(H_U)$ (which is considered a shear rate in our fluid mechanical concept of the universe) in accordance with the Smolukowski relationship [12, 13, 14] where $J_U$ represents the collision rate of galaxies of two number concentrations $(n_i, n_j)$ and diameters$(d_i, d_j)$:

$$J_U = \frac{1}{6}n_i n_j H_U\left(d_i + d_j\right)^3 \qquad (2.11.19)$$

The superconducting Reynolds' number $(R_e)$, which is defined as the ratio of inertial forces to viscous forces is presented in equation 2.11.20 and has the value of one(1). In terms of conventional fluid mechanics this would indicate a condition of laminar flow [18].

$$R_e = \frac{m_t}{cavity}v_{dx}\ cavity\frac{1}{section}\frac{1}{\mu} = \frac{m_t v_{dx}}{\mu\ section} = 1 \qquad (2.11.20)$$

$$a_U = \pi e^{-Euler}\left(\frac{v_{dy}^2}{2P}\right)\Bigg|_{\substack{T_c=8.11E-16\\ T_b=2.729}}$$

$$= cH_U = \frac{R_U H_U^2}{\sqrt{3}} = \sqrt{\frac{P_U c^2}{\mu_U V_U}} = 6.93x10^{-8}\,\frac{cm}{\sec^2} \qquad (2.11.21)$$

The numerical value of the universal deceleration $(a_U)$ expressed in equation 2.11.21 is nearly equal to that observed with Pioneer 10 & 11 [53], but this near equality is viewed as consequence of equation 2.11.22 for the particular dimensional characteristics (cross section $(A_c)$, Mass $(M)$ and thrust coefficient $(C_t)$) of the Pioneer 10 & 11 space crafts.

$$a_U = cH_U = \frac{H_U}{c}c^2 \approx C_t \frac{A_c}{M}\rho_U c^2 = 6.93x10^{-8}\,\frac{cm}{\sec^2} \qquad (2.11.22)$$

This acceleration component of the universal dark matter/energy may provide the mechanism for observed universe expansion and primarily at the boundary of the elastic space CPT lattice and Baryonic matter (star systems, galaxies, etc) and under the universal gravity and quantum mechanical relationship as follows:

$$m_t c^2 = G\frac{m_t M_U}{R_U} = \frac{\sqrt{3}}{8\pi}\frac{M_U}{m_t}\frac{h^2}{2m_t}\frac{1}{R_U^2}$$

$$= \frac{2}{3}\left(\frac{\sqrt{3}}{8\pi}\right)^2 \frac{h^2}{2m_t}\frac{1}{B_{proton}^2} = 3\pi k_m \varepsilon g_s^2 \frac{h^2}{2m_t}\frac{1}{A^2} \qquad (2.11.23)$$

Which fits well into the relativistic derivations in Appendix F.

## 2.12 Superconducting Resonant Variance At T/Tc

Establish Gap and Critical Fields as a Function of $T/T_c$ using Statistical Mechanics [19]

$$z_s = e^{\frac{-\hbar^2 K_B^2}{2m_e k_b T_c}\frac{chain}{cavity}} = e^{-\frac{2}{3}} \qquad (2.12.1)$$

$$cavity = \frac{3\pi^3\hbar^3}{6^{\frac{1}{2}} m_t^{\frac{3}{2}} \left(\frac{B}{A}\right)_t k_b^{\frac{3}{2}} T_c^{\frac{3}{2}}} \qquad (2.12.2)$$

$$z_t = \frac{(2\pi m_t k_b T_c)^{\frac{3}{2}}\, cavity}{8\pi^3\hbar^3} \overset{1.3\%}{\approx} \left(\frac{T}{T_c}\right)^{\frac{3}{2}} \qquad (2.12.3)$$

$$k_t = \frac{z_t}{z_s} e^{\frac{k_b T_c}{k_b T}} = \left(\frac{T}{T_c}\right)^{\frac{3}{2}} e^{\frac{2}{3}} e^{-\frac{T_c}{T}} = \left(\frac{T}{T_c}\right)^{\frac{3}{2}} e^{\left(\frac{2}{3}-\frac{T_c}{T}\right)} \qquad (2.12.4)$$

$$k_s = \frac{2}{3} - k_t = \frac{chain}{cavity} - k_t \qquad (2.12.5)$$

$$\Delta_T = 1 - \frac{1}{3}\frac{T}{T_c}\frac{1}{\ln\left(\frac{1}{k_t}\right)} = 1 + \frac{1}{3}\frac{T}{T_c}\frac{1}{\ln(k_t)} \qquad (2.12.6)$$

$$H_{T1}^2 = H_o^2\left[1 - \frac{2}{3}\pi^2\left(\frac{T}{\pi e^{-Euler}T_c}\right)^2\right] \quad \text{at} \quad \frac{T}{T_c} \sim 0 \qquad (2.12.7)$$

$$H_{T2}^2 = H_o^2 k_s \quad \text{at} \quad \frac{T}{T_c} \text{ near } 1 \qquad (2.12.8)$$

$$AvgH_{cT}^2 = \begin{bmatrix} \text{if } H_{cT1}^2 > H_{cT2}^2 \\ \text{then } H_{cT1}^2 \text{ else } H_{cT2}^2 \end{bmatrix} \qquad (2.12.9)$$

## 2.13 Superconductor Resonant Energy Content

Table 2.13.1 presents predictions of superconductivity energy content in various units that may be useful in anticipating the uses of these materials.

**Table 2.13.1** Superconductor Resonant Energy in Various Units.

| $T_c\,(^0K)$ | 8.95E+13 | 2,135 | 1,190 | 447 | 93 | 8.11E-16 |
|---|---|---|---|---|---|---|
| $T_b\,(^0K)$ | 9.07E+14 | 4.43E+09 | 3.31E+09 | 2.03E+09 | 9.24E+08 | 2.729 |
| erg/cm$^3$ | 7.19E+38 | 2.00E+12 | 4.64E+11 | 4.01E+10 | 7.92E+08 | 1.78E-34 |
| joule/liter | 7.19E+34 | 2.00E+08 | 4.64E+07 | 4.01E+06 | 7.92E+04 | 1.78E-38 |
| BTU/ft$^3$ | 1.93E+33 | 5.37E+06 | 1.25E+06 | 1.08E+05 | 2.13E+03 | 4.77E-40 |
| $\frac{\text{kwhr}}{\text{ft}^3}$ | 5.66E+29 | 1.57E+03 | 3.65E+02 | 3.16E+01 | 6.23E-01 | 1.40E-43 |
| $\frac{\text{hp-hr}}{\text{gallon}}$ | 1.01E+29 | 2.82E+02 | 6.54E+01 | 5.66E+00 | 1.12E-01 | 2.51E-44 |
| gasoline equivalent | 1.55E+27 | 4.30E+00 | 9.98E-01 | 8.63E-02 | 1.70E-03 | 3.82E-46 |

Table 2.13.2 represents energy level of trisine energy states and would be key indicators material superconductor characteristics at a particular critical temperature. The BCS gap is presented as a reference energy.

**Table 2.13.2** Energy (mev) associated with Various Trisine Wave Vectors. The energy is computed by $(\hbar^2/2m_t)K^2$ and expressing as electron volts(ev). The BCS gap is computed as $(\pi/\exp(Euler))(\hbar^2/2m_t)K_B^2$.

| $T_c\,(^0K)$ | 8.95E+13 | 2,135 | 1,190 | 447 | 93 | 8.11E-16 |
|---|---|---|---|---|---|---|
| $T_b\,(^0K)$ | 9.07E+14 | 4.43E+09 | 3.31E+09 | 2.03E+09 | 9.24E+08 | 2.729 |
| $K_B^2$ | 267,138 | 183.98 | 102.55 | 38.52 | 8.01 | 6.99E-17 |
| $K_C^2$ | 896,191 | 617.22 | 344.02 | 129.22 | 26.89 | 2.34E-16 |
| $K_{Ds}^2$ | 831,836 | 572.89 | 319.32 | 119.95 | 24.96 | 2.18E-16 |
| $K_{Dn}^2$ | 320,119 | 220.47 | 122.88 | 46.16 | 9.60 | 8.37E-17 |
| $K_P^2$ | 356,184 | 245.31 | 136.73 | 51.36 | 10.69 | 9.31E-17 |
| $K_{Ds}^2 + K_B^2$ | 1,098,974 | 756.87 | 421.86 | 158.46 | 32.97 | 2.87E-16 |
| $K_{Dn}^2 + K_B^2$ | 587,257 | 404.45 | 225.43 | 84.68 | 17.62 | 1.54E-16 |
| $K_{Ds}^2 + K_C^2$ | 1,728,027 | 1190.11 | 663.34 | 249.17 | 51.84 | 4.52E-16 |
| $K_{Dn}^2 + K_C^2$ | 1,216,310 | 837.68 | 466.91 | 175.38 | 36.49 | 3.18E-16 |
| $K_{Ds}^2 + K_P^2$ | 1,188,020 | 818.20 | 456.05 | 171.30 | 35.64 | 3.11E-16 |
| $K_{Dn}^2+K_P^2$ | 676,303 | 465.78 | 259.61 | 97.52 | 20.29 | 1.77E-16 |
| $K_A^2$ | 6,049,289 | 4166.20 | 2322.15 | 872.27 | 181.48 | 1.58E-15 |
| BCS gap | 471,198 | 324.52 | 180.88 | 67.94 | 14.14 | 1.23E-16 |

## 2.14 Superconductor Resonant Gravitational Energy

As per reference[25], a spinning nonaxisymmetric object rotating about its minor axis with angular velocity $\omega$ will radiate gravitational energy in accordance with equation 2.14.1

$$\dot{E}_g = \frac{32G}{5c^5} I_3^2 \varsigma^2 \omega^6 \qquad (2.14.1)$$

This expression has been derived based on weak field theory assuming that the object has three principal moments of inertia I1, I2, I3, respectively, about three principal axes $a > b > c$ and $\varsigma$ is the ellipticity in the equatorial plane.

$$\varsigma = \frac{a-b}{\sqrt{ab}} \qquad (2.14.2)$$

Equations 2.14.1 with 2.14.2 were developed from first principles (based on a quadrapole configuration) for predicting gravitational energy emitted from pulsars [25], but we use them to calculate the gravitational energy emitted from rotating Cooper CPT Charge conjugated pairs with resonant trisine mass ($m_t$) in the trisine resonant superconducting mode.

Assuming the three primary trisine axes *C*>*B*>*A* corresponding to a>b>c in reference [25] as presented above, then the trisine ellipticity (or measure of *trisine* moment of inertia) would be:

$$\varsigma = \frac{C-B}{\sqrt{CB}} = \frac{chain}{cavity}\frac{A}{B} = .2795 \qquad (2.14.3)$$

Assuming a mass $(m_t)$ rotating around a nonaxisymmetric axis of resonant radius $B$ to establish moment of inertia$(I_3)$ then equation 2.14.1 becomes a resonant gravitational radiation emitter with no net energy emission outside of resonant *cavity*:

$$\frac{E_g}{time_\pm} = \frac{32G}{5}\frac{(\varsigma)^2 (m_t B^2)^2}{v_{dx}^5}\left(\frac{2}{time_\pm}\right)^6 \qquad (2.14.4a)$$

where angular velocity $\omega = 2/time_\pm$, and replacing the speed of light $(c)$ with resonant consequential sub- and super-luminal de Broglie speed of light $(v_{dx})$, then equation 2.14.4 in terms of dimension *'B'* becomes the Newtonian gravitational energy relationship in equation 2.14.5:

$$E_g = -G\frac{m_t^2}{B} \qquad (2.14.5)$$

All of this is consistent with the Virial Theorem.

Gravitational Potential Energy/2 = Kinetic Energy

$$\frac{1}{2}Gm_t^2 \frac{3}{\Delta x+\Delta y+\Delta z}\bigg|_{\substack{T_c=8.11E-16\\T_b=2.729}} = \frac{1}{2}m_r v_{dx}^2\bigg|_{\substack{T_c=8.11E-16\\T_b=2.729}} \qquad (2.14.7)$$

and

$$hH_U\big|_{\substack{T_c=8.11E-16\\T_b=2.729}} = \frac{cavity}{chain} G \frac{m_t^2}{B_p}\bigg|_{\substack{T_c=8.95E13\\T_b=9.07E14}} \qquad (2.14.7a)$$

Which is the same form as 2.14.5 and in conjunction with equation 2.11.13 implies the following:

$$\hbar K_B\big|_{\substack{T_c=8.95E13\\T_b=9.07E14}} = 8\pi \frac{chain}{cavity} m_t c \qquad (2.14.7b)$$

also:

$$\left\{\begin{matrix}\dfrac{1}{2\varepsilon_x k_m} m_t v_{dx}^2 \\ \dfrac{chain}{cavity}\dfrac{1}{2\varepsilon_x k_m}\dfrac{1}{\varepsilon}\dfrac{e^2}{B_{CMBR}}\end{matrix}\right\}\Bigg|_{\substack{T_c=8.11E-16\\T_b=2.729}} = \frac{1}{2}m_r v_{dx}^2\bigg|_{\substack{T_c=8.11E-16\\T_b=2.729}}$$

$$where: \quad 2\varepsilon_x k_m = \frac{2c^2}{v_{dx}^2}\bigg|_{\substack{T_c=8.11E-16\\T_b=2.729}} \qquad (2.14.7c)$$

Where $m_t$ and $m_r$ are related as in Equation 2.1.25 and $\Delta x, \Delta y, \Delta z$ are Heisenberg Uncertainties as expressed in equation 2.1.20.

In the case of a superconductor, the radius R is the radius of curvature for pseudo particles described in this report as the variable B. All of the superconductor pseudo particles move in asymmetric unison, making this virial equation applicable to this particular engineered superconducting situation. This is consistent with Misner Gravitation [35] page 978 where it is indicated that gravitational power output is related to "power flowing from one side of a system to the other" carefully incorporating "only those internal power flows with a time-changing quadrupole moment". The assumption in this report is that a super current flowing through a trisine superconducting CPT lattice has the appropriate quadrupole moment at the dimensional level of B. The key maybe a linear back and forth trisine system. Circular superconducting systems such as represented by Gravity Probe B gyroscopic elements indicate the non gravitational radiation characteristic of such circular systems reinforcing the null results of mechanical gyroscopes [21, 22, 23].

## 2.15 Critical Optical Volume Energy (COVE).

This section considers the concept of critical volume, which would be analogous to critical mass in atomic fission. The critical volume would establish a parameter beyond which the output energy would exceed the input energy. It is anticipated that this energy would have a benign character and not be usable in explosive applications. The basic principle is that a coherent beam length is not related to input power creating that beam while virtual particle generation (in accordance with the CPT theorem) is a function of length. Given that volume is length cubed, then there must be a critical volume at which the energy of created virtual particles exceeds the input power to the volume. The following is a derivation of this concept.

$$Poytning\ Vector(S) = c\left(\frac{1}{2}\right)\left(\frac{m_t v_{\varepsilon x}^2}{cavity}\right) \qquad (2.15.1)$$

then

$$Input\ Power\left(2\sqrt{3}B^2 S\right) = \left(\frac{c}{2}\right)\left(\frac{m_t v_{\varepsilon x}^2}{A}\right) \qquad (2.15.2)$$

Now consider the out put power.

$$Output\ Power = \left(\frac{1}{2}\right)\left(\frac{m_t v_{dx}^2}{time}\right) = \left(\frac{1}{2}\right)\left(\frac{m_t v_{dx}^3}{2B}\right) \qquad (2.15.3)$$

Now calculate output energy.

$$Output\ Energy\ (Length) = \left(\frac{1}{2}\right)\left(\frac{m_t v_{dx}^3}{2B}\right)(Length) \qquad (2.15.4)$$

The input energy is calculated as follows:

$$Input\ Energy\ (2B) = \left(\frac{c}{2}\right)\left(\frac{m_t v_{\varepsilon x}^2}{A}\right)(2B) \qquad (2.15.5)$$

Now define the critical condition:

$$let\ \ \frac{Energy\ Out}{Energy\ In} = 1 \qquad (2.15.6)$$

Now a critical *Length* (L) can be established.

$$L = \left(\frac{c}{2}\right)\left(\frac{m_t v_{\varepsilon x}^2}{A}\right)2B\frac{2\cdot 2B}{m_t v_{dx}^3} = 4\left(\frac{c}{v_{dx}}\right)\left(\frac{v_{\varepsilon x}^2}{v_{dx}^2}\right)\left(\frac{B}{A}\right)B \qquad (2.15.7)$$

It is envisioned that this length would be a measure of trisine lattice size necessary for criticality and in particular a defined Critical Optical Volume defined at the of intersection of three lengths. To retrieve the power, one side (+ or -) of the virtual pair would have to be pinned (perhaps optically). In this case power could be retrieved by electrically coupling lattice to a pick up coil or arranging to intercept 56 Mev radiation.

## 3. Discussion

This effort began as an effort towards formulating a theoretical frame work towards explaining the Podkletnov's and Nieminen's [7] results but in the process, relationships between electron/proton mass, charge and gravitational forces were established which appear to be unified under this correlation with links to the cosmological constant, the dark matter of the universe, and the universe black body radiation as a continuation of its big bang origins and nuclear forces.

The overall correlation is an extension to what Sir Arthur Eddington proposed during the early part of the 20th century. He attempted to relate proton and electron mass to the number of particles in the universe and the universe radius. Central to his approach was the fine structure constant:

$$\left(\hbar c/e^2\right)$$

which he knew to be approximately equal to 137. This relationship is also found as a consequence of our development in equation 2.1.19, the difference being that a material dielectric$(\varepsilon)$ modified velocity of light$(v_\varepsilon)$ is used. To determine this dielectric$(v_\varepsilon)$, displacement$(D)$ and electric fields$(E)$ are determined for Cooper CPT Charge conjugated pairs moving through the trisine CPT lattice under conditions of the trisine CPT lattice diamagnetism equal to $-1/4\pi$ (the Meissner effect). Standard principles involving Gaussian surfaces are employed and in terms of this model development the following relationship (3.1) is deduced from equation 2.1.19 repeated here

$$\frac{1}{2}g_e g_s \frac{m_t}{m_e}\frac{e_\pm \hbar}{2m_e v_\varepsilon} = cavity\ H_c$$

with equation 3.1 containing the constant $\left(\hbar c/e^2\right)$ and relating electron mass$(m_e)$, trisine resonant transformed mass $(m_t)$ and the characteristic trisine geometric ratio$(B/A)$.

$$\frac{1}{2}\frac{m_t^2}{m_e^2}g_e g_s = 4\pi 3^{-\frac{1}{4}}\left(\frac{\hbar c}{e_\pm^2}\right)\left(7\frac{A^2}{B^2}+12\frac{B^2}{A^2}+19\right) \qquad (3.1)$$

or

$$\frac{m_t^2}{m_e^2} = 3\left(\frac{\hbar c}{g_e g_s e_\pm^2}\right)\left(7\frac{A^2}{B^2}+12\frac{B^2}{A^2}+19\right) \qquad (3.1a)$$

and

$$\frac{m_t^2}{m_e^2} = \left(\frac{8\pi\hbar G M_U H_U}{g_s^3 e_\pm^2 c^2}\right)\left(\frac{A}{B}+\frac{B}{A}\right) \qquad (3.1b)$$

and also

$$\frac{G M_U H_U}{c^3} = 3^{1/2} \qquad (3.1c)$$

The extension to traditional superconductivity discussion is postulated wherein the underlying principle is resonance expressed as a conservation of energy and momentum (elastic character). The mechanics of this approach are in terms of a cell (*cavity*) defined by momentum states:

$$K_1,\ K_2,\ K_3\ and\ K_4$$

and corresponding energy states:

$$K_1^2,\ K_2^2,\ K_3^2\ and\ K_4^2$$

by a required new mass called trisine mass $(m_t)$, which is intermediate to the electron and proton masses. The cell dimensions can be scaled from nuclear to universe dimensions with the trisine mass remaining constant. The triangular hexagonal cell character is generally accepted by conventional solid-state mechanics to be equivalent in momentum and real space making the energy *(~$K^2$)* and momentum *(~K)* cell (*cavity*) identity possible. Also, the hexagonal or triangulation energy *(~$K^2$)* and momentum *(~K)* have a 1/3, 2/3 character inherent to the standard model 1/3, 2/3 quark definitions.

Observation of the universe from the smallest (nuclear particles) to the largest (galaxies) indicates a particularly static condition making our elastic assumption reasonable and applicable at all dimensional scales. Indeed, this is the case. This simple cellular model is congruent with proton density and radius as well as interstellar space density and energy density.

In order to scale this cellular model to these dimensions varying by 15 orders of magnitude, special relativity Lorentz transforms are justifiably suspended due to resonant cancellation in the context of the de Broglie hypothesis. Superluminous velocities, which occur above 3,100,000 K and less than $10^{-10}$ cm or within nuclear cell dimensions are then theoretically acceptable.

Most importantly, the gravity force is linked to the Hubble constant through Planck's constant. This is a remarkable result. It indicates that gravity is linked to other forces at the proton scale and not the Planck scale. Indeed, when the model trisine model is scaled to the Planck scale (B= 1.62E-33 cm), a mass of 1.31E+40 Gev/$c^2$ (2.33E+16 g) is obtained. This value would appear arbitrary. Indeed the model can be scaled to the Universe mass ($M_U$) of 3.02E56 grams at a corresponding dimension (*B*) of 1.42E-53 cm.

## 4. Conclusion

A momentum and energy conserving (elastic) CPT lattice called trisine and associated superconducting theory is postulated whereby electromagnetic and gravitational forces are mediated by a particle of resonant transformed mass (110.12275343 x electron mass or 56.2726/$c^2$ Mev) such that the established electron/proton mass is maintained, electron and proton charge is maintained and the universe radius as computed from Einstein's cosmological constant is 2.25E28 cm, the universe mass is 3.0E56 gram, the universe density is 6.38E-30 g/$cm^3$ and the universe time is 1.37E10 years.

The cosmological constant is based on a universe surface tension directly computed from a superconducting resonant energy over surface area.

The calculated universe mass and density are based on an isotropic homogeneous media filling the vacuum of space analogous to the 'aether' referred to in the 19th century (but still in conformance with Einstein's relativity theory) and could be considered a candidate for the 'cold dark matter' in present universe theories and as developed by Primack [80]. Universe density is 2/3 of conventionally calculated critical density. This is in conformance with currently accepted inflationary state of the universe.

The 'cold dark matter', a postulated herein, emits a CMBR of 2.729 $^oK$ as has been observed by COBE and later satellites although its temperature in terms of superconducting critical temperature is an extremely cold 8.11E-16$^oK$.

Also, the reported results by Podkletnov and Nieminen wherein the weight of an object is lessened by .05% are theoretically confirmed where the object is the superconductor, although the gravitational shielding phenomenon of $YBa_2Cu_3O_{7-x}$ is yet to be explained. It is understood that the Podkletnov and Nieminen experiments have not been replicated at this time. It could be that the very high super currents that are required and theoretically possible have not be replicated from original experiment.

The trisine model provides a basis for considering the phenomenon of superconductivity in terms of an ideal gas with 1, 2 or 3 dimensions (degrees of freedom). The trisine model was developed primarily in terms of 1 dimension, but experimental data from $MgB_2$ would indicate a direct translation to 3 dimensions.

These trisine structures have an analog in various resonant structures postulated in the field of chemistry to explain properties of delocalized $\pi$ electrons in benzene, ozone and a myriad of other molecules with conservation of energy and momentum conditions defining a superconductive (elastic) condition.

Other verifying evidence is the trisine model correlation to the ubiquitous 160-minute resonant oscillation [68,69] in the universe and also the model's explanation of the Tao effect [65, 66, 67].

Also dimensional guidelines are provided for design of room temperature superconductors and beyond. These dimensional scaling guidelines have been verified by Homes' Law and generally fit within the conventionally described superconductor operating in the "dirty" limit. [54]

The deceleration observed by Pioneer 10 & 11 as they exited the solar system into deep space appear to verify the existence of an space energy density consistent with the amount theoretically presented in this report. This translational deceleration is independent of the spacecraft velocity. Also, the same space density explains the Pioneer spacecraft rotational deceleration. This is remarkable in that translational and rotational velocities differ by three (3) orders of magnitude.

General conclusions from study of Pioneer deceleration data are itemized as follows:

1. Dark Matter and Dark Energy are the same thing and are represented by a Trisine Elastic Space CPT lattice which is composed of virtual particles adjacent to Heisenberg Uncertainty

2. Dark Matter and Dark Energy are related to Cosmic Microwave Background Radiation (CMBR) through very large space dielectric (permittivity) and permeability constants.

3. A very large Trisine dielectric constant (6.16E10) and CPT equivalence makes the Dark Matter/Energy invisible to electromagnetic radiation.

4. A velocity independent thrust force acts on all objects passing through the Trisine Space CPT lattice and models the translational and rotational deceleration of Pioneer 10 & 11 spacecraft.

5. The Asteroid Belt marks the interface between solar wind and Trisine, the radial extent an indication of the dynamics of the energetic interplay of these two fields

6. The Trisine Model predicts no Dust in the Kuiper Belt. The Kuiper Belt is made of large Asteroid like objects.

7. The observed Matter Clumping in Galaxies is due to magnetic or electrical energetics in these particular areas at such a level as to destroy the Coherent Trisine.

8. The apparent high observed speed of outer star rotation in galaxies is due to forced redistribution with R of orbital velocities with time due to disproportionate change in R of inner faster moving stars relative to R of outer slower orbiting stars in accordance with orbital velocity (v) relationship:

$$v^2 = \frac{GM}{R}$$

This is a residual effect left over from the time when galaxies were primarily dust or small particles when the formula:

$$deceleration = C_d A \rho_U c^2$$

$\rho_U$ = conventionally accepted space vacuum density

$$\rho_U = \frac{2H_U^2}{8\pi G}$$

M = object mass

A = object cross sectional area

$C_t$ = thrust coefficient

was primarily applicable and is still applicable in as much as galaxies still consist of small particles (< a few meters) gravitationally coupled to large objects such as stars.

9. The observed Universe Expansion is due to Trisine Space CPT lattice internal pressure and is at rate defined by the speed of light (*c*) escape velocity.

10. When the Trisine is scaled to molecular dimensions, superconductor resonant parameters such as critical fields, penetration depths, coherence lengths, Cooper CPT Charge conjugated pair densities are modeled.

11. A superconductor is consider electrically neutral with balanced conjugated charge as per CPT theorem. Observed current flow is due to one CPT time frame being pinned relative to observer.

12. The nuclear Delta particles $\Delta^{++}$, $\Delta^{+}$. $\Delta^{0}$, $\Delta^{-}$ with corresponding quark structure uuu, uud, udd, ddd, with masses of 1232 Mev/$c^2$ are generally congruent with the trisine prediction of *(2/3)hc/B* or 1242 Mev/$c^2$ where *B* reflects the proton radius. Also, the experimentally determined mean lifetime of the delta particles (6E-24 seconds) is consistent and understandably longer than the resonant time for the proton (2.68E-25 second).

13. It is noted that the difference in neutron ($m_n$) and proton ($m_p$) masses about two electron ($m_e$) masses ($m_n$-$m_p$~2$m_e$).

14. The nuclear weak force particles $W^-W^+$ $Z^0$ are associated with the superconductor specific heat jump (2.10.5) with corresponding masses 78.07 Gev/$c^2$ for $W^-W^+$ and 78.07/$\cos^2(\theta)$ Gev/$c^2$ or 89.3 Gev/$c^2$ for $Z^0$ and at proton dimension (B). (experimental mass values for particles $W^-$ and $W^+$ 80.4 {91.187/ $\cos^2(\theta)$} and particle $Z^0$ 91.187 Gev/$c^2$). This is consistent with observed beta decay related to nuclear weak force wherein parity is broken or other words, the chiral trisine symmetry is broken.

15. The top quark (Higgs Particle) is associated with energy ($K_AK_A$) at a corresponding mass of 174.62 Gev/$c^2$.

16. The up (u), down (d) and charm (c) quarks are associated with energies (2/3 $m_e/m_t$ $\hbar K_B c$, 1/3 $m_e/m_t$ $\hbar K_B c$ & 2/3 $\hbar K_B c$) or corresponding masses .00564, .00282 & 1.24 Gev/$c^2$.

17. The strange quark (s) is associated with hypotenuse energy ($K_AK_A$ & $K_AK_A$) with a mass of 0.10 Gev/$c^2$.

18. The bottom (b) quark is associated with BCS superconducting 'gap' energy (2.10.8) with a corresponding mass of 4.53 Gev/$c^2$.

19. The Gluon ratio is 60/25 = 2.4 ≅ B/A=2.379760996.

Other experimental evidence of the ubiquitous character of the 56 Mev radiation is the NASA Voyager I spacecraft as it passes through our solar system terminal shock and detected what are called anomalous cosmic rays (ACRs) (Figure 4.1http://imagine.gsfc.nasa.gov/docs/features/bios/christian/anomalous.html

**Figure 4.1** Spectra for Voyager 1 (left column) and Voyager 2 (right column) for the time periods 1994 days 157-261 (a and b), 1994 days 1 - 105 (c and d) and 1993 days 53 - 157 (e and f).

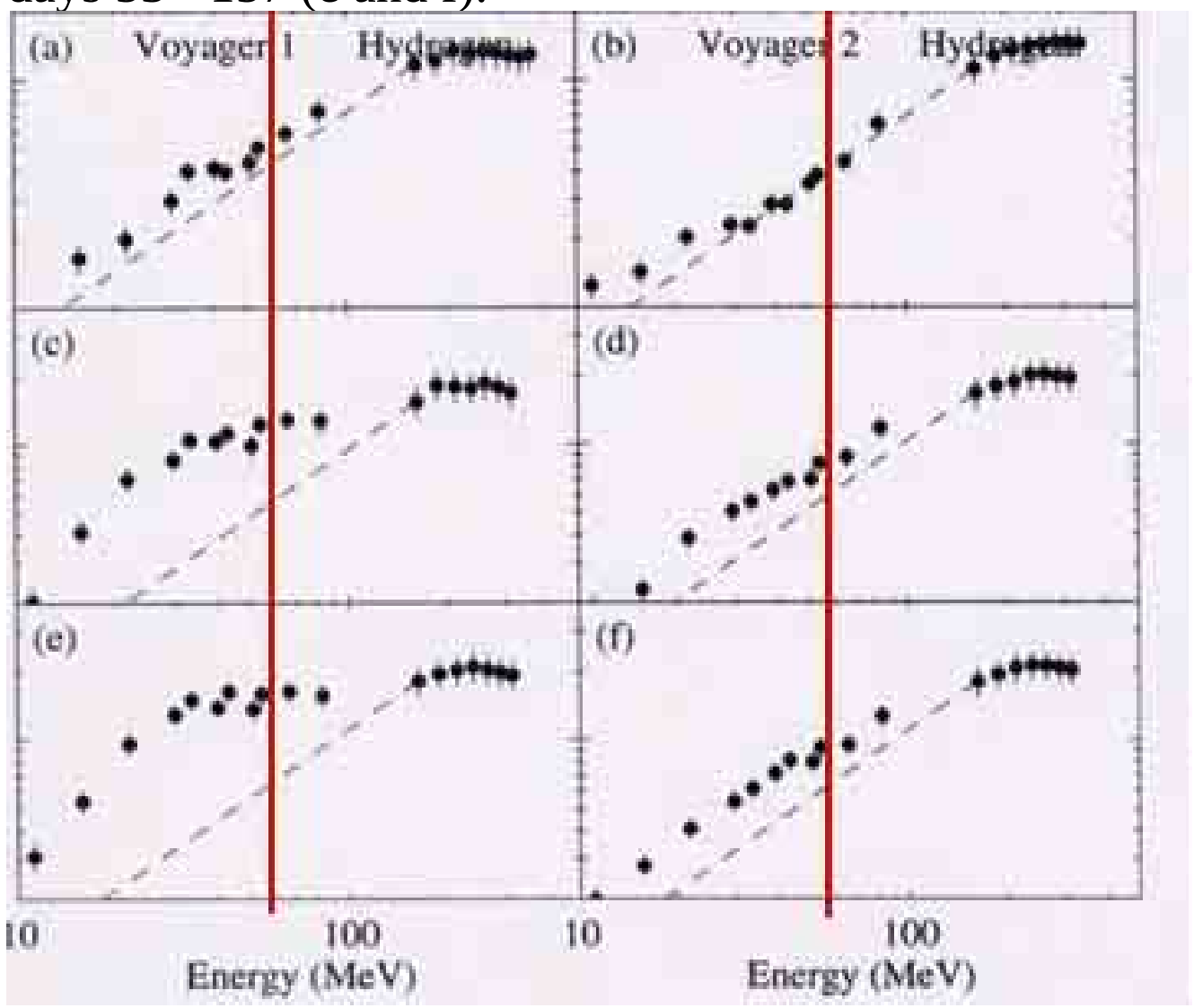

It is assumed that this terminal shock zone generates ACR 56 Mev radiation in a mechanism similar to the Pioneer satellites deceleration and expresses itself in a black body curve. Voyager II will enter our solar system terminal shock zone in 2008. 56 Mev radiation verification is anticipated

Also there is an indication of a 56 Mev peak in black body curve from the EGRET instrument aboard The Compton Gamma Ray Observatory Mission (Comptel) spacecraft, which generally observed the extra galactic diffuse background x-ray radiation. The data graphic is replicated below in Figure 4.2 by author permission [73] with the 56 Mev resonant energy superimposed. Verification is anticipated with the launch of The Gamma-ray Large Area Space Telescope (GLAST) spacecraft platform in 2008 which will provide a much more detailed delineation of the extra galactic diffuse X-ray background radiation and potentially the trisine space lattice 56 Mev/$c^2$ dark matter component.

References [76,79] provide evidence that the dark matter is observed to be spatially separate for visual matter. This can be explained in terms of the correlations presented in this report which indicate that dark matter is essentially a rigid quantum state each with its own mass density (KK/c^2) which is observed within the boundary of these states which occur in the vicinity of galaxies or residual historical areas of collision events and not generally observed in the vastness of space because of its homogeneity congruent with CMBR.

A question logically develops from this line of thinking and that is "What is the source of the energy or mass density?". After the energy is imparted to the spacecraft (or other objects) passing through space, does the energy regenerate itself locally or essentially reduce the energy of universal field? And then there is the most vexing question based on the universal scaled superconductivity resonant hypothesis developed in this report and that is whether this phenomenon could be engineered at an appropriate scale for beneficial use.

**Figure 4.2 Diffuse X-ray Background radiation including that from EGRET [73] with 56 Mev imposed.**

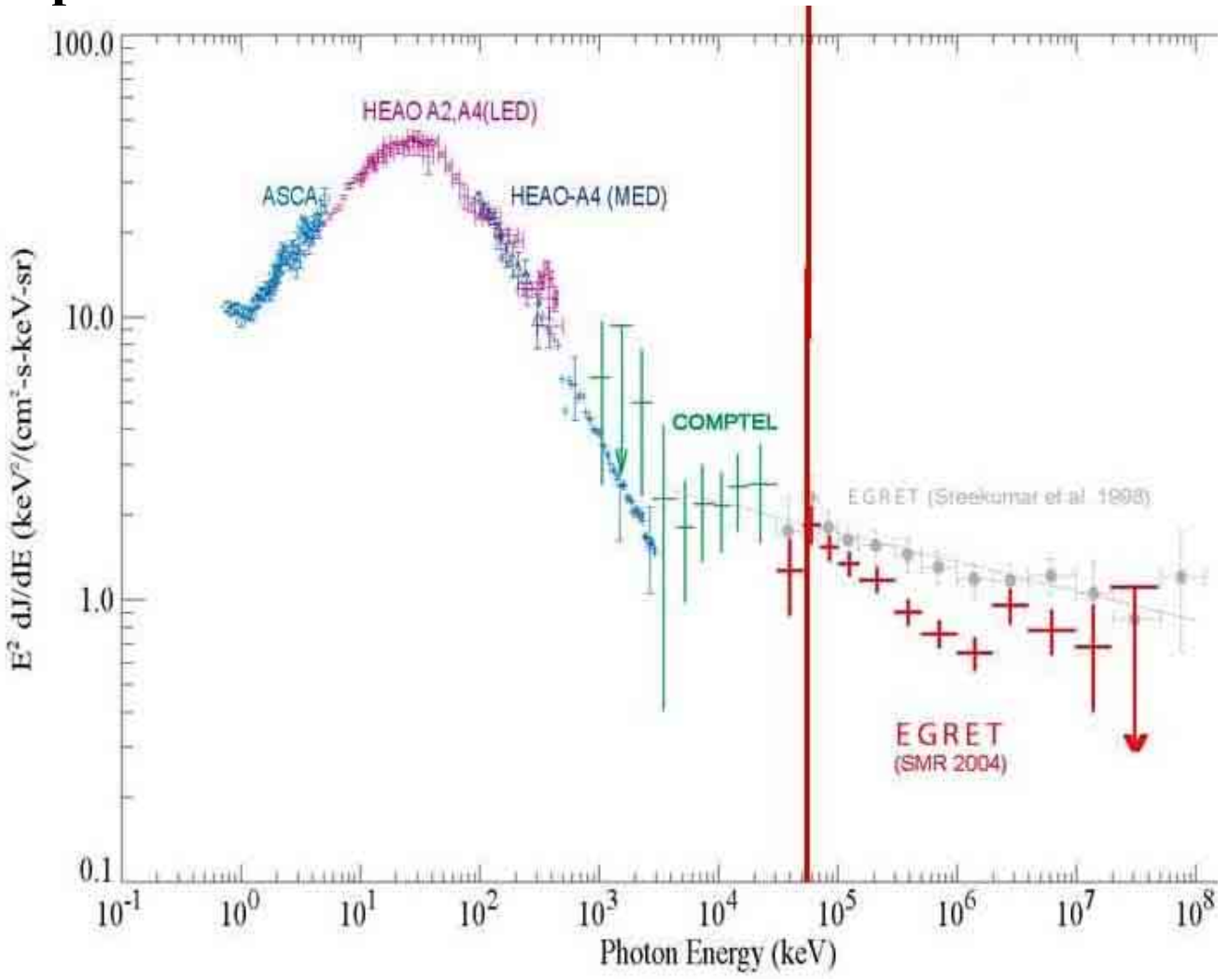

One concept for realizing this goal, is to engineer a wave crystal reactor. Ideally, such a wave crystal reactor would be made in the technically manageable microwave (cm wave length). Perhaps such microwaves would interact at the $\hbar\omega/2$ energy level in the Trisine CPT lattice pattern. Conceptually a device as indicated in Figures 4.3 and 4.4, may be appropriate where the green area represents reflecting, generating and polarizing surfaces of coherent electromagnetic radiation creating standing waves which interact in the intersecting zone defined by the trisine characteristic angle of (90 – 22.8) degrees and 120 degrees of each other.

A similar approach is made with the National Ignition Facility(NIF) at Livermore, CA. The question immediately arises whether the NIF could be run in a more continuous less powerful mode with selected laser standing wave beams of the trisine characteristic angle of (90 – 22.8) degrees and 120 degrees of each other with no material target and study the resultant interference pattern or lattice for virtual particles.

Above and beyond the purely scientific study of such a lattice, it would be of primary importance to investigate the critical properties of such a lattice as the basis of providing useful energy generating mechanisms for the good of society. The major conclusion of this report is that an interdisciplinary team effort is required to continue this effort involving physicists, chemists and engineers. Each has a major role in achieving the many goals that are offered if more complete understanding of the superconducting resonant phenomenon is attained in order to bring to practical reality and bring a more complete understanding of our universe.

**Figure 4.3 Conceptual Trisine Generator (Elastic Space CPT lattice)**

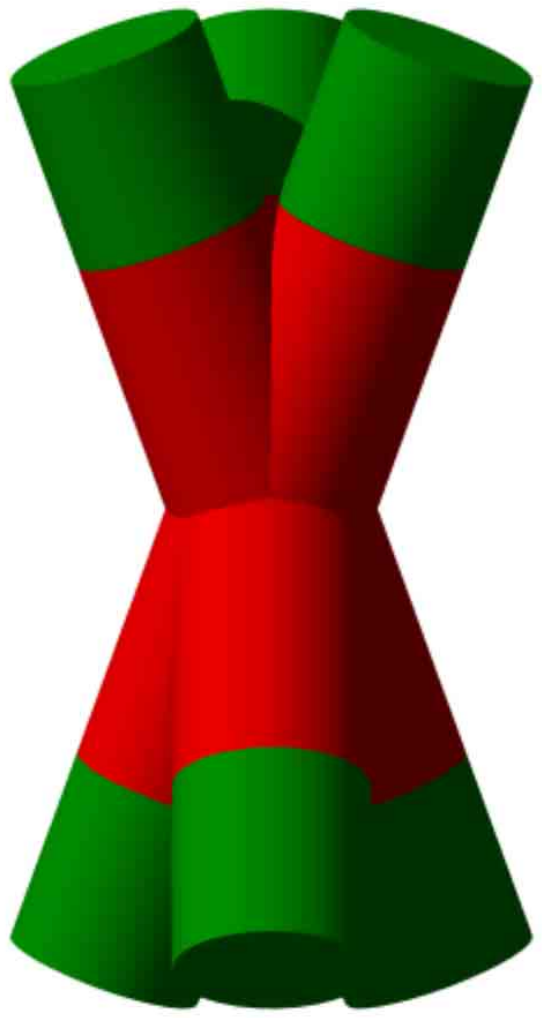

**Figure 4.4 Conceptual Trisine Generator (Elastic Space CPT lattice) Within Articulating Motion Control Mechanism**

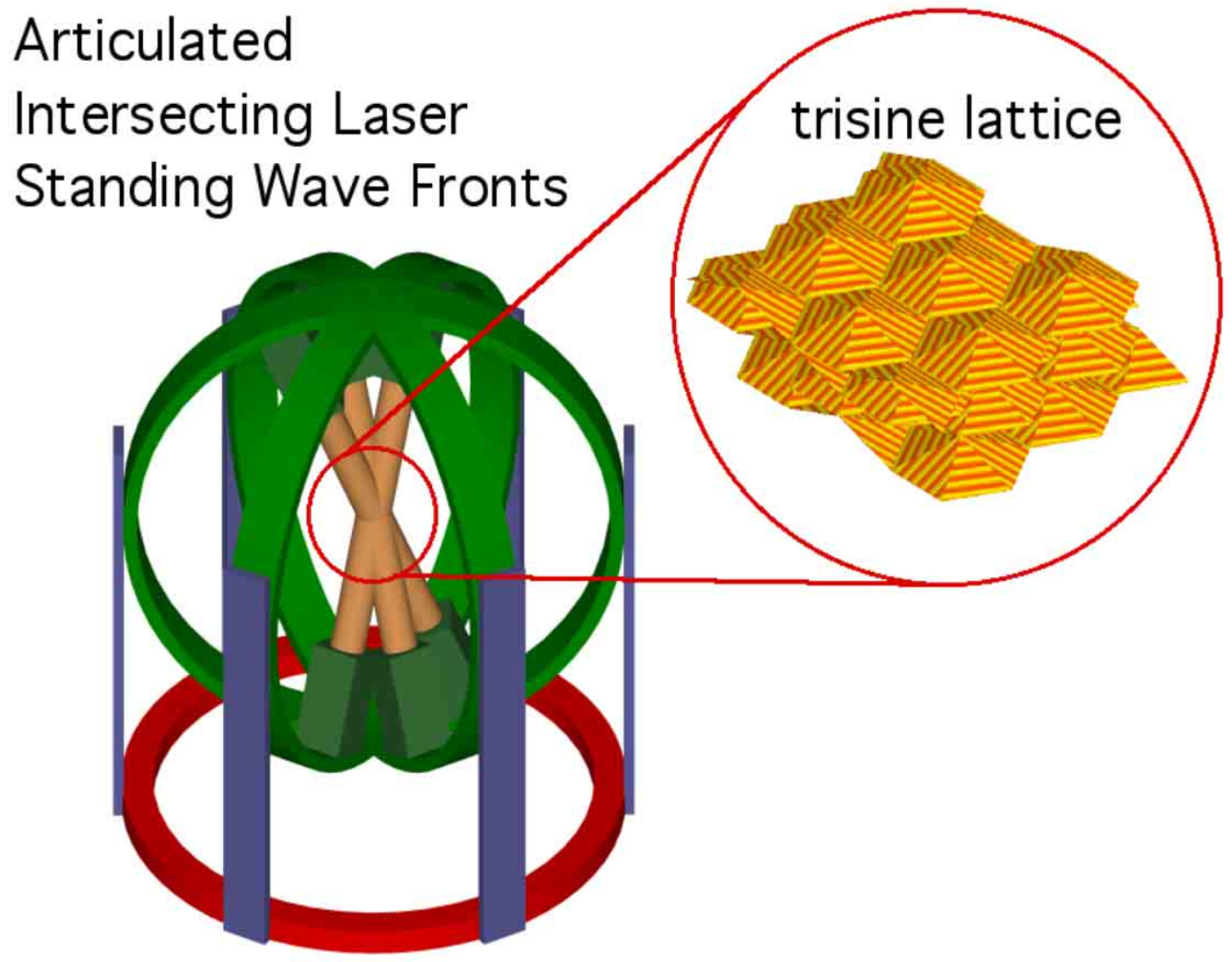

## 5. Variable And Constant Definitions

Fundamental Physical Constants are from National Institute of Standards (NIST).

| name | symbol | value | units |
|---|---|---|---|
| Attractive Energy | V | | $g\ cm^2 \sec^{-2}$ |
| Boltzmann constant | $k_b$ | 1.380658120E-16 | $g\ cm^2 \sec^{-2}\ Kelvin^{-1}$ |
| Bohr magneton | $e\hbar/(2m_e c)$ $\mu_B$ | 9.27400949E-21 | *erg/gauss (gauss cm³)*<br>$g^{1/2}\ cm^{5/2} \sec^{-1}$ |
| Bohr radius | $a_o$ | 5.2917724924E-09 | $cm$ |
| de Broglie velocity | $v_{dx}, v_{dy}, v_{dz}, v_{dT}$ | | $cm\ \sec^{-1}$ |
| Cooper CPT conjugated pair | $\mathbb{C}$ | 2 | *unitless* |
| correlation length | $\xi$ | | $cm$ |
| critical magnetic fields(internal) | $H_{c1}, H_{c2}, H_c$ | | $g^{1/2}\ cm^{-1/2} \sec^{-1}$ |
| critical temperature | $T_c$ | | $^oK$ |
| Diffusion coefficient | $D_c$ | 6.60526079E-2 | $cm^2 \sec^{-1}$ |
| Dirac's number | $g_d$ | 1.001159652186 | *unitless* |
| electric field (internal) | $E$ | | $g^{1/2}\ cm^{-1/2} \sec^{-1}$ |
| displacement field (internal) | $D$ | | $g^{1/2}\ cm^{-1/2} \sec^{-1}$ |
| electron mass | $m_e$ | 9.10938975E-28 | $g$ |
| earth radius | $R_E$ | 6.371315E+08 | $cm$ |
| earth mass | $M_E$ | 5.979E+27 | $g$ |
| electron charge | $e$ | 4.803206799E-10 | $cm^{3/2} g^{1/2} \sec^{-1}$ |
| electron gyromagnetic-factor | $g_e$ | 2.0023193043862 | *unitless* |
| energy | E | | $g\ cm^2 \sec^{-2}$ |
| fine structure constant | $hc/e^2$ $1/\alpha$ | 137.03599936 | *unitless* |
| fluxoid | $\Phi$ | 2.0678539884E-7 | $g^{1/2} cm^{3/2} \sec^{-1}$ *(gauss cm²)* |
| fluxoid dielectric $(\varepsilon)$ mod | $\Phi_\varepsilon$ | | $g^{1/2} cm^{3/2} \sec^{-1}$ *(gauss cm²)* |
| gravitational constant | $G$ | 6.6725985E-8 | $cm^3 \sec^{-2} g^{-1}$ |
| London penetration depth | $\lambda$ | | $cm$ |
| momentum vectors | $p_x,\ p_y,\ p_z$ | | $g\ \mathrm{cm}^{-1} \sec^{-1}$ |
| natural log base | e | 2.718281828459050 | *unitless* |
| nuclear magneton | $e\hbar/(2m_p c)$ $\mu_p$ | 5.05078342E-24 | *erg/gauss (gauss cm³)*<br>$g^{1/2} cm^{5/2} \sec^{-1}$ |
| permeability | $k_m$ | | *unitless* |
| permittivity (dielectric constant) | $\varepsilon$ | | *unitless* |
| Planck's constant | $\hbar(h/2\pi)$ | 1.054572675E-27 | $g\ cm^2 \sec^{-1}$ |
| Planck length | $(\hbar G/c^3)^{1/2}$ | 1.61624E-33 | $cm$ |
| Planck time | $(\hbar G/c^5)^{1/2}$ | 5.39121E-44 | sec |
| Planck mass | $(\hbar c/G)^{1/2}$ | 2.17645E-05 | $g$ |
| Planck energy | $(\hbar c^5/G)^{1/2}$ | 1.95610E+16 | $g\ cm^2 \sec^{-2} (erg)$ |
| Planck momentum | $(\hbar c^3/G)^{1/2}$ | 6.52483E+05 | $g\ cm\ \sec^{-1}$ |
| Planck force | $c^4/G$ | 1.21027E+49 | $g\ cm\ \sec^{-2}\ (dyne)$ |
| Planck density | $c^5/(\hbar G^2)$ | 5.15500E+93 | $g\ cm^{-3}$ |
| Planck acceleration | $c^6/(\hbar G)$ | 1.03145E+97 | $cm\ \sec^{-2}$ |
| Planck kinematic viscosity | $(c^7/(\hbar G))^{1/2}$ | 5.56077E+53 | $cm^2\ \sec$ |
| Planck absolute viscosity | $(c^9/(\hbar G^3))^{1/2}$ | 2.49779E+71 | $g\ cm^{-1}\ \sec^{-1}$ |
| proton mass | $m_p$ | 1.67262311E-24 | $g$ |
| proton radius | $B_p/(2\sin(\theta))$ | 8.59E-14 (8 E-14) | $cm$ |
| proton density | | 2.31E-14 | $g\ cm^{-3}$ |
| resonant mass | $m_r$ | | $g$ |
| resonant CPT time$_\pm$ | $time_\pm$ | | *second* |
| superconductor gyro-factor | $g_s$ | 1.00971389902 | *unitless* |
| trisine angle | $\theta$ | 22.80 | angular degrees |
| trisine areas | *section, approach, side* | | $cm^2$ |
| trisine unit | *cell* | | *unitless* |
| trisine constant | *trisine* | 2.010916597 | *unitless* |
| trisine density of states | $D(\in_t)$ | | $\sec^2 g^{-1} cm^{-2}$ |
| trisine dimensions | *A, B, C, P* | | $cm$ |
| trisine magneton | $e\hbar/(2m_t c)$ $\mu_t$ | 8.42151973E-23 | *erg/gauss (gauss cm*$^3$*)* <br> $g^{1/2} cm^{5/2} \sec^{-1}$ |
| trisine mass | $m_t$ | 1.0031502157E-25 | g |
| transformed mass | $m_t$ | 110.122753426 x $m_e$ | $g$ |
| trisine size ratio | $(B/A)_t$ | 2.379760996 | *unitless* |
| trisine volume per cell | *cavity* | | $cm^3$ |
| trisine volume per chain | *chain* | | $cm^3$ |
| trisine wave vectors | $K_A, K_B, K_C, K_{Dn}, K_{Ds}$ | | $cm^{-1}$ |
| universe absolute viscosity | $\mu_U$ | | $g$ cm$^{-1}$ sec$^{-1}$ |
| universe density | $\rho_U$ | 6.38E-30 | $g$ cm$^{-3}$ |
| universe Hubble constant | $H_U$ | 2.31E-18 | sec$^{-1}$ |
| universe kinematic viscosity | $\nu_U$ | 1.39E-02 | cm$^2$ sec$^{-1}$ |
| universe radius | $R_U$ | 2.25E+28 | $cm$ |
| universe mass | $M_E$ | 3.02E+56 | $g$ |
| universe pressure | $p_U$ | | $g$ cm$^{-1}$ sec$^{-2}$ |
| universe time | $T_U$ | 4.32E+17 | sec |
| velocity of light | $c$ | 2.997924580E+10 | $cm\ \sec^{-1}$ |
| $\varepsilon$ modified velocity of light | $v_\varepsilon$ | | $cm\ \sec^{-1}$ |

## Appendix A. Trisine Number $m_t$ And B/A Ratio Derivation

BCS approach [2]

$$\Delta_o^{BCS} = \frac{\hbar\omega}{\sinh\left(\frac{1}{D(\in_T)V}\right)} \qquad (A.1)$$

Kittel approach [4]

$$\Delta_o = \frac{2\hbar\omega}{e^{\frac{1}{D(\in_T)V}} - 1} \qquad (A.2)$$

Let the following relationship where BCS (equation A.1) and Kittel (equation A.2) approach equal each other define a particular superconducting condition.

$$k_b T_c = \begin{cases} \dfrac{e^{Euler}}{\pi} \dfrac{\hbar\omega}{\sinh\left(\frac{1}{D(\in_T)V}\right)} \\ \dfrac{\hbar\omega}{e^{\frac{1}{D(\in_T)V}} - 1} \end{cases} \qquad (A.3)$$

Define:

$$\frac{1}{D(\in_T)V} = trisine \qquad (A.4)$$

Then:

$$trisine = -\ln\left(\frac{2e^{Euler}}{\pi} - 1\right) \qquad (A.5)$$
$$= 2.01091660$$

Given:

$$\frac{2m_t k_b T_c}{\hbar^2} = \begin{Bmatrix} \dfrac{(KK)_{\hbar\omega}\, e^{Euler}}{\pi \sinh(trisine)} \\ \dfrac{(KK)_{\hbar\omega}}{e^{trisine} - 1} \end{Bmatrix} = K_B^2 \qquad (A.6)$$

Ref: [2,4]

And within Conservation of Energy Constraint:

$$\left(\frac{1}{g_s}\right)\left(K_B^2 + K_C^2\right) = \left(K_{Ds}^2 + K_{Dn}^2\right) \qquad (A.7)$$

And Within Conservation of Momentum Constraint:

$$(g_s)(K_B + K_C) = (K_{Ds} + K_{Dn}) \qquad (A.8)$$

And from $(KK)$ from table A.1:

$$(KK) = K_C^2 + K_{Ds}^2 \qquad (A.9)$$

Also from table A.1:

$$\frac{2\Delta_o^{BCS}}{k_b T_c} = \frac{2\pi}{e^{Euler}} = \frac{2K_{Ds}}{K_B} \overset{.065\%}{\approx} 3.527754 \qquad (A.10)$$

Also:

$$\frac{\Delta_o}{k_b T_c} = 2 \qquad (A.11)$$

Also from table A.1:

$$trisine = \frac{K_B^2 + K_P^2}{K_P K_B} \qquad (A.12)$$

**Table A.1** Trisine Wave Vector Multiples $(KK)$ Compared To $K_B K_B$

| | | $\frac{(KK)e^{Euler}}{\pi \sinh(trisine)}$ | |
|---|---|---|---|
| $K$ | $\frac{(KK)}{K_B K_B}$ | $\frac{(KK)}{K_B K_B} \frac{1}{e^{trisine} - 1}$ | $\frac{K}{K_B}$ |
| $K_{Dn}$ | 1.1981 | 0.1852 | 1.0946 |
| $K_P$ | 1.3333 | 0.2061 | 1.1547 |
| $K_{Ds}$ | 3.1132 | 0.4812 | 1.7644 |
| $K_C$ | 3.3526 | 0.5182 | 1.8310 |
| $K_A$ | 22.6300 | 3.4976 | 4.7571 |

But also using the following resonant condition in equation A.13 and by iterating equations A.7, A.8 and A.13.

$$k_b T_c = \begin{Bmatrix} \dfrac{\hbar^2 K_B^2}{2m_t} \\ \dfrac{\hbar^2}{\frac{m_e m_t}{m_e + m_t}(2B)^2 + \frac{m_t m_p}{m_t + m_p}(A)^2} \end{Bmatrix} \qquad (A.13)$$

This trisine model converges at:

$$\begin{aligned} m_t &= 110.122753426 \text{ x } m_e \\ B/A &= 2.379760996 \end{aligned} \qquad (A.14)$$

## Appendix B. Debye Model Normal And Trisine Reciprocal Lattice Wave Vectors

This appendix parallels the presentation made in ref [4 pages 121 - 122].

$$N = \begin{Bmatrix} \left(\frac{L}{2\pi}\right)^3 \frac{4\pi}{3} K_{Dn}^3 \\ \left(\frac{L}{2\pi}\right)^3 K_{Ds}^3 \end{Bmatrix} = \begin{Bmatrix} \text{Debye Spherical Condition} \\ \text{Debye Trisine Condition} \end{Bmatrix} \qquad (B.1)$$

where:

$$K_{Ds} = (K_A K_B K_P)^{\frac{1}{3}} \qquad (B.2)$$

$$\frac{dN}{d\omega} = \begin{bmatrix} \frac{cavity}{2\pi^2} K_{Dn}^2 \frac{dK_{Dn}}{d\omega} \\ \frac{cavity}{8\pi^3} 3K_{Ds}^2 \frac{dK_{Ds}}{d\omega} \end{bmatrix} = \begin{Bmatrix} \text{Debye Spherical Condition} \\ \text{Debye Trisine Condition} \end{Bmatrix} \qquad (B.3)$$

$$K = \frac{\omega}{v} \qquad (B.4)$$

$$\frac{dK}{d\omega} = \frac{1}{v} \tag{B.5}$$

$$\frac{dN}{d\omega} = \begin{Bmatrix} \frac{cavity}{2\pi^2}\frac{\omega^2}{v^3} \\ \frac{3\cdot cavity}{8\pi^3}\frac{\omega^2}{v^3} \end{Bmatrix} = \begin{Bmatrix} \text{Debye spherical} \\ \text{Debye trisine} \end{Bmatrix} \tag{B.6}$$

$$N = \begin{Bmatrix} \frac{cavity}{6\pi^2}\frac{\omega^3}{v^3} \\ \frac{cavity}{8\pi^3}\frac{\omega^3}{v^3} \end{Bmatrix} = \begin{Bmatrix} \frac{cavity}{6\pi^2}K_{Dn}^3 \\ \frac{cavity}{8\pi^3}K_{Ds}^3 \end{Bmatrix} = \begin{Bmatrix} \text{Debye Spherical Condition} \\ \text{Debye Trisine Condition} \end{Bmatrix} \tag{B.7}$$

Given that $N = 1$ Cell

$$\begin{Bmatrix} K_{Dn} \\ K_{Ds} \end{Bmatrix} = \begin{Bmatrix} \left(\frac{6\pi^2}{cavity}\right)^{\frac{1}{3}} \\ \left(\frac{8\pi^3}{cavity}\right)^{\frac{1}{3}} \end{Bmatrix} = \begin{Bmatrix} \text{Debye Spherical Condition} \\ \text{Debye Trisine Condition} \end{Bmatrix} \tag{B.8}$$

## Appendix C. One Dimension And Trisine Density Of States

This appendix parallels the presentation made in ref [4 pages 144-155].

Trisine

$$\frac{2K_C^3}{\left(\frac{2\pi}{L}\right)^3} = N \tag{C.1}$$

One Dimension

$$\in = \frac{\hbar^2}{2m_t}N^2\frac{\pi^2}{(2B)^2} \tag{C.2}$$

Now do a parallel development of trisine and one dimension density of states.

$$N = \begin{bmatrix} \frac{2K_C^3\cdot cavity}{8\pi^3} \\ \left(\frac{2m_t}{\hbar^2}\right)^{\frac{1}{2}}\frac{2B}{\pi}\in^{\frac{1}{2}} \end{bmatrix} = \begin{bmatrix} \text{Trisine Dimensional Condition} \\ \text{One Dimensional Condition} \end{bmatrix} \tag{C.3}$$

$$\frac{dN}{d\in} = \begin{Bmatrix} \frac{2\cdot cavity}{8\pi^3}\left(\frac{2m_t}{\hbar^2}\right)^{\frac{3}{2}}\frac{3}{2}\in^{\frac{1}{2}} \\ \left(\frac{2m_t}{\hbar^2}\right)^{\frac{1}{2}}\frac{2B}{\pi}\frac{1}{2}\in^{-\frac{1}{2}} \end{Bmatrix} = \begin{Bmatrix} \text{Trisine Dimensional Condition} \\ \text{One Dimensional Condition} \end{Bmatrix} \tag{C.4}$$

$$\frac{dN}{d\in} = \begin{Bmatrix} \frac{2\cdot cavity}{8\pi^3}\left(\frac{2m_t}{\hbar^2}\right)^{\frac{3}{2}}\frac{3}{2}\in^{\frac{1}{2}} \\ \left(\frac{2m_t}{\hbar^2}\right)^{\frac{1}{2}}\frac{2B}{\pi}\frac{1}{2}\in^{-\frac{1}{2}} \end{Bmatrix} = \begin{Bmatrix} \text{Trisine Dimensional Condition} \\ \text{One Dimensional Condition} \end{Bmatrix} \tag{C.5}$$

$$\frac{dN}{d\in} = \begin{Bmatrix} 2\frac{3}{2}\frac{cavity}{8\pi^3}\left(\frac{2m_t}{\hbar^2}\right)^{\frac{3}{2}}\left(\frac{2m_t}{\hbar^2}\right)^{-\frac{1}{2}}K_C \\ \left(\frac{2m_t}{\hbar^2}\right)^{\frac{1}{2}}\frac{2B}{\pi}\frac{1}{2}\left(\frac{2m_t}{\hbar^2}\right)^{\frac{1}{2}}\frac{1}{K_B} \end{Bmatrix} = \begin{Bmatrix} \text{Trisine Dimensional Condition} \\ \text{One Dimensional Condition} \end{Bmatrix} \tag{C.6}$$

$$\frac{dN}{d\in} = D(\in) = \begin{Bmatrix} \frac{3}{8}\frac{cavity}{\pi^3}\left(\frac{2m_t}{\hbar^2}\right)K_C \\ \frac{2m_t}{\hbar^2K_B^2} \end{Bmatrix} = \begin{Bmatrix} \text{Trisine Dimensional Condition} \\ \text{One Dimensional Condition} \end{Bmatrix} \tag{C.7}$$

Now equating trisine and one dimensional density of states results in equation C.8.

$$\begin{aligned} K_C &= \frac{8\pi^3}{3\cdot cavity\cdot K_B^2} \\ &= \frac{4\pi}{3\sqrt{3}A} \end{aligned} \tag{C.8}$$

$\in =$ energy

$D(\in) =$ density of states

## Appendix D. Ginzburg-Landau Equation And Trisine Structure Relationship

This appendix parallels the presentation made in ref [4 Appendix I].

$$|\psi|^2 = \frac{\alpha}{\beta} = \frac{\mathbb{C}}{Cavity} \tag{D.1}$$

$$\frac{\alpha^2}{2\beta} = \frac{k_bT_c}{chain} = \frac{\hbar^2K_B^2}{2m_t\cdot chain} = \frac{H_c^2}{8\pi} \tag{D.2}$$

$$\lambda = \frac{m_tv_\varepsilon^2}{4\pi(\mathbb{C}\cdot e)^2|\psi|^2} = \frac{m_tv_\varepsilon^2\beta}{4\pi q^2\alpha} = \frac{m_tv_\varepsilon^2 cavity}{4\pi(\mathbb{C}\cdot e)^2\mathbb{C}} \tag{D.3}$$

$$\beta = \frac{\alpha^2 2m_t chain}{2\hbar^2K_B^2} = \frac{\alpha^2 chain}{2k_bT_c} \tag{D.4}$$

$$\alpha = \beta\frac{\mathbb{C}}{cavity} = \frac{\alpha^2 2m_t chain}{2\hbar^2K_B^2}\frac{\mathbb{C}}{cavity} = \frac{\alpha^2 chain}{2k_bT_c}\frac{\mathbb{C}}{cavity} \tag{D.5}$$

$$\alpha = k_bT_c\frac{cavity}{chain} = \frac{\hbar^2K_B^2}{2m_t}\frac{cavity}{chain} \tag{D.6}$$

$$\xi^2 \equiv \frac{\hbar^2}{2m_e\alpha} = \frac{\hbar^2}{2m_e}\frac{2m_t \cdot chain}{\hbar^2 K_B^2 \cdot cavity} = \frac{m_t}{m_e}\frac{1}{K_B^2}\frac{chain}{cavity} \qquad \text{(D.7)}$$

$$H_{c2} = \frac{\phi_\varepsilon \cdot 2 \cdot 4\pi}{section} = \frac{\phi_\varepsilon}{\pi\xi^2}\frac{m_t}{m_e}\frac{cavity}{chain} \qquad \text{(D.8)}$$

## Appendix E. Heisenberg Uncertainty within the Context of De Broglie Condition

Given Heisenberg uncertainty:

$$\Delta p \Delta x \geq \frac{h}{4\pi} \qquad \text{(E.1)}$$

And the assumption that mass 'm' is constant then:

$$m\Delta v \Delta x \geq \frac{h}{4\pi} \qquad \text{(E.2)}$$

Given de Broglie Condition:

$$p\ x = h \qquad \text{(E.3)}$$

And again assuming constant mass 'm':

$$mvx = h \qquad \text{(E.4)}$$

Rearranging Heisenberg Uncertainty and de Broglie Condition:

$$m \geq \frac{h}{4\pi}\frac{1}{\Delta v \Delta x}$$
$$m = h\frac{1}{vx} \qquad \text{(E.5)}$$

This implies that:

$$m \geq m \qquad \text{(E.6)}$$

but this cannot be (constant mass cannot be greater than itself) therefore:

$$m = m \qquad \text{(E.7)}$$

therefore:

$$h\frac{1}{vx} = \frac{h}{4\pi}\frac{1}{\Delta v \Delta x} \qquad \text{(E.8)}$$

And:

$$vx = 4\pi \Delta v \Delta x \qquad \text{(E.9)}$$

Both the Heisenberg Uncertainty and the de Broglie condition are satisfied. Uncertainty is dimensionally within the de Broglie condition. Schrödinger in the field of quantum mechanics further quantified this idea.

## Appendix F. Equivalence Principle In The Context Of Work Energy Theorem

Start with Newton's second law

$$F = \frac{dp}{dt} \qquad \text{(F.1)}$$

and for constant mass

$$F = \frac{dp}{dt} = \frac{d(mv)}{dt} = m\frac{dv}{dt} \qquad \text{(F.2)}$$

then given the 'work energy theorem'

$$F = \frac{dp}{dt} = \frac{d(mv)}{dt} = m\frac{dv}{dt}\frac{ds}{ds} = mv\frac{dv}{ds}$$

where $\frac{ds}{dt} = v$ (F.3)

and therefore:

$$Fds = mvdv$$

where $ds$ is test particles incremental distance along path $'s'$

Now given Newton's second law in relativistic terms:

$$F = \frac{dp}{dt} = \frac{d}{dt}\frac{mv}{\sqrt{1-\frac{v^2}{c^2}}} \qquad \text{(F.4)}$$

let Lorentz transform notation as '$b$' :

$$b = \frac{1}{\sqrt{1-\frac{v^2}{c^2}}} \qquad \text{(F.5)}$$

therefore:

$$F = \frac{dp}{dt} = \frac{d(mvb)}{dt} \qquad \text{(F.6)}$$

Now transform into a work energy relationshipwith a $K$ factor of dimensional units 'length/mass'

$$a\left(\frac{1}{K}\right)ds = mbvdv \qquad \text{(F.7)}$$

acceleration '$a$' can vary and the path '$s$' is arbitrary within the dimensional constraints of the equation.

Define acceleration *'a'* in terms of change in Volume *'V'* as:

$$\frac{d^2V}{dt^2}\frac{1}{V} = \text{constant} \qquad \text{(F.8)}$$

and incremental path 'ds' as:

$$ds = 2(area)dr = 2(4\pi r^2)dr \qquad \text{(F.9)}$$

*'area'* is the surface of the spherical volume *'V'*

Now work-energy relationship becomes:

$$\frac{d^2V}{dt^2}\frac{1}{V}\frac{1}{K}2(4\pi r^2)dr = mbvdv \qquad \text{(F.10)}$$

Now integrate both sides:

$$\frac{d^2V}{dt^2}\frac{1}{V}\frac{1}{K}2(\frac{4}{3}\pi r^3) = -\left(mc^2 - mv^2\right)b \qquad \text{(F.11)}$$

for $v << c$ then $b \sim 1$

and $V = \frac{4}{3}\pi r^3$

then:

$$-\frac{d^2V}{dt^2}\frac{1}{V}\frac{2}{K}=\left(\frac{mc^2}{V}-\frac{mv^2}{V}\right) \qquad (F.12)$$

and:

$$-\frac{d^2V}{dt^2}\frac{1}{V}=\frac{K}{2}\left(\frac{mc^2}{V}-\frac{mv^2}{V}\right) \qquad (F.13)$$

and:

$$-\frac{d^2V}{dt^2}\frac{1}{V}=\frac{K}{2}\begin{pmatrix}\frac{mc^2}{V}-\frac{mv}{area_x\ time}\\ -\frac{mv}{area_y\ time}\\ -\frac{mv}{area_z\ time}\end{pmatrix} \qquad (F.14)$$

This is the same as the Baez narrative interpretation except for the negative signs on momentum. Since momentum *'mv'* is a vector and energy '$mv^2$' is a scalar perhaps the relationship should be expressed as:

$$-\frac{d^2V}{dt^2}\frac{1}{V}=\frac{K}{2}\begin{pmatrix}\frac{mc^2}{V}\pm\frac{mv}{area_x time}\\ \pm\frac{mv}{area_y time}\\ \pm\frac{mv}{area_z time}\end{pmatrix} \qquad (F.15)$$

or perhaps in terms of spherical surface area 'arear'

$$-\frac{d^2V}{dt^2}\frac{1}{V}=\frac{K}{2}\left(\frac{mc^2}{V}\pm\frac{mv}{area_r time}\right) \qquad (F.16)$$

Lets look again at the equation F.7 and define another path *'s'*:

$$ds=16\pi rdr \qquad a=\frac{d^2r}{dt^2}=\text{constant} \qquad (F.17)$$

then

$$a\left(\frac{2}{K}\right)8\pi rdr=mbvdv \qquad (F.18)$$

Then integrate:

$$a\left(\frac{2}{K}\right)4\pi r^2=-\left(mc^2-mv^2\right)b \qquad (F.19)$$

Given that:

$$K=\frac{8\pi G}{c^2} \qquad (F.20)$$

and solve equation F.19 for *'a'*

$$a=-\frac{Gm}{r^2}\ \frac{1}{\text{b}} \qquad (F.21)$$

or

$$a=-\frac{Gm}{r^2}\ \sqrt{1-\frac{\text{v}^2}{\text{c}^2}} \qquad (F.22)$$

Does this equation correctly reflect the resonant gravitational acceleration such as a satellite orbiting the earth??

Of course this equation reduces to the standard Newtonian gravitational acceleration at 'v << c'

$$a=-\frac{Gm}{r^2} \qquad (F.23)$$

The equation F.7 would appear to be general in nature and be an embodiment of the equivalence principle. Various candidate accelerations 'a' and geometric paths 'ds' could be analyzed within dimensional constraints of this equation.

A simple case defined by a path *'mK ds'* may be appropriate for the modeling the Pioneer deceleration anomaly.

$$a\left(\frac{1}{K}\right)mKds=mads=Fds=mbvdv \qquad (F.24)$$

## Appendix G. Koide Constant for Leptons - tau, muon and electron

Koide Constant for Lepton masses follows naturally from nuclear condition as defined in paragraph 1.1.

$$\begin{aligned}
&\mathrm{m}_\tau = 1.777 \quad \mathrm{Gev}/c^2\\
&\mathrm{m}_\mu = .1057 \quad \mathrm{Gev}/c^2\\
&\mathrm{m}_e = .000511\ \mathrm{Gev}/c^2\\
&\mathrm{m}_\tau=\frac{2}{3}\hbar c\left(K_C+K_B\right)\frac{1}{c^2}\\
&\mathrm{m}_\mu=\frac{1}{3}\frac{m_e}{m_t}\frac{\hbar^2}{2m_t}\left(K_C^2+K_B^2\right)\frac{1}{c^2}\\
&\mathrm{m}_e=\frac{\pi}{\sqrt{2}}\frac{m_e}{m_t}\left(g_s-1\right)\hbar c\left(K_C+K_B\right)\frac{1}{c^2}\\
&\mathrm{m}_\tau=\frac{2}{3}\hbar c\left(K_{Ds}+K_{Dn}\right)\frac{1}{c^2}\\
&\mathrm{m}_\mu=\frac{1}{3}\frac{m_e}{m_t}\frac{\hbar^2}{2m_t}\left(K_{Ds}^2+K_{Dn}^2\right)\frac{1}{c^2}\\
&\mathrm{m}_e=\frac{\pi}{\sqrt{2}}\frac{m_e}{m_t}\left(g_s-1\right)\hbar c\left(K_{Ds}+K_{Dn}\right)\frac{1}{c^2}\\
&\frac{\mathrm{m}_\tau+\mathrm{m}_\mu+\mathrm{m}_e}{\left(\sqrt{\mathrm{m}_\tau}+\sqrt{\mathrm{m}_\mu}+\sqrt{\mathrm{m}_e}\right)^2}=\frac{chain}{cavity}=\frac{2}{3}
\end{aligned} \qquad (G.1)$$

## Appendix H. Authority for Expenditure (Cost Estimate)

Project Name____ Three Dimensional Laser Energetics Relating Trisine Geometry to the CPT theorem — Date ___ January 15, 2007

Location ________ Corpus Christi, Texas

Description ______ Apparatus and Support Facilities for 10 years

| | Completed Cost |
|---|---|
| PROJECT - INTANGIBLE COSTS | |
| Project Physicists | $10,000,000 |
| Project Chemists | $10,000,000 |
| Project Engineers | $10,000,000 |
| Technicians Draftsmen Electronic Electrical Specifications | $10,000,000 |
| Salaries management and office staff | $5,000,000 |
| Architectural Services | $500,000 |
| Landscaping | $250,000 |
| Maintenance | $400,000 |
| Office Library Supplies | $750,000 |
| Library Reference Searches | $750,000 |
| Patent Trademark Attorney Services | $1,000,000 |
| Legal Services | $500,000 |
| Financial Services | $500,000 |
| Travel | $1,000,000 |
| Publication Services | $500,000 |
| Marketing | $500,000 |
| Infrastructure maintainance | $500,000 |
| Utilities | $400,000 |
| Security | $300,000 |
| ______ | $0 |
| SubTotal | $52,850,000 |
| Contingencies 50.00% | $26,425,000 |
| PROJECT - INTANGIBLE COSTS | $79,275,000 |
| | |
| PROJECT - TANGIBLE COSTS | |
| Vibration Attenuation foundation | $1,000,000 |
| Laser Reactor Area 2500 sq ft $5,000 per sq ft | $12,500,000 |
| Assembly Area 10000 sq ft $500 per sq ft | $5,000,000 |
| Office Laboratory Library Space 10000 sq ft $200 per sq ft | $2,000,000 |
| Office Laboratory Library Furniture | $500,000 |
| Office Laboratory Computers CAD Software Plotters Displays Printers LAN | $500,000 |
| Heating Air Conditioning | $1,000,000 |
| Heavy Equipment movement and placement trollleys and lifts | $1,000,000 |
| Test Benches and Equipment Storage | $1,000,000 |
| Test and Measurement Equipment | $1,000,000 |
| Faraday and Optical Cage | $500,000 |
| Three Dimensional Positioning System | $5,000,000 |
| Laser Generators | $25,000,000 |
| Laser Optics, Polarizers and Rectifiers | $9,000,000 |
| Vacuum Enclosure for Laser and Optics | $5,000,000 |
| Vacuum Pumping Equipment | $5,000,000 |
| Cryogenic Equipment | $5,000,000 |
| Data Sensors and Acquisition Equipment | $5,000,000 |
| Computers and Servers for Analysing and Storing Data | $5,000,000 |
| Laser Interference Reactor | $25,000,000 |
| PROJECT - TANGIBLE COSTS | $115,000,000 |
| TOTAL PROJECT COSTS | $194,275,000 |
| | |
| LEASE EQUIPMENT | |
| truck and sedan Vehicle | $200,000 |
| Out Sourcing Services | $5,000,000 |
| ______ | $0 |
| ______ | $0 |
| TOTAL LEASE EQUIPMENT | $5,200,000 |
| TOTAL COST | $199,475,000 |

## 6. References

[1] William Melis and Richard Saam, "U. S. Patent 4,526,691", Separator Apparatus, 02 July 1985.

[2] John Bardeen, Leon Neil Cooper and John Robert Schrieffer, "Theory of Superconductivity", Physical Review, Vol 28, Number 6, December 1, 1957, pages 1175-1204.

[3] Erwin Schrödinger, "An Undulatory Theory of the Mechanics of Atoms and Molecules", Physical Review, Vol 108, Number 5, December, 1926, pages 1049-1069.

[4] Charles Kittel, Introduction to Solid State Physics, Seventh Edition, John Wiley & Sons, Inc., New York, N.Y., 1996.

[5] William A. Little, "Experimental Constraints on Theories of High Transition Temperature Superconductors", Department of Applied Physics, Stanford University, 1987.
Phys. Rev. 134, A1416-A1424 (1964)

[6] Leonard Eyges, "The Classical Electromagnetic Field", Dover Edition, New York, New York, 1980.

[7] Eugene Podkletnov and R. Nieminen, "A Possibility of Gravitational Force Shielding by Bulk $YBa_2Cu_2O_{7-x}$ Superconductor", Physica C 203, (1992), 414 - 444.

[8] Sir Arthur Eddington, "The Universe and the Atom" from The Expanding Universe, Cambridge University Press.

[9] Giovanni Modanese, "Theoretical Analysis of a Reported Weak Gravitational Shielding Effect", MPI-PhT/95-44, May 1995, Max-Planck-Institut für Physik, Werner-Heisenberg-Institut, Föhringer Ring 6, D 80805 München (Germany) to appear in Europhysical Letters.

[10] Giovanni Modanese, "Role of a 'Local' Cosmological Constant in Euclidian Quantum Gravity", UTF-368/96 Jan 1996, Gruppo Collegato di Trento, Dipartimento di Fisica dell'Universita I-38050 POVO (TN) - Italy

[11] J. E. Sonier et al, "Magnetic Field Dependence of the London Penetration Depth in the Vortex State of $YBa_2Cu_2O_{6.95}$", Triumf, Canadian Institute for Advanced Research and Department of Physics, University of British Columbia, Vancouver, British Columbia, Canada, V6T.

[12] T. R. Camp and P. C. Stein, "Velocity Gradients and Internal Work in Fluid Motion", Journal of the Boston Society Civil Engineers. 30, 219 (1943).

[13] Miron Smoluchowski, Drei Vorträge über Diffusion, Brownsche Molekularbewegung und Koagulation von Kolloidteilchen (Three Lectures on Diffusion, Brownian Motion, and Coagulation of Colloidal Particles), Phys. Z., 17, 557 (1916); Versuch einer Mathematischen Theorie der Koaguationskinetik Kolloider Lösugen (Trial of a Mathematical Theory of the Coagulation Kinetics of Colloidal Solutions), Z. Physik. Chem., 92, 129, 155 (1917).

[14] Gordon M. Fair, John C. Geyer and Daniel A. Okun, Water and Wastewater Engineering, John Wiley & Sons, Inc. New York, 1968, pages 22-9 to 22-14.

[15] Georg Joos, Theoretical Physics, translated from the first German Edition by Ira Freeman, Hafner Publishing Company, Inc., New York, 1934.

[16] Robert C. Weast and Samuel M. Selby, Handbook of Chemistry and Physics, 53 Edition, The Chemical Rubber Company, Cleveland, Ohio, 1972-73.

[17] D. R. Harshman and A. P. Mills, Jr., "Concerning the Nature of High-Tc Superconductivity: Survey of Experimental Properties and Implications for Interlayer Coupling", Physical Review B, 45, 18, 01 May 92.

[18] John K. Vennard, Fluid Mechanics, John Wiley & Sons, Inc., New York, NY, 1961.

[19] Walter J. Moore, Physical Chemistry, Prentice-Hall, Inc., Englewood Cliffs, NJ, 1962.

[20] Takeshi Fukuyama, "Late-Time Mild Inflation-a possible solution of dilemma: cosmic age and the Hubble parameter", Department of Physics, Ritsumeikan University, Kusatsu Shiga, 525-77 JAPAN, August 14, 1996.

[21] Hayasaka and Takeuchi, "Anomalous Weight Reduction on a Gyroscope's Right Rotations around the Vertical Axis on the Earth", Physical Review Letters, 63, 25, December 18, 1989, p 2701-4.

[22] Faller, Hollander, and McHugh, "Gyroscope-Weighing Experiment with a Null Result", Physical Review Letters, 64, 8, February 19, 1990, p 825-6.

[23] Nitschke and Wilmarth, "Null Result for the Weight Change of a Spinning Gyroscope", Physical Review Letters, 64, 18, April 30, 1990, p 2115-16.

[24] Donald J. Cram and George S. Hammond, Organic Chemistry, McGraw-Hill Book Company, New York, NY, 1964.

[25] Kimberly, Chanmugan, Johnson, & Tohline, Millisecond Pulsars Detectable Sources of Continuous Gravitational Waves, Astrophysics J., 450:757, 1995.

[26] Staggs, Jarosik, Meyer and Wilkinson, Enrico Fermi Institue, University of Chicago, Chicago, Illinois 60637, Sept 18, 1996.

[27] Copi, Schramm and Turner, Science, 267, January 13, 1995, pp 192-8.

[28] Albert Einstein, The Meaning of Relativity, Fifth Edition, MJF Books, New York, NY, 1956.

[29] Albert Einstein, The Special and the General Theory, Three Rivers Press, New York, NY, 1961.

[30] Wolfgang Pauli, The Theory of Relativity, Dover Publications, New York, NY, 1958.

[31] Robert M. Wald, General Relativity, University of Chicago Press, Chicago, 1984.

[32] Ray D'Inverno, Introducing Einstein's Relativity, Clarendo Press, Oxford, 1995.

[33] Paul A. M. Dirac, General Theory of Relativity, Princeton University Press, Princeton, NJ, 1996.

[34] Barrett O'Neill, The Geometry of Kerr Black Holes, A K Peters, Wellesley, MA, 1995.

[35] Charles W. Misner, Kip S. Thorne & John A. Wheeler, Gravitation, W. H. Freeman and Company, New York, 1998.

[36] Rutherford Aris, Vectors, Tensors, and the Basic Equations of Fluid Mechanics, Dover Publications, New York, 1962.

[37] Heinrich W. Guggenheimer, Differential Geometry, Dover Publications, New York, 1977.

[38] Hans Schneider, George P. Barker, Matrices and Linear Algebra, Dover Publications, New York, 1962.

[39] Bernard Schutz, Geometrical Methods of Mathematical Physics, Cambridge University Press, Cambridge, 1980.

[40] Richard P. Feynman, Six Easy Pieces, Addison-Wesley Publishing Company, New York, 1995.

[41] Jed Z. Buchwald, From Maxwell to Microphysics, University of Chicago Press, Chicago, 1973.

[42] Leonard Eyges, The Classical Electromagnetic Field, Dover Publications, New York, 1972.

[43] L. D. Landau and E. M. Lifshitz, The Classical Theory of Fields, Volume 2, Butterworth-Heinenann, Oxford, 1998.

[44] C. K. Birdsall and A. B. Langdon, Plasma Physics via Computer Simulation, Institute of Physics Publishing, Philadelphia, 1991.

[45] Peter W. Milonni, The Quantum Vacuum, An Introduction to Quantum Electrodynamics, Academic Press, New York, 1994.

[46] Claude Cohen-Tannoudju, Jacques Dupont-Roc & Gilbert

Grynberg, Photons & Atoms, Introduction to Quantum Electrodynamics, John Wiley & Sons, Inc., New York, 1989.

[47] Eugen Merzbacher, Quantum Mechanics, Third Edition, John Wiley & Sons, Inc., New York, 1998.

[48] James P. Runt and John J. Fitzgerald, Dielectric Spectroscopy of Polymeric Materials, Fundamentals and Applications, American Chemical Society, Washington, DC, 1997.

[49] Michael Tinkham, Introduction to Superconductivity, McGraw-Hill Companies, New York, NY, 1996.

[50] Eugene Hecht, Optics Third Edition, Addison-Wesley, New York, NY, 1998.

[51] C. C. Homes et al, Universal Scaling Relation in High-Temperature Superconductors. Nature, Vol 430, 29 July 2004. http://xxx.lanl.gov/abs/cond-mat/0404216

[52] Jan Zaanen, Why the Temperature is High, Nature, Vol 430, 29 July 2004.

[53] Study of the anomalous acceleration of Pioneer 10 and 11, John D. Anderson, Philip A. Laing, Eunice L. Lau, Anthony S. Liu, Michael Martin Nieto, and Slava G. Turyshev, 09 July, 2004
http://arxiv.org/abs/gr-qc/0104064

[54] Christopher C. Homes et al, Are high-temperature superconductors in the dirty limit
http://xxx.lanl.gov/abs/cond-mat/0410719

[55] N. Klein, B.B. Jin, J. Schubert, M. Schuster, H.R. Yi, Forschungszentrum Jülich, Institute of Thin Films and Interfaces, D-52425 Jülich, Germany, A. Pimenov, A. Loidl, Universität Augsburg, Experimentalphysik V, EKM, 86135 Augsburg, Germany, S.I. Krasnosvobodtsev, P.N. Lebedev Physics Institute, Russian Academy of Sciences, 117924 Moscow, Russia, http://xxx.lanl.gov/abs/cond-mat/0107259, Energy gap and London penetration depth of $MgB_2$ films determined by microwave resonator, Submitted to Physical Review Letters, June 29, 2001. $MgB_2$ superconductivity originally discovered by Nagamatsu et al. 2001 Superconductivity at 39 K in $MgB_2$, Nature 410 63

[56] Directly Measured Limit on the Interplanetary Matter Density from Pioneer 10 and 11 Michael Martin Nieto, Slava G. Turyshev and John D. Anderson
http://arxiv.org/abs/astro-ph/0501626

[57] Study of the Pioneer Anomaly: A Problem Set, Slava G. Turyshev, Michael Martin Nieto, and John D. Anderson http://arxiv.org/abs/astro-ph/0502123

[58] Heliocentric Trajectories for Selected Spacecraft, Planets, and Comets, NSSDC Goddard National Spaceflight Center, http://nssdc.gsfc.nasa.gov/space/helios/heli.html

[59] Scaling of the superfluid density in high-temperature superconductors, Christopher C. Homes, S. V. Dordevic, T. Valla, M. Strong
http://xxx.lanl.gov/abs/cond-mat/0410719

[60] Frank M. White, Viscous Fluid Flow, Second Edition, McGraw Hill. NY, NY, 1974

[61] Francis W. Sears and Mark W. Zemansky, University Physics, Third Edition, Addison-Wesley. Reading, MA, 1964.

[62] James L. Anderson, Thompson Scattering in an Expanding Universe,
http://xxx.lanl.gov/abs/gr-qc/9709034

[63] Michael Martin Nieto, Slava G. Turyshev, Measuring the Interplanetary Medium with a Solar Sail,
http://xxx.lanl.gov/abs/gr-qc/0308108

[64] Transport of atoms in a quantum conveyor belt, A. Browaeys, H. H¨affner, C. McKenzie, S. L. Rolston, K. Helmerson, and W. D. Phillips, National Institute of Standards and Technology, Gaithersburg, MD 20899, USA (Dated: April 23, 2005)
http://arxiv.org/abs/cond-mat/0504606

[65] R. Tao, X. Zhang, X. Tang and P. W. Anderson, Phys. Rev. Lett. 83, 5575 (1999).

[66] R. Tao , X. Xu , Y.C. Lan , Y. Shiroyanagi , Electric-field induced low temperature superconducting granular balls, Physica C 377 (2002) 357-361

[67] R. Tao, X. Xu X and E. Amr, Magnesium diboride superconducting particles in a strong electric field, Physica C 398 (2003) 78-84

[68] Didkovsky L. V., Rhodes E. J., Jr., Dolgushin A.I., Haneychuk V. I., Johnson N. M., Korzennik S.G., Kotov V. A., Rose P. J., Tsap T. T. : "The first results of solar observations made in the Crimean Astrophysical Observatory using a magneto-optical filter", 1996, Izv. Vuzov, ser. RADIOFIZIKA, Tom 39, No. 11-12, p. 1374-1380.

[69] Kotov V.A., Lyuty V. M., Haneychuk V. I. : "New evidences of the 160-minute oscillations in active galactic nuclei", 1993, Izv. Krym. Astrofiz. Obs., Tom 88, p. 47-59.

[70] Andrew Huxley, Critical Breakthrough, Science 309 1343, CEA laboratory in Grenoble and the Grenoble High Magnetic Field Laboratory (GHMFL).

[71] R. F. Klie, J. P. Buban, M. Varela, A. Franceschetti, C. Jooss, Y. Zhu, N. D. Browning, S. T. Pantelides and S. J. Pennycook, Enhanced current transport at grain boundaries in high-Tc superconductors, Nature p475, Vol 435, 26 May 2005

[72] Enhanced flux pinning in YBa2Cu3O7– films by nanoscaled substrate surface roughness Zu-Xin Ye, Qiang Li, Y. Hu, W. D. Si, P. D. Johnson, and Y. Zhu Appl. Phys. Lett. 87, 122502 (2005) (3 pages), BNL.

[73] Andrew W. Strong, Igor V. Moskalenko & Olaf Reimer, A New Determination Of The Diffuse Galactic and Extragalactic Gamma-Ray Emission
http://lanl.arxiv.org/abs/astro-ph/0506359

[74] F. L. Pratt, S. J. Blundell, Universal scaling relations in molecular superconductors, Phys. Rev. Lett. 94, 097006 (2005), http://arxiv.org/abs/cond-mat/0411754

[75] A. Einstein, Ann. Der Physik, 17, p. 549 (1905)

[76] Marusa Bradac, Douglas Clowe, Anthony H. Gonzalez, Phil Marshall, William Forman, ChristineJones, Maxim Markevitch, Scott Randall, Tim Schrabback, and Dennis Zaritsky, Strong And Weak Lensing United Iii: Measuring The Mass Distribution Of The Merging Galaxy Cluster 1e0657_56 arXiv:astro-ph/0608408 v1 18 Aug 2006

[77] Louis-Victor de Broglie (1892-1987) Recherches Sur La Theorie Des Quanta (Ann. de Phys., 10 serie, t. III (Janvier-Fevrier 1925). Translation by A. F. Kracklauer

[78] Merav Opher, Edward C. Stone, Paulett C. Liewer and Tamas Gombosi, Global Asymmetry of the Heliosphere, arXiv:astro-ph/0606324 v1 13 Jun 2006

[79] Massey R., et al. Nature, advance online publication, doi:10.1038/nature05497 (2007).

[80] Joel R. Primack. Precision Cosmology: Successes and Challenges, http://arxiv.org/abs/astro-ph/0609541, 19 Sep 06

Acknowledgements: Jiri Kucera, Joseph Krueger
Chris Tolmie, William Rieken,
Alan Schwartz